\documentclass{aa}
\usepackage{natbib}
\bibpunct{(}{)}{;}{a}{}{,}

\usepackage{graphicx}
\usepackage{subfigure}
\usepackage{txfonts}

\usepackage[normalem]{ulem}

\begin{document}

\title{MUSE observations of the blue compact dwarf galaxy Haro~14\thanks{Based on observations made with ESO Telescopes at Paranal Observatory under programme 
ID 60.A-9186(A)}}

\subtitle{Data analysis and first results on morphology and stellar populations}

\author{L. M. Cair\'os
    \inst{1}
    \and
    J.N. Gonz\'alez-P\'erez
    \inst{2}
    \and
    P. M. Weilbacher
    \inst{3}
    \and
    R. Manso Sainz
    \inst{4}
   }
\institute{Institut f{\"u}r Astrophysik, Georg-August-Universit{\"a}t,
Friedrich-Hund-Platz 1, D-37077 G{\"o}ttingen, Germany \\
           \email{luzma@astro.physik.uni-goettingen.de}
           \and Hamburger Sternwarte,
            Gojenbergsweg 112,
21029 Hamburg, Germany
\and Leibniz-Institut f{\"u}r Astrophysik (AIP),
An der Sternwarte 16, 14482 Potsdam, Germany 
\and Max Planck Institute for Solar System Research,
      Justus-von-Liebig-Weg 3,
       D-37077 G{\"o}ttingen, Germany 
}

\date{Januar 2021}

\abstract{Investigations of blue compact galaxies (BCGs) are essential to advancing our understanding of galaxy formation and evolution. BCGs are  low-luminosity, low-metallicity, gas-rich objects that form stars at extremely high rates, meaning they are good analogs to the high-redshift star-forming galaxy population. Being low-mass starburst systems, they 
 also constitute excellent
 laboratories
in which to investigate the star formation process and the interplay between massive stars and their surroundings. 
This work presents results from integral field spectroscopic observations of the BCG Haro\,14 taken with the Multi Unit Spectroscopic Explorer (MUSE) at the Very Large Telescope in wide-field adaptive optics mode. The large MUSE  field of view ($1\arcmin\times1\arcmin$=$3.8\times3.8$~kpc$^{2}$ at the adopted distance of 13~Mpc)  enables simultaneous observations of the central starburst  and the low-surface-brightness host galaxy, which is a huge improvement with respect to previous integral field spectroscopy of BCGs. From these data we built galaxy maps in continuum and in the brightest emission lines. We also generated synthetic broad-band images in the VRI bands, from which we produced color index maps and surface brightness profiles. We detected numerous clumps spread throughout the galaxy, both in continuum and in emission lines, and produced a catalog with their position, size, and photometry. This analysis allowed us to study the morphology and stellar populations of Haro 14  in detail. The stellar distribution shows a  pronounced asymmetry; the intensity peak in continuum is not centered with respect to the underlying stellar host but is displaced by about 500~pc southwest. At the position of the continuum peak we find a bright stellar cluster that with M$_{v}=-12.18$ appears as a strong 
super stellar cluster candidate. We also find a highly asymmetric, blue, but nonionizing  stellar component that occupies almost the whole eastern part of the galaxy. We conclude that there are at least three different stellar populations in Haro~14: the current starburst of about 6~Myr; an intermediate-age component of between ten and several hundred million years; and a red and regular host of several gigayears.  The pronounced lopsidedness in the continuum and also in the color maps, and the presence of numerous stellar clusters,  are consistent with a scenario of mergers or interactions acting in Haro~14.}


\maketitle
%

\section{Introduction}

Among the rich variety of galaxies,  blue compact galaxies (BCGs), which are low-mass systems 
 experiencing vigorous episodes of star formation (SF), constitute a relatively rare but 
extremely interesting galaxy type \citep{ThuanMartin1981,Papaderos1996a,Cairos2001b,Cairos2001a,GildePaz2003}.

\smallskip 

First, in the widely accepted scenario of a cold dark matter Universe, larger galaxies form hierarchically  from the assembly of smaller systems 
\citep{Kauffmann1993,Springel2006,Loeb2013}. Therefore, dwarf objects are essential to our understanding of galaxy formation and evolution. Within dwarfs, BCGs are particularly relevant because they have low chemical abundances  \citep{Izotov1999, Kunth2000}, high amounts of neutral gas \citep{ThuanMartin1981,Gordon1981,Salzer2002}, high specific  star formation rates  (SFR, \citealp{HunterElmegreen2004}), and clumpy morphology \citep{Cairos2001b,GildePaz2003}, meaning that these systems constitute precious analogs  to the
star forming galaxies observed at higher redshifts.

\smallskip 

Furthermore, BCGs are excellent 
targets with which to investigate the SF process as well as its trigger and propagation mechanism(s). The absence of spiral density waves or strong shear forces makes it possible to examine SF in a relatively simple environment \citep{Hunter1997} and, at the same time, provides evidence that an alternative mechanism must be in place. A lot of effort has been invested into finding out  how the SF in BCGs is triggered and maintained, 
but no conclusive findings have been found so far \citep{Brinks1990, Taylor1997,HunterElmegreen2004,Pustilnik2001,Brosch2004}. However, recent observational results indicate that dwarf--dwarf mergers, 
interactions, or infalling gas may play a major role in the ignition of the burst in low-mass systems  \citep{LopezSanchez2012,Nidever2013, Stierwalt2015,Privon2017,Pearson2018,Zhang2020a,Zhang2020b,Cairos2020}. 

\smallskip 

Finally, studies focusing on BCGs will help to understand the complex interplay between massive 
stellar feedback and 
the interstellar medium (ISM). The impact of stellar feedback 
is expected to be especially dramatic  in a low-mass system: the lack of density waves and significant shear forces allows the expanding shells created by stellar winds and supernovae (SNe) to grow to larger sizes and live longer than in typical spirals \citep{Tomisaka1981,McCray1987,Walter1999}, and in a shallow potential well, these shells can easily break out of the galaxy disk, or even the halo \citep{Heckman1990,DeYoung1994,MacLow1999,Veilleux2005}. Starburst-driven winds are indeed often invoked to explain fundamental questions such as 
the luminosity--metallicity relation  \citep{Garnett2002,Tremonti2004,Chisholm2018}, the chemical enrichment of the intergalactic medium \citep{Theuns2002,Springel2003}, or the formation of galaxies \citep{Larson1974,Dekel1986,Kauffmann1994,Naab2017}. However, in spite of its unquestionable importance, massive stellar  feedback remains a poorly understood phenomenon: its multi-phase nature and the complexity of the physics involved make it quite difficult to tackle,  both analytically and numerically \citep{McKee2007,Somerville2015,Naab2017}. High-quality data of nearby starburst galaxies  can  mitigate this problem  by providing observational constraints on the feedback parameters \citep{Martin1997,Martin1998,Calzetti1999,Calzetti2004}.

\smallskip 

Motivated by the relevance of these topics,  we initiated a project centered on the analysis of BCGs by means of integral field spectroscopy. The main goal of this project is to characterize the ongoing SF episode  in a large sample of starbursting dwarf galaxies and  to investigate its impact on the environment. Our  first results have convincingly demonstrated  that integral field spectrographs (IFS), providing simultaneously spectrophotometric and kinematic information for a large galaxy region, are the most advantageous  approach to probing small, highly asymmetrical, and compact systems such as BCGs  \citep{Cairos2009b,Cairos2009a,Cairos2010, Cairos2012,Cairos2015,Cairos2017a,Cairos2017b,Cairos2020}. 
The  advent of a new generation of wide-field IFSs in 8-m class telescopes considerably
broadens the possibilities in this research field. In particular, the 
Multi Unit
Spectroscopic Explorer (MUSE; \citealp{Bacon2010}), with its unique combination of high spatial and spectral resolution, large field of view (FoV), and extended wavelength range, has proven to be a powerful tool in BCG research \citep{Bik2015,Bik2018,Cresci2017,Kehrig2018,Menacho2019,Menacho2021}.

\smallskip 

This is the first in a series of publications presenting a spectrophotometric study of the BCG Haro~14 based on  MUSE observations. Haro~14, with a luminosity of M$_{B}=-16.92$  \citep{GildePaz2003} and a metallicity of  12+log(O/H) =8.2$\pm$0.1 (\citealp{Cairos2017a}, hereafter CGP17a), is a good representative of the BCG class. In this first paper, we describe our observations and data processing, and discuss our results on the morphology, structure, and stellar populations of the galaxy.

\section{Observations and data processing}

\subsection{The target galaxy: the BCG Haro~14}

Haro~14 is a nearby dwarf galaxy designated as a BCG  in the
pioneering work by  \cite{ThuanMartin1981}.
According to the morphological
classification by  \cite{Loose1987}, it belongs to the most common BCG type,
namely the iE BCGs---these comprise about 70\% of the BCG population.
Optical and near-infrared (NIR) imaging  reveals that Haro~14,  as the great majority of BCGs, is made of an irregular high-surface-brightness (HSB) 
region placed atop a smooth low-surface-brightness (LSB) underlying stellar component
\citep{Marlowe1997,Marlowe1999, Doublier1999,Doublier2001,GildePaz2003,GildePaz2005,Noeske2003,HunterElmegreen2006}. One distinct peculiarity of our galaxy  is its pronounced asymmetry: the HSB region is clearly off-center with  respect to the  outer isophotes---the intensity peak is displaced by about 500~pc southwest from the geometrical center of the galaxy (see Figure~\ref{Figure:Haro14_continuum} in this work and \citealp{Noeske2003}).  Integrated spectrophotometry in the optical has been carried out by \cite{Hunter1999} and \cite{Moustakas2006}. 

\smallskip

We included Haro~14 in our  study  of a sample of eight BCGs 
with the VIsible Multi-Object Spectrograph (VIMOS; \citealp{Cairos2015},CGP17a). These works showed  that the stellar and ionized gas distribution of the galaxy are significantly different: while in continuum the intensity peaks in two knots close to the galaxy center, the emission-line maps reveal  an extended ongoing starburst resolved in numerous SF knots distributed over the whole mapped area (about 1.7$\times$1.7~kpc$^{2}$; see Figure~\ref{Figure:Haro14_continuum}). The peaks in emission and continuum maps, which are considerably separated from each other, trace two spatially decoupled episodes of SF in the galaxy.  Unfortunately,  the VIMOS FoV does not completely cover the extended starburst and we find a plethora of
incipient filaments and shells, a large concentration of dust, and evidence of shocked regions at the edge of the observed field, calling for integral field spectroscopy in a larger FoV.

\begin{table}
\caption{Basic parameters for Haro\,14
\label{tab:data}}
\begin{center}
\begin{tabular}{lcc}
\hline
Parameter   & Value & Reference \\ 
\hline
\\
RA (J2000)                                         &  00$^h$45$^m$46$\fs$4 &    \\
Dec (J2000)                                         &   $-15\degr$35$\arcmin$49$\arcsec$  &     \\
 Distance                        &  13.0$\pm$0.1       Mpc                                             &     \\
 Spatial scale                  & 63~pc arcsec$^{-1}$ &                                     \\
  D$_{25}$                   & 4.65~kpc                                            &    RC3 \\
 m$_{B}$                     & 13.65$\pm$0.05$^{a}$ &  GP03 \\
 A$_{B}$                         &       0.075                                                          &     \\
 M$_{B}$          & $-16.92^{b}$              &  \\
 M$_{HI}$         & 3.2$\times$10$^{8}$M$\odot$ & TM81  \\
 M$_{T}$          & 3.8$\times$10$^{8}$M$\odot$ & TM81  \\ 
 12+log(O/H)      & 8.2$\pm$0.1     & CG17      \\
Morphology                       & SOpec, BCG, iE BCD                                                   & \\

Alternatives names             & NGC 0244, UGCA 10 & \\
                                          & VV~728, PGC~2675                                  & \\
 \hline
\end{tabular}
\end{center}
\small Notes:
RA, DEC, distance, apparent major isophotal diameter D$_{25}$ measured
at a surface brightness level of 25.0 mag~arcsec$^{-2}$ and Galactic extinction are
taken from NED (http://nedwww.ipac.caltech.edu/). The distance was
calculated using a Hubble constant of 73~km s$^{-1}$ Mpc$^{-1}$, and
taking into account the influence of the Virgo Cluster, the Great
Attractor and the Shapley supercluster.  (a) integrated magnitude from
\cite{GildePaz2003}, corrected for Galactic extinction; (b) absolute
magnitude in the B-band computed from the tabulated B magnitude and
distance. References: CG17: \cite{Cairos2017a}; RC3: \cite{deVaucouleurs1991}; GP03: \cite{GildePaz2003};
TM81:\cite{ThuanMartin1981}.  
\end{table}

\subsection{Data acquisition}

Haro\,14 was observed with 
MUSE \citep{Bacon2010} 
at the Very Large Telescope (VLT; ESO Paranal Observatory, Chile). MUSE is a panoramic IFS which, operating in its  wide field mode (WFM), provides  a FoV of $1\arcmin\times1\arcmin$ with a  
spatial sampling of 0\farcs2.

\smallskip 

The observations were performed on September 2017 as part of the 
MUSE Adaptive Optics Facility (AOF)  science verification run \citep{Leibundgut2017}.  With the AOF, the MUSE WFM is supported by ground layer adaptive optics (GLAO) through four artificial  laser guide stars and the adaptive optics system GALACSI (ground atmospheric optics for spectroscopic imaging). 

\smallskip 
The data were obtained  in nominal mode (wavelength range 4750-9300~\AA) with a spectral sampling of  about 1.25\AA\,pix$^{-1}$ in
dispersion direction and an average  resolving power of R$\sim$3000. There is a gap between $\sim$5800 and 5950~\AA\ caused by the Na~D blocking filter, because the AOF works with four sodium lasers and, otherwise, the detector would saturate. 
We took
four 1370\,s exposures of Haro\,14, each rotated by 
90$^{\circ}$ with respect to the previous one (a total of 5480~s on source). Because the target fills most of the MUSE FoV, we took separate sky fields of 120~s between the science exposures.

\subsection{Data reduction}

The data were processed using the standard MUSE pipeline
\citep{Weilbacher2016,Weilbacher2020} working within the
{\sc esorex} environment with the default set of calibrations. The reduction 
follows the standard steps: bias and
flat-field correction,  tracing of the data on the CCD, and wavelength calibration. We applied geometric and astrometric calibrations from April 2017
and carried out a secondary spatial flat-field
correction using a twilight cube. The flux calibration was carried out  using exposures of the  
 standard star
EG\,274, taken  with the same instrumental mode the evening of the observations.

\smallskip

We determined the sky continuum from the offset sky field. In order to extract the sky lines we used  the science exposures: we fit the lines in the outskirts of the field using a line spread function from June 2017 obtained for the same instrumental mode.

\smallskip 

After correcting to
barycentric velocities, we created cubes for all four science exposures.
We used the white-light images to manually determine the centroid for the
brightest continuum source in the field, and computed offsets relative to its
Gaia DR1 position \citep{Lindegren2016}. We then combined all exposures using FWHM-based weighting.
The final  cube was sampled at the natural MUSE binning of
$0\farcs2\times0\farcs2\times1.25~\AA\,\text{pixel}^{-1}$, resulting in a cube
of $323\times322\times3681$ voxels.
The effective image quality is difficult to estimate because no bright foreground
stars are visible in the field.  From a fit of a Moffat profile to the brightest continuum source,
we estimate that the spatial resolution is 
$\sim0\farcs80$ in the red 
and $\sim1\farcs00$ in the blue part of the spectrum.

\subsection{Sky subtraction}

The sky subtraction of the standard pipelines of integral field instruments often leaves significant systematic residuals. These residuals, which arise mainly from the imperfect estimation of the profiles of the telluric lines,  can distort or even dominate the  spectrum when the source is faint. An accurate modeling of the contribution of telluric lines is difficult because of the intrinsic spatio-temporal variability
of the sky emission lines, imperfections of the spectrograph optics (e.g., distortions, flexures, and aberrations;  \citealp{Davies2007,Hart2019}), the different characteristics of the optical fibers (in multi-fiber spectrographs), or the different optical paths from slice to slice (in spectrographs like 
MUSE with an image slicer). If the amplitude of the sky emission is underestimated (overestimated), emission (absorption) line residuals appear, whereas imperfect wavelength calibration produces P-Cygni-like profiles.
Such miscorrections are particularly severe when the spectral lines are poorly sampled by the spectrograph, as is the case in MUSE (see, e.g., Figure~3 in \citealp{McLeod2015} or Figure~2 in \citealp{James2020}).

\smallskip

Because we aim to accurately measure the flux in the LSB regions of Haro~14,  we need to mitigate systematic residuals from telluric lines.
The systematic errors left by the sky subtraction are correlated across extended wavelength ranges and  
the principal component analysis (PCA) has proven to be  an excellent tool for the correction of residuals of telluric lines both for multi-fiber and image slicer IFS \citep{Kurtz2000,Wild2005,SharpParkinson2010,Hart2019,Soto2016}.
We implemented a PCA-based method for the removal of telluric lines that is similar to the ZAP tool developed by \cite{Soto2016}. Our routine consists of the following steps:

\smallskip

1)  {\em Selection of the spaxels that contain only sky background}. We search for sky-dominated spaxels, namely those without continuum or nebular line emission. These are defined as having a maximum H$\alpha$ flux of 1.5$\times$10$^{-20}$~erg~s$^{-1}$~cm$^{-2}$ and a spectral flux density below 1.5$\times$10$^{-20}$~erg~s$^{-1}$~cm$^{-2}$~\AA$^{-1}$ in the continua adjacent to H$\alpha$ and [\ion{O}{iii}]$\lambda5007$. To this end, we first run a test trial line fit to the data cube. We found 1322 spaxels with fluxes dominated by the sky emission.

\smallskip

2)  {\em Subtraction of the continuum to keep only the lines}. We remove the continuum of all spaxels in the image, including those dominated by the galaxy emission,
applying a running-median filter to all spectra   and then subtracting the original one from the median filtered spectra. The box-size of the median filter is 100 pixels (125 \AA) which is small enough to estimate the shape of the continuum but large enough not to be affected by the spectral lines.

\smallskip

3) {\em Calculation of the spectral eigenvectors}. The principal components, or eigenvectors, are calculated using the {\sc Matlab} routine {\tt PCA} with the singular value decomposition algorithm. We carefully  check the resulting eigenvalues and
remove from the pool of sky spaxels those whose eigenvalue dominates over all the others---we thus removed 22 ill-behaved spaxels that are dominating only one principal component.

\smallskip

4) {\em Selection of the principal components to be used for the reconstruction}. We select the number of components to be used following the same approach as \cite{Soto2016}. We use the explained variance for an increasing number of components and search for the point at which it reaches the linear regime. As a result, we choose the first 20 eigenvectors for the telluric correction.

\smallskip

5) {\em Calculation of the correction to be applied to all spaxels}.  We estimate the correction to be applied to all spaxels by first calculating the dot product of the continuum-subtracted spectrum of the spaxel with each eigenspectrum. This gives us the components that are then multiplied by the eigenspectra to obtain the telluric correction. This correction is finally subtracted from the original spectrum. 

\begin{figure*}
\centering
\begin{subfigure}{}
\includegraphics[width=8cm]{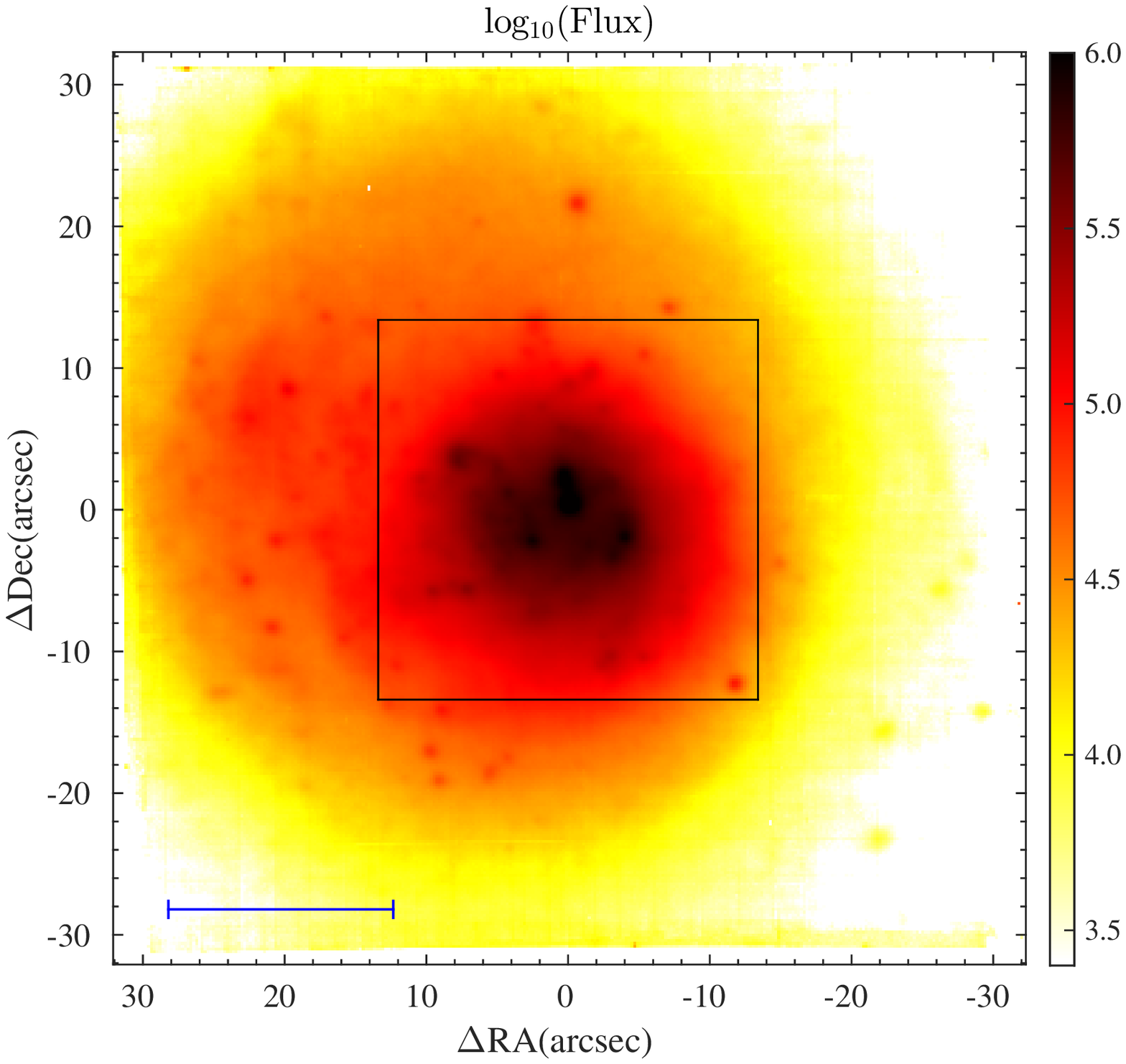}
 \end{subfigure}
\begin{subfigure}{}
\includegraphics[width=8cm]{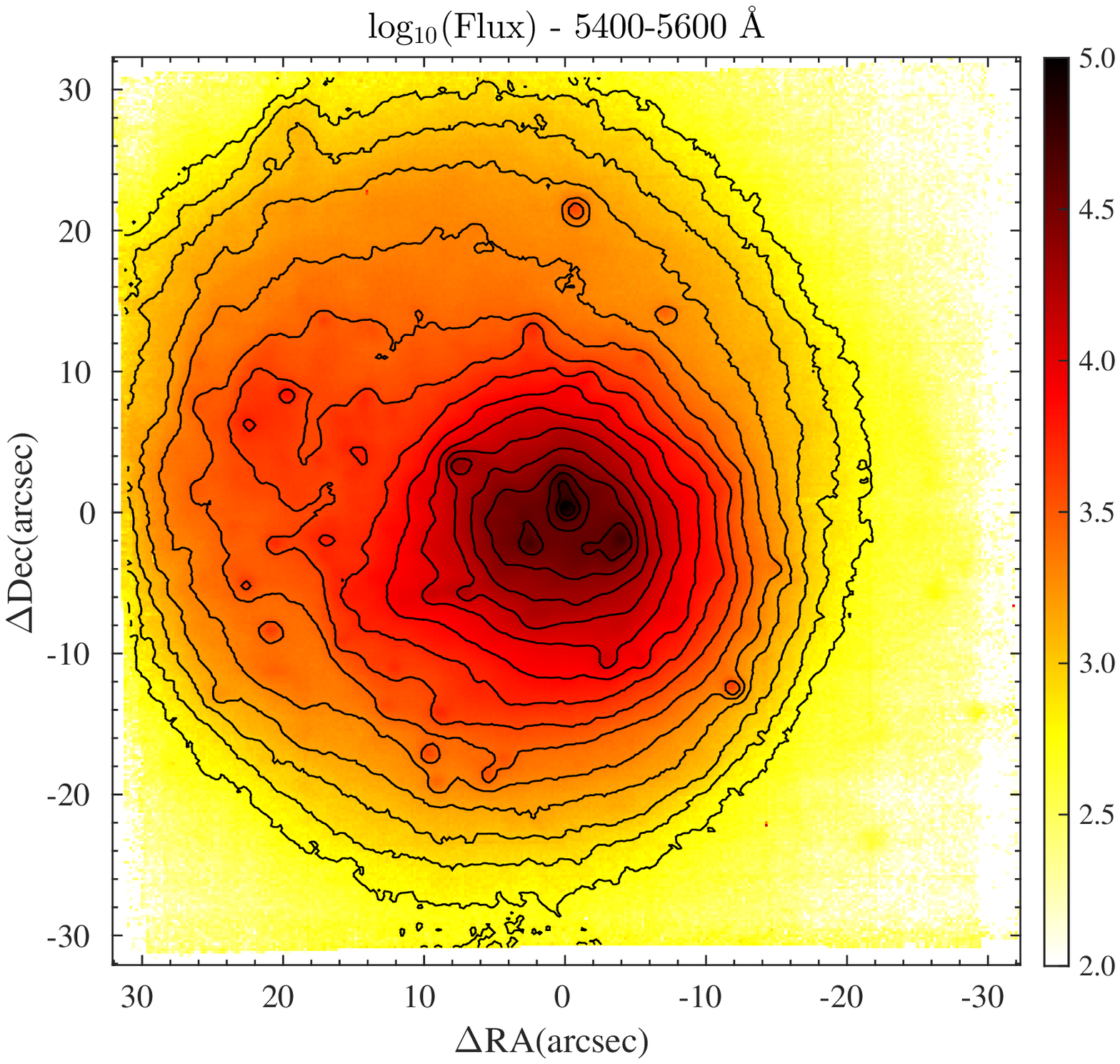}
 \end{subfigure}
\begin{subfigure}{}
\includegraphics[width=8cm]{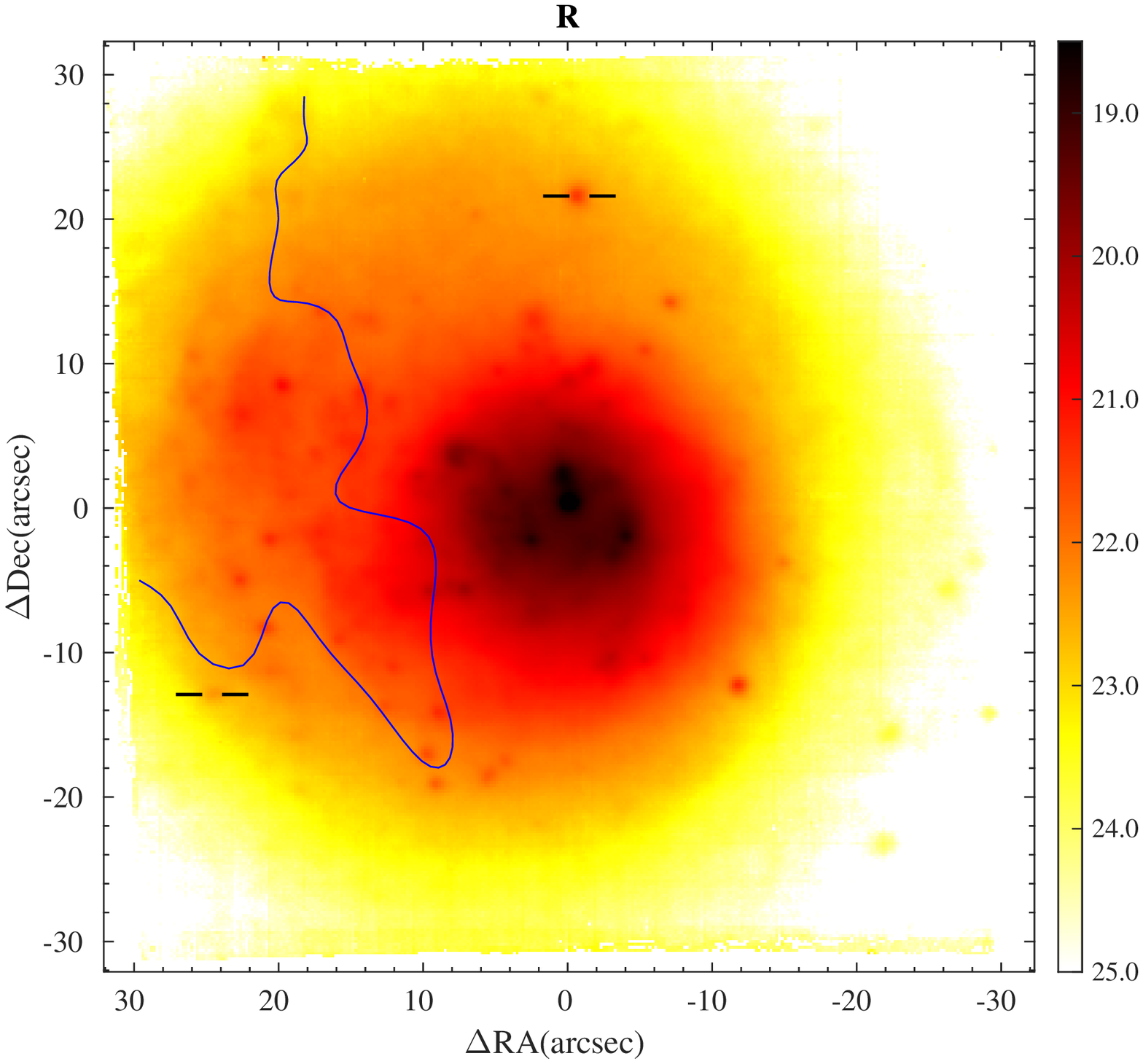}
 \end{subfigure}
 \begin{subfigure}{}
\includegraphics[width=8cm]{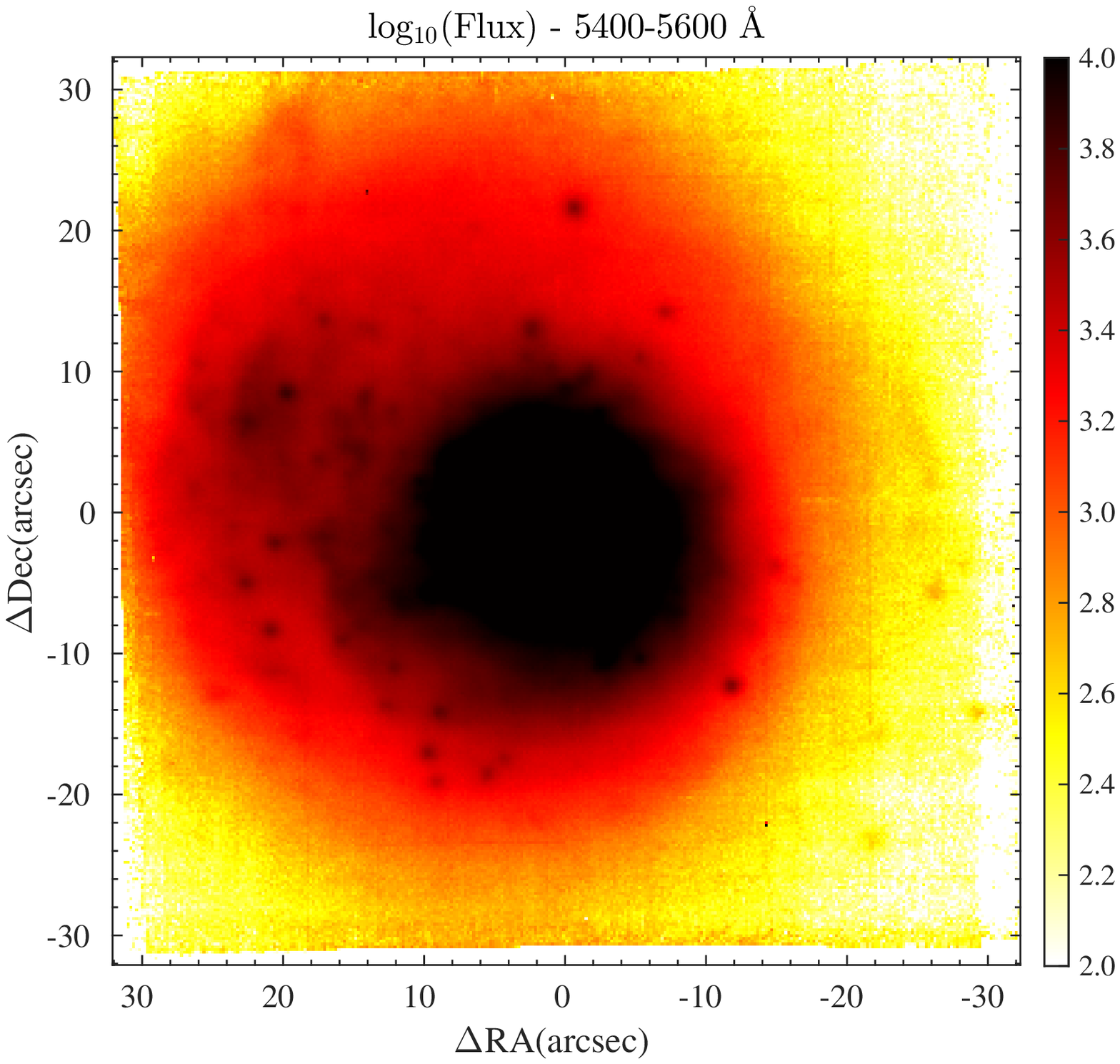}
\end{subfigure}
\caption{Haro\,14 as seen in broadband by MUSE. 
The FoV is $\sim$ 3.8$\times$3.8~kpc$^{2}$ and 
north is up and east to the left in all the maps presented in this paper.
{\em Top-left:} Image obtained by integrating over the whole observed spectral range.
The blue bar corresponds to 1~kpc at the distance of Haro 14.
The central square marks the FoV of our previous VIMOS observations.
{\em Top-right:} ``Pure continuum'' image in the wavelength range  5400 to 5600~\AA\ 
with contours over-plotted. The center of the outermost isophotes lies roughly 
around the position $\Delta$RA$\simeq+5"$, $\Delta$Dec$\simeq+2"$.
{\em Lower-right:} Saturated pure continuum image 
to better visualize the dimmer external parts of the galaxy.
{\em Lower-left:} Synthetic image in the filter $R$. 
The color bar shows the calibrated surface brightness in units of mag~arcsec$^{-2}$, 
the ticks mark the position of two background emission galaxies (see Appendix) and 
the blue line roughly outlines the bluish regions of the galaxy at intermediate intensity levels (see Section~\ref{Section:colormaps}).
There are no foreground stars in the observed FoV.
}
\label{Figure:Haro14_continuum} 
\end{figure*}

\subsection{Generating the galaxy maps}
\label{Section:making_maps}

The next step  in the data process is to associate each of the wavelength- and flux-calibrated spectra with its corresponding spatial position in the galaxy, that is,  to  generate  reconstructed images (galaxy maps); we can create maps in any selected spectral window within the observed wavelength range. 

\smallskip

We built  an image of Haro~14 by integrating over the whole spectral range (see Figure~\ref{Figure:Haro14_continuum}) and continuum maps by integrating the flux in specific spectral windows selected as to avoid strong emission lines or residuals from telluric lines (e.g., 5400-5600~\AA, 6100-6300~\AA, or 7800-8000~\AA).  These "pure" continuum maps are particularly suited to disentangling  the different stellar populations  in the inner galaxy regions, where a significant fraction of the light is due to lines originating from the warm ionized gas.

\smallskip 

We also generated maps of Haro~14 in the brighter emission lines, namely H$\alpha$ and [\ion{O}{iii}]$\lambda5007$. We computed the 
emission-line fluxes by fitting Gaussians to the line profiles. 
The fit
was carried out with the {\em Trust-region} algorithm for nonlinear least squares
using the {\em fit}  function of {\sc Matlab}; reasonable initial values and lower and
upper bounds for the parameters were given. The fitting algorithm provides the 
relevant parameters of the emission lines (line flux, centroid position, line
width, and continuum)  as well as their associated errors.

\smallskip 

 A known problem when deriving  emission line fluxes of the Balmer lines is the presence of stellar absorption \citep{McCall1985,Olofsson1995}. This contribution becomes increasingly important for higher terms of the Balmer series \citep{GonzalezDelgado1999a,GonzalezDelgado1999b}. As here we deal only with emission line maps in H$\alpha$, we have not attempted to correct for underlying stellar absorption.

\section{Morphology and stellar populations}
\label{Section:results}

The spectra of star forming galaxies are rich in emission lines generated 
in the warm ionized gas, 
mostly recombination lines of hydrogen and helium and forbidden lines of trace species \citep{Aller1984,Osterbrock2006}.
In small starburst systems, such as BCGs, the optical morphology is often dominated by the nebular emission,  which very much complicates the study of their stellar populations  \citep{Izotov1997,Izotov2011,Cairos2001b,GildePaz2003,Papaderos1998,Papaderos2002,Guseva2001}. This problem is  
substantially mitigated  by means of integral field observations: 
IFS data enable the generation of "pure continuum" maps, namely maps in spectral windows free from bright nebular lines, and emission line maps, separately; 
this allows us to locate the youngest star-forming regions and distinguish them  from more evolved post-starburst and old stellar components.

\subsection{Continuum maps}
\label{Section:continuum_morphology}

Figure~\ref{Figure:Haro14_continuum}  shows a map of 
Haro~14 generated by integrating the light over the full observed range (left panel), and a pure continuum map built by integrating the light in the spectral window  5400 to 5600 \AA\ (right panel). Both maps are very similar: a HSB region with three major emitters appears, surrounded by an extended LSB component--- differences between the two maps are only evident in the central HSB area.
The inner regions reproduce the  continuum map presented in CGP17a, but  more emission knots (presumably stellar clusters) are resolved in
the higher spatial resolution MUSE maps.   The much larger FoV of MUSE ($1\arcmin\times1\arcmin$=3.8$\times$3.8~kpc$^{2}$ at the adopted distance of 13.0~Mpc) allows us  to reach the galaxy outskirts and trace regions of very low surface brightness.

\smallskip

Haro~14 presents  a peculiar morphology. 
The LSB  host, which is mostly smooth and reasonably well described  by elliptical  isophotes (see section~\ref{Section:structure} below), almost fills   the MUSE FoV.  The HSB
region is off-centered with respect to the elliptical host: the intensity peak is not located at the geometrical center of the outermost isophotes, but is significantly displaced ($\sim$0.5~kpc) to the southwest. From this peak, the light decreases unevenly, so that the surrounding region of intermediate intensity level  has a rather irregular, diffuse and filamentary appearance. A structure resembling a tail spreads northeast, extending about $\sim$30$\arcsec$ ($\approx$1.9~kpc) from the continuum peak. This structure is highly visible in the pure continuum
images, indicating that it is not due to filaments of ionized gas but traces mostly a nonrelaxed stellar component (see right panels of Figure~\ref{Figure:Haro14_continuum}).

\smallskip 

A swarm of blobs is visible over the whole field. They are unevenly distributed and more numerous to the east and in the region of intermediate surface brightness. Such spatial distribution suggests that most of them  belong to the galaxy---they are neither foreground stars nor background objects (but see section~\ref{Section:clumps}).

%
\subsection{Broad band photometry}
\label{Section:bb_photometry}

The spectral range of MUSE makes it possible to generate synthetic broad-band images in various passbands of standard photometric systems. We can build, for example, images of Haro~14 in the V, R, and I filters of the Johnson-Cousins UBVRI  system \citep{Bessell1990}, or in the {\em r'i'} of the Sloan Digital Sky Survey photometric system \citep{Fukugita1996}.  This is quite a useful feature, because it  makes the photometric calibration of the maps straightforward, which  facilitates the comparison with  results from the literature as well as with model predictions. 

\smallskip

We generated Johnson-Cousins VRI images of Haro~14 integrating 
the fluxes in each spaxel, taking into account the transmission curve of the corresponding filter. 
Before this, we calculated the  flux in the spectral region 5800-5970 \AA\ (the region filtered out due to the  Na AO laser) performing a  linear interpolation of the spectrum. The synthetic image of Haro~14 in the R band is displayed in 
Fig.~\ref{Figure:Haro14_continuum} (lower left panel) . 

\smallskip 

The apparent and absolute magnitudes of Haro~14 are presented in Table~\ref{tab:photometry}---integrated magnitudes are corrected from Galactic extinction using the reddening coefficients from \cite{Schlafly2011}. As zero-point fluxes, we used the values presented in  the Table~A2 of \cite{Bessell1998}. 
The errors in the photometry are dominated by the flux calibration, which can contribute up to
5\% \citep{Weilbacher2020}; this translates to errors of up to 0.03~mag 
when integrating over a broad-band filter. Table~\ref{tab:photometry} also provides available photometric measurements from the literature: there is reasonably good agreement between the magnitudes derived from the MUSE data cube and those in other works.
To facilitate the comparison, we have not corrected for emission lines. 
This correction can be performed following the procedure explained in the following section but
the increment of integrated magnitude is very low (0.01 in V, and 0.02 in R; there are no important emission lines in I).

\begin{figure*}
\centering
\begin{subfigure}{}
\includegraphics[width=8cm]{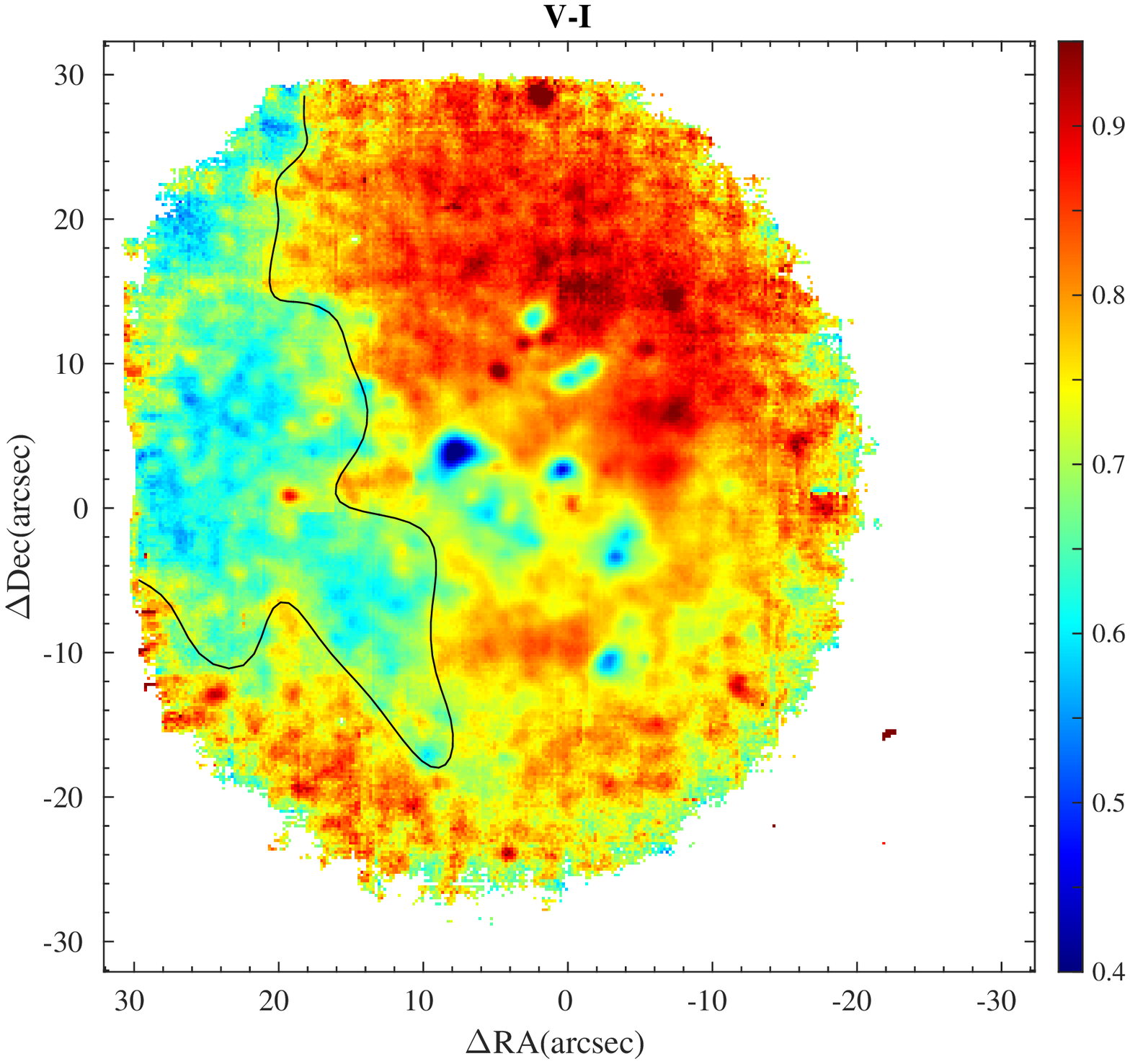}
\end{subfigure}
\begin{subfigure}{}
\includegraphics[width=8cm]{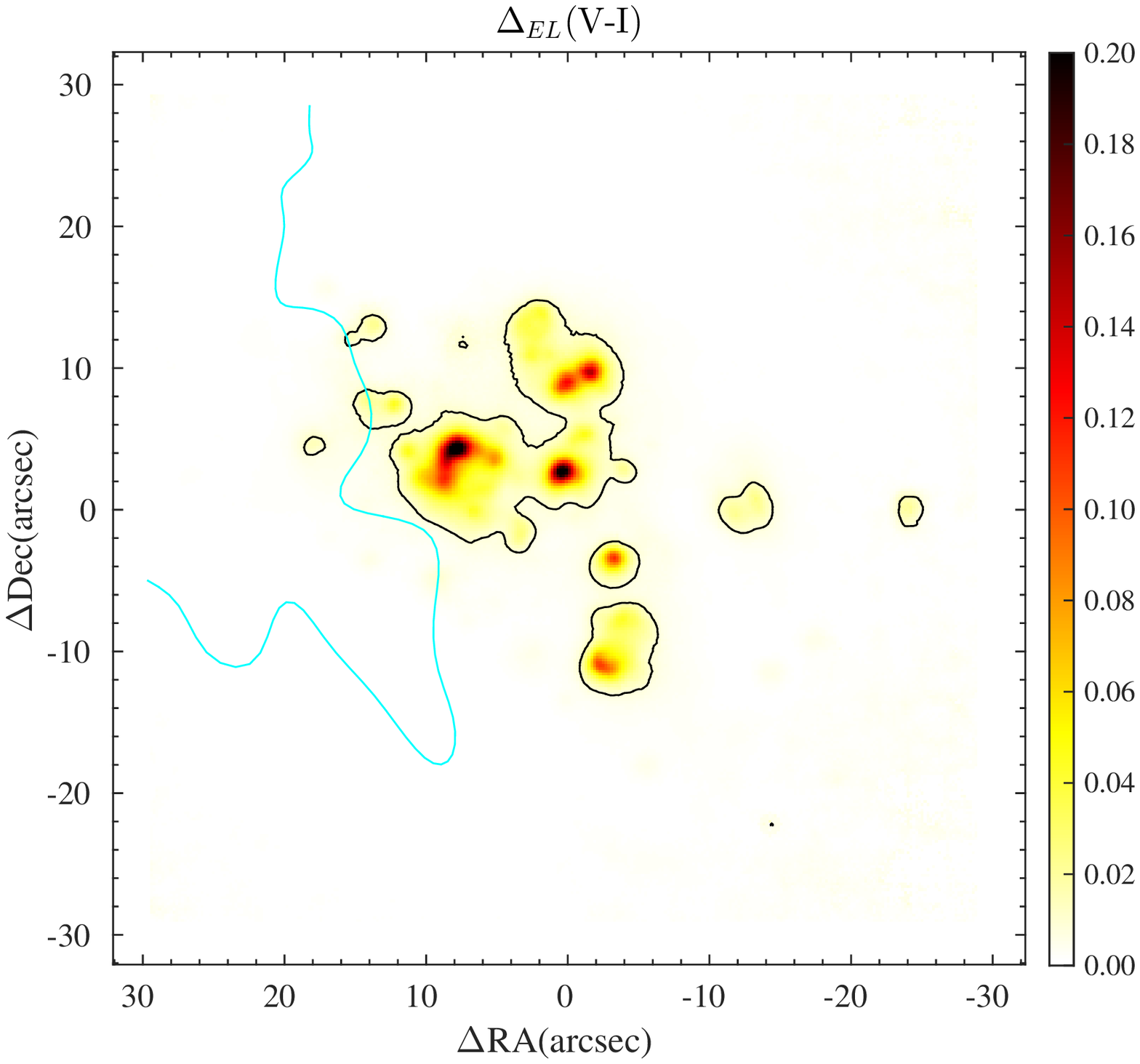}
 \end{subfigure}
\caption{Color spatial pattern in Haro\,14.  {\em Left:}  $(V-I)$ color index map; bluer colors indicate the bluer (younger) SF regions  and red corresponds to the redder (older) stars in the galaxy. The map is not corrected for the contribution of emission lines. {\em Right:} Increment in the (V-I) color index due to the contribution of emission lines. The blue line roughly delineates the bluish but not ionizing stellar component  unveiled in the color map and the black contour corresponds to  $\Delta_{L}(V-I)=0.01$. }
\label{Figure:Haro14_broadband} 
\end{figure*}

\begin{table}
\caption{Integrated photometry of Haro~14.\label{tab:photometry}}
\begin{center}
\begin{tabular}{ccccc}
\hline
Param.   & V & R  & I & Reference \\ 
 & (mag) & (mag) & (mag) & \\
\hline
m                                                          &  13.24                      & 12.90                  & 12.53    &  This work\\
M                                                           & \!\!\!\!$-17.33$        & \!\!\!\!$-17.67$    & \!\!\!\!$-18.04$   & This work \\
\hline
m                                                           & 13.21                    & ---                        & 12.49 & \citealp{Marlowe1997} \\
m                                                           & ---                          & 12.96                & ---       & \citealp{GildePaz2003}\\
m                                                           & ---                          & 12.76                 & ---       & \citealp{Doublier1999}\\
\hline
\end{tabular}
\end{center}
\small Notes. All magnitudes are corrected from Galactic extinction using the reddening coefficients from \cite{Schlafly2011}. The bottom lines provide photometry from the literature for comparison. \cite{Marlowe1997}  computed magnitudes in a circular aperture of diameter 78~$\arcsec$,   \cite{GildePaz2003}  computed magnitudes in a polygonal aperture containing the total integrated light  in the filter,  and  \cite{Doublier1999} derived asymptotic magnitudes.
\end{table}

\subsection{Color index maps}
\label{Section:colormaps}

Color index maps of a galaxy are created by dividing two galaxy frames in different spectral regions.  These  maps  
are effective tools with which to investigate compound systems as starburst galaxies, because they permit a clear identification of the \ion{H}{ii}-regions and stellar clusters and thus the spatial discrimination between the younger stellar populations and those that are more evolved. In addition, they can be used to delineate the dust patches and lanes that are frequently found in areas of active SF. 

\smallskip

We built color index maps of Haro~14 from the flux-calibrated broadband  frames. The $(V-I)$ color map  shown in the left panel of Figure~\ref{Figure:Haro14_broadband} presents a very peculiar spatial pattern.  There is, as expected, a strong color gradient, but the 
spatial distribution is unlike that typical of BCGs. 
While the color maps of BCGs usually exhibit one or more blue SF regions on top of a redder, much more extended host \citep{Cairos2001b, Cairos2003, Cairos2007,Janowiecki2014,Koleva2014}, 
in Haro~14 almost the entirety of the eastern part of the galaxy is nearly uniformly blue. Such  large-scale  asymmetries in the color maps of galaxies  (excluding the starburst knots)  are indeed unusual and are suggestive of mergers and/or interactions \citep{Zheng2004,Bassino2017,Xu2020}. 

\smallskip

A well-known problem when interpreting the broadband photometry of starbursts is the contamination by emission lines: the light from strong lines generated in the warm ionized gas adds to the flux in broadband filters and can significantly  modify the integrated colors and color patterns of galaxies \citep{Krueger1995,Zackrisson2001,Anders2003,Cairos2002,Cairos2007}. Hence, a reliable photometry for the stellar component requires the quantification of the gas emission. As this gaseous contribution presents high spatial variability,  the correction by means of traditional observing techniques is troublesome and observationally demanding: in addition to the broad-band frames, we need narrow-band imaging in each bright emission line or, alternatively, a sequence of long-slit spectra sweeping the starburst region. This, in turn, introduces uncertainties associated to the combination of frames taken with different instrumental setups and/or atmospheric conditions, or from the positioning of the slit \citep{Guseva2001,Guseva2003a,Guseva2003b,Guseva2003c,Guseva2004,Cairos2007}.

\smallskip 
Integral field observations with MUSE  overcome these problems in an efficient and precise manner. MUSE provides wide field imaging and spectroscopy simultaneously for every element of spatial resolution in a large spectral range. This offers several ways to get rid of the  emission line contribution: we can generate
images in the brightest lines and delimit the areas affected by gas; we can build color maps using galaxy images in spectral regions free of emission lines (the "pure continuum" maps presented in Section~\ref{Section:making_maps}); and  also,  we can precisely quantify  the contribution of the emission lines in each standard broadband filter.

\smallskip

That the blue color in the  eastern sector of Haro~14 is intrinsic and not due to gas emission into the broadband filters can be readily proven from the analysis of the spatial distribution of  emission lines. The emission line maps (Figure~\ref{Figure:emission}) clearly  show that the ionized gas is concentrated in the central SF regions and its influence on the colors of the outer galaxy regions is minor.  Furthermore, we generated color maps from pure continuum slices, that is, using spectral windows free from emission lines (see Section~\ref{Section:making_maps}). We compared color maps with and without emission lines and find that, as expected, the nebular contribution is only important in the SF regions and the color spatial pattern remains the same in both cases. 

\smallskip 

We also computed the total flux of strong emission lines in every synthetic broadband filter and thus the contribution of the gas to the color maps. This is a simple task working with integral field data,
which provide the full spectrum $F_{\rm obs}(\lambda)$ for each spatial resolution element. 
We obtained the flux, $F_\ell(\lambda)$, due to the emission in the $\ell=1, ..., N_\ell$ strongest lines by fitting their profiles (we considered here the two strongest recombination lines of H~{\sc i} and the 
[O~{\sc iii}], [N~{\sc ii}], and [S~{\sc ii}] forbidden lines; $N_\ell=8$ in all). The corrected flux
for every element of spatial resolution is %
\begin{equation*}
F_{\rm corr}(\lambda)=F_{\rm obs}(\lambda)-\sum_\ell F_{\ell}(\lambda),
\end{equation*}
which was then used to calculate the corrected synthetic colors and to derive the color increment due to emission lines, $\Delta_{\rm EL}(V-I)$. The color map corrected from emission lines can be directly obtained by adding the map of the increments to the original $(V-I)_{\rm obs}$ map,
\begin{equation*}
(V-I)_{\rm corr}=(V-I)_{\rm obs}+\Delta_{\rm EL}(V-I).
\end{equation*}

\smallskip

The right panel of Figure~\ref{Figure:Haro14_broadband} displays  $\Delta_{\rm EL}(V-I)$ for Haro~14. Only the central SF regions are substantially affected by the ionized gas emission; whereas in the brighter knots 
$\Delta_{\rm EL}(V-I)$ is as high as 0.2,  it is
negligible outside of the SF regions. In particular, the blue region detected to the east is free of emission lines and its intrinsically blue color is a reliable indicator of an intermediate age stellar population.

\smallskip

Therefore,  the color map of Haro~14  displays three well differentiated zones which trace three distinct stellar populations: 
the  bluest knots  in the inner part of the galaxy, which delineate the \ion{H}{ii} regions; the blue 
wide band extending northeast,  which spatially coincides with the diffuse tail structure detected in continuum and has $(V-I) \sim$0.5-0.7; and the redder areas in the outskirts with $(V-I)\sim$0.8-1.0, which trace the LSB galaxy host. In addition to the unresolved stellar components, blobs with a broad range of colors are distinguished throughout the map, suggesting stellar clusters at different evolutionary stages.

Color maps are also useful to detect dust. In the $(V-I)$ color map of Haro~14, we distinguish several red patches in the inner galaxy regions, quite prominent east and south-east of the central SF knots, that are likely due to dust. But without additional observational constrains, it is difficult to ascertain whether they trace dust or are just gaps in the blue stellar distribution.

\subsection{Surface brightness photometry}
\label{Section:structure}

Using the synthetic VRI images we also built the surface brightness profiles (SBPs) of Haro~14. Deriving the SBPs of irregular objects, such as BCGs, is not a straightforward task. The standard methods employed  to perform surface brightness photometry, which  assume some sort of galaxy symmetry, are not appropriate for describing the HSB regions of a BCG---this topic has been widely discussed in the literature and alternative procedures have been proposed 
\citep{Papaderos1996a,Papaderos2002,Doublier1997,Cairos2001a,Cairos2003,Noeske2003,GildePaz2005,Micheva2013a,Micheva2013b,Janowiecki2014}.

\smallskip

\begin{figure}
\centering
\includegraphics[angle=0, width=\linewidth]{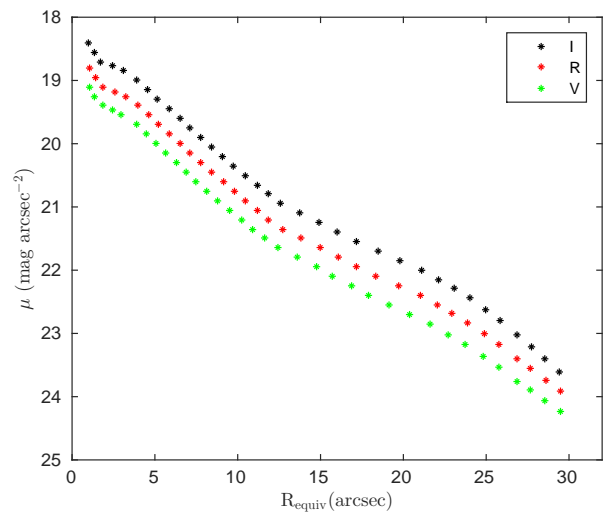}
\caption{Surface brightness profiles of Haro~14 built on the synthetic  V, R, and I images from MUSE.  
}
\label{Figure:SBP}
\end{figure}

\begin{figure*}
\centering
\begin{subfigure}{}
\includegraphics[width=8cm]{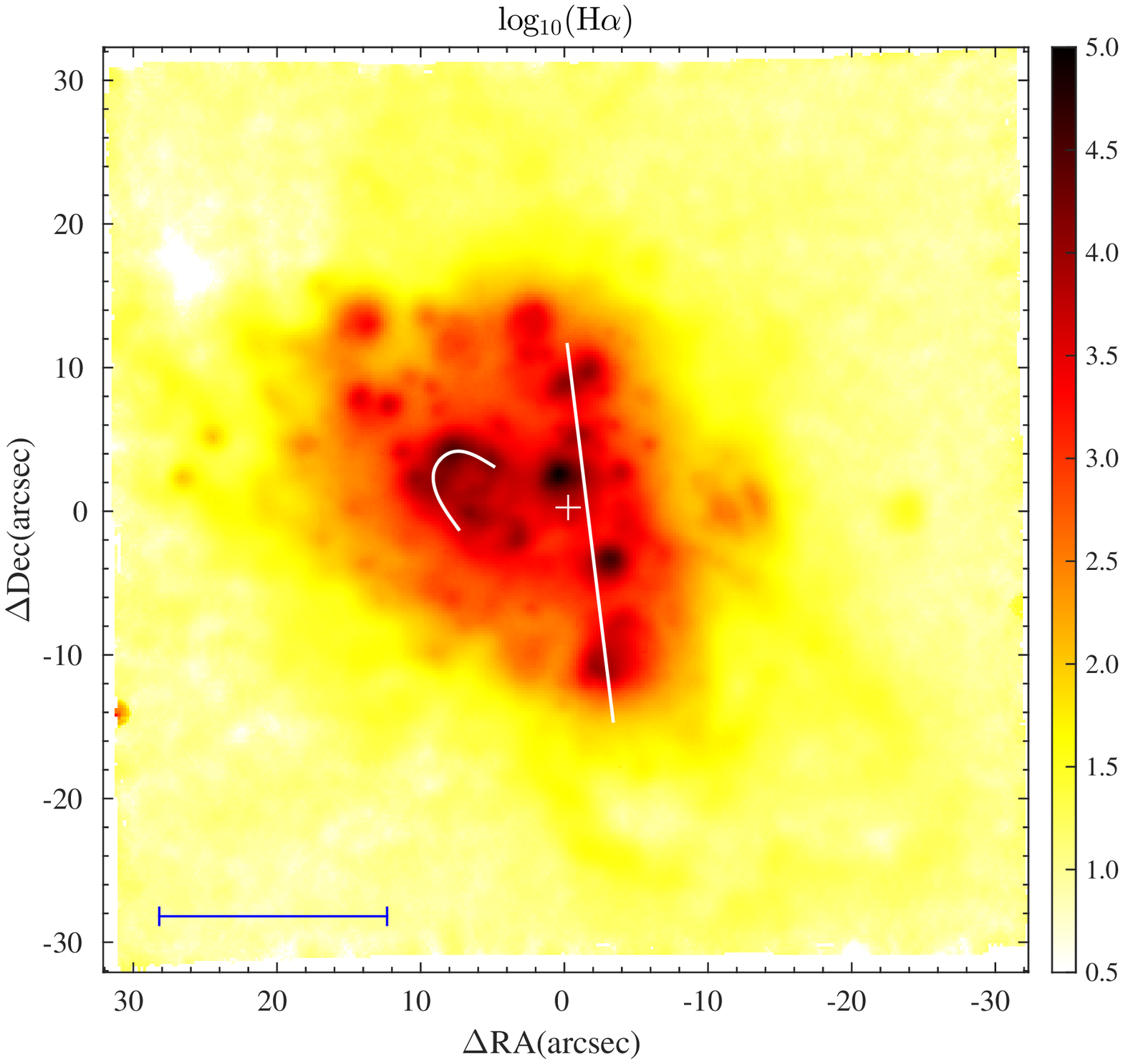}
\end{subfigure}
\begin{subfigure}{}
\includegraphics[width=8cm]{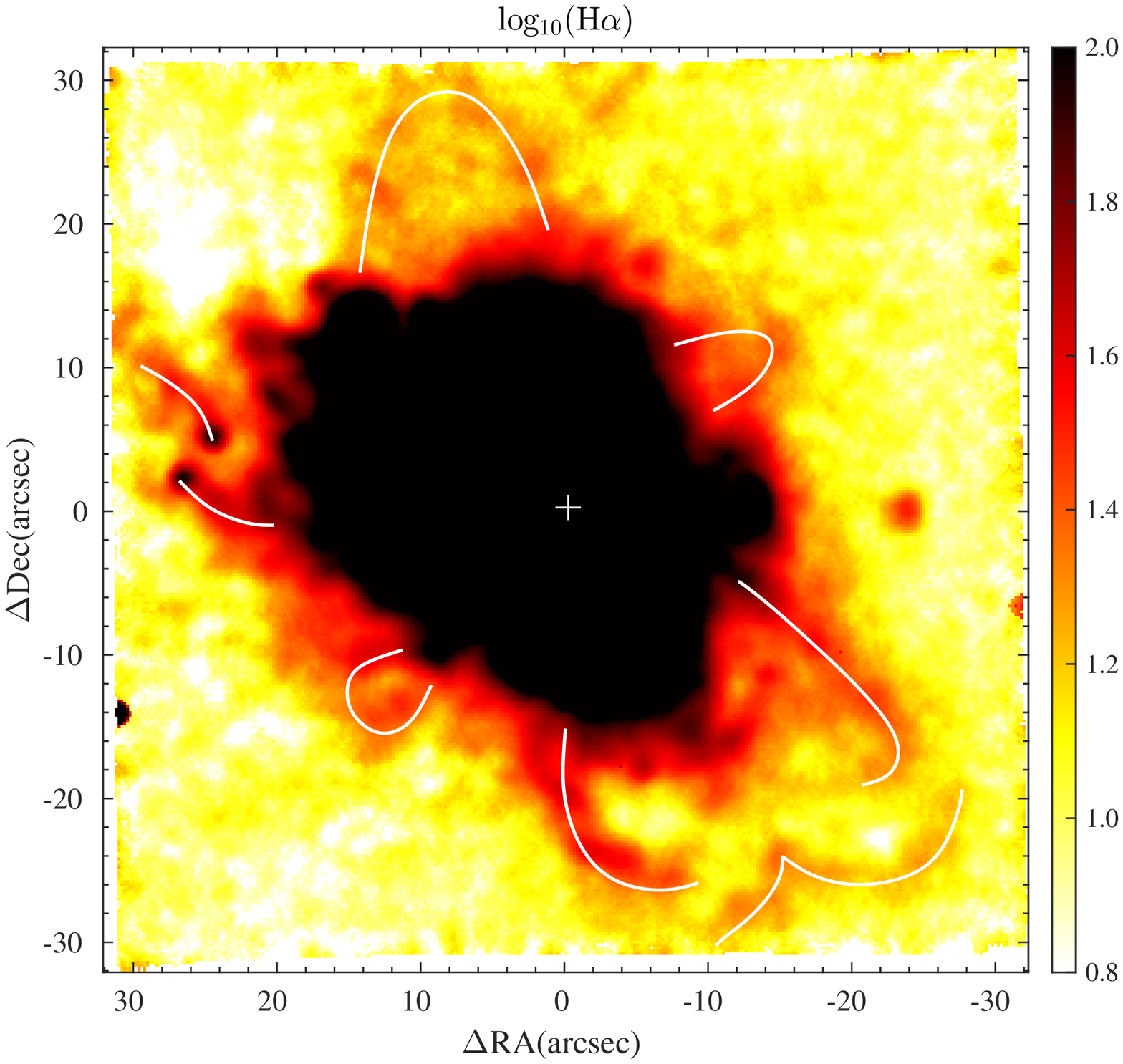}
\end{subfigure}
\begin{subfigure}{}
\includegraphics[width=8cm]{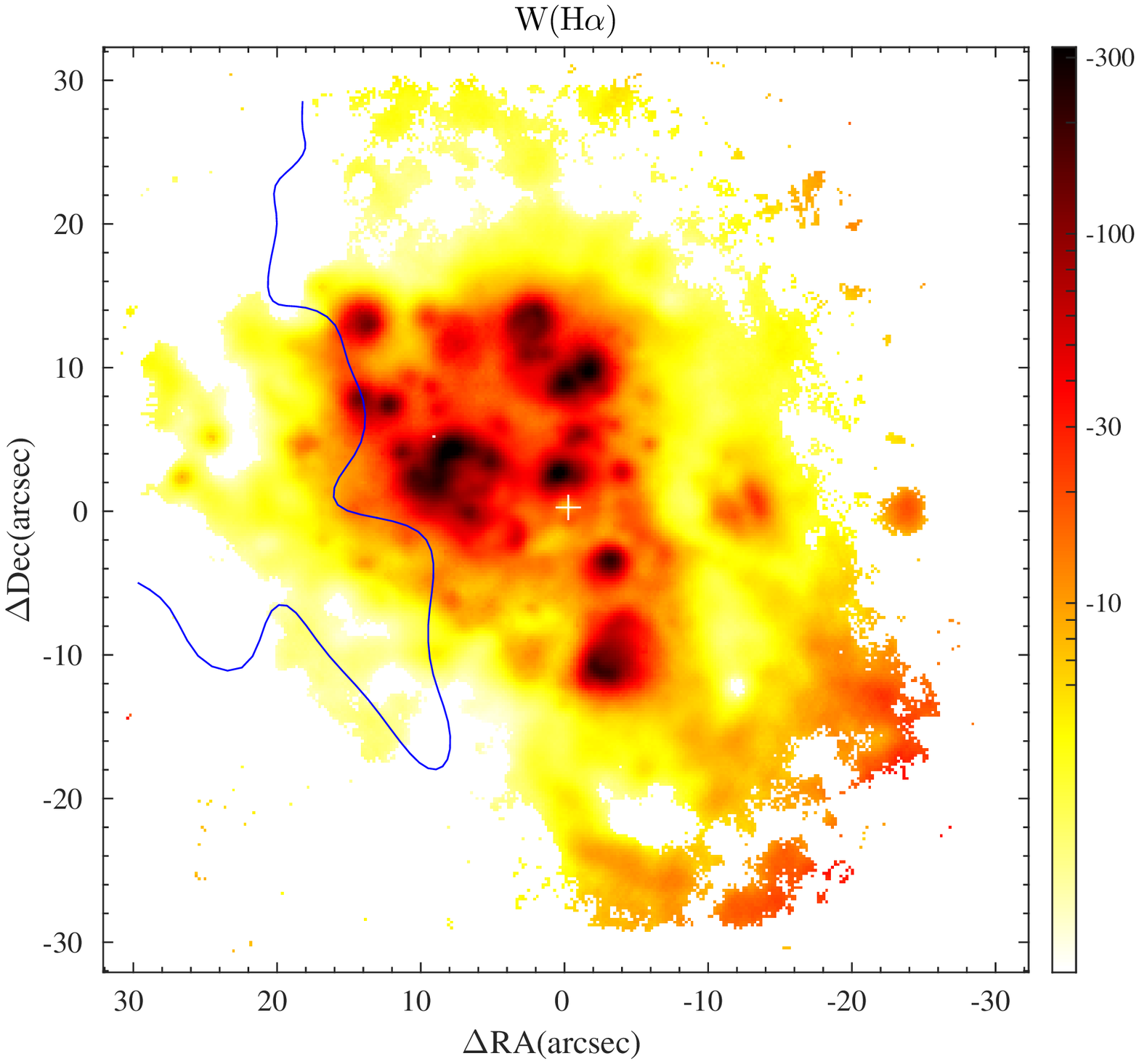}
\end{subfigure}
\begin{subfigure}{}
\includegraphics[width=8cm]{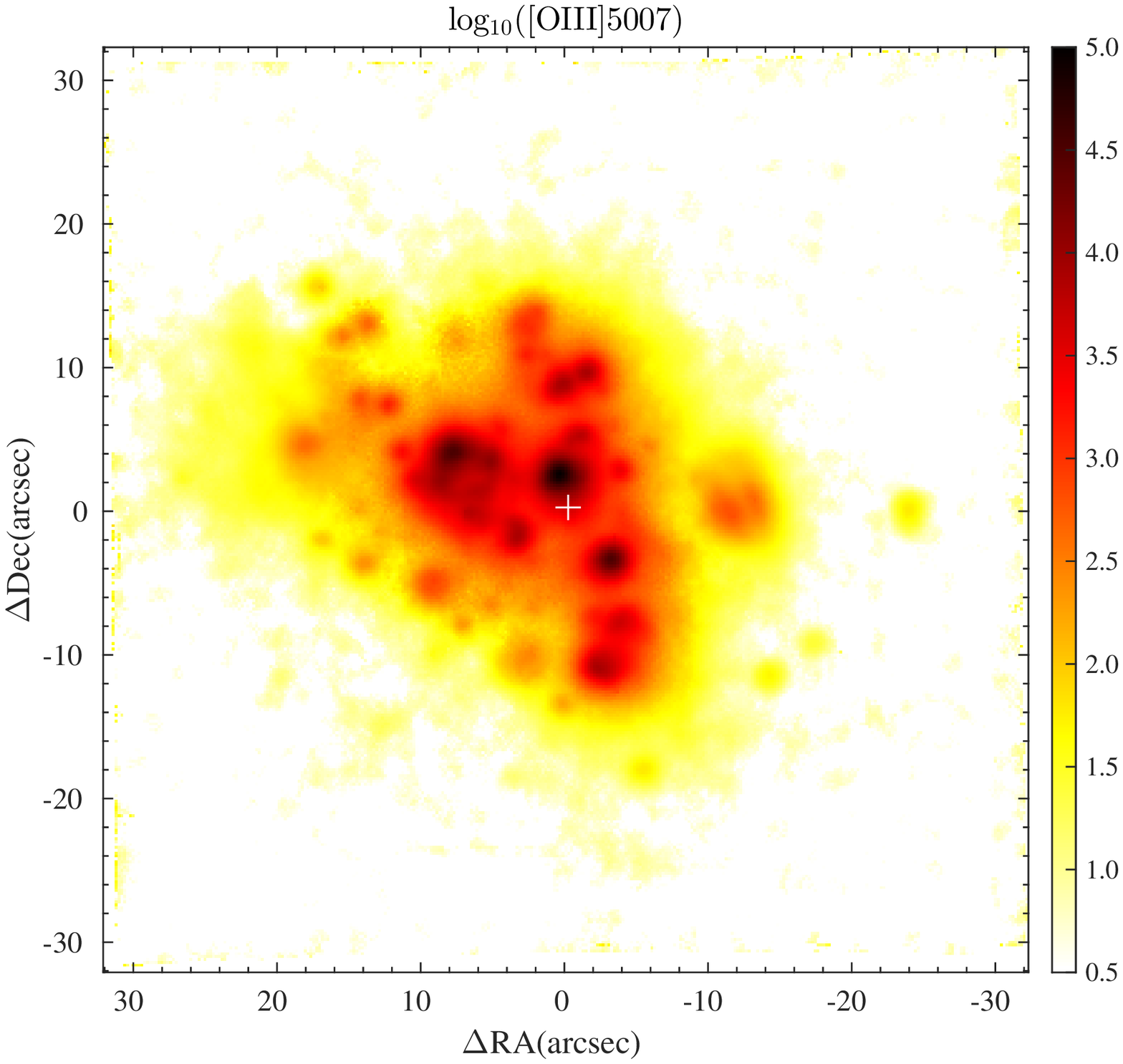}
\end{subfigure}
\caption{Haro~14 maps in emission lines:  {\em Top-left}: H$\alpha$ emission line flux map of Haro\,14; the chain-like structure and the curvilinear feature in the central regions are indicated with white lines. The blue line  (bottom-left) corresponds to 1~kpc. {\em Top-right}: H$\alpha$ emission line flux map  with the intensity in the inner galaxy regions saturated  in order to  enhance the faint surface brightness features in the outer galaxy parts; the white lines indicate the most prominent shells and filaments in the galaxy periphery. {\em Bottom-left}: H$\alpha$ equivalent width map (in \AA); the blue line roughly delineates the bluish regions of the galaxy at intermediate intensity levels. {\em Bottom-right}: [\ion{O}{iii}]~$\lambda5007$  flux map. Flux units in the emission line maps are  are 10$^{-20}$\,erg\,s$^{-1}$\,cm$^{-2}$; the central white cross marks the position of the galaxy peak in continuum. }
\label{Figure:emission} 
\end{figure*}

We built the SBPs of Haro~14 
following the approach described in  \cite{Cairos2001b,Cairos2003}.  In the HSB regime, 
where the galaxy is markedly irregular, isophotal radial profiles are derived, that is, we compute the  equivalent radii inside isophotes of decreasing intensity without making any assumptions about their geometrical shapes. In the outermost regions, we fit ellipses  to the images  using
the {\sc iraf} task {\tt ellipse}  \citep{Jedrzejewski1987}. The final SBP is created  as a combination of these two profiles:  the isophotal radial profiles in the high- and intermediate-intensity areas and elliptical fitting in the outer parts (see 
Figure~\ref{Figure:SBP}). In the central galaxy regions, the  equivalent  radius (R$_\mathrm{equiv}$) is defined as the radius of the circle whose area is equal to the corresponding isophotal surface; when fitting ellipses, the  
 R$_\mathrm{equiv}$ is computed as the square root of the product of the lengths of the semi-axes. 
\smallskip 

The light profiles of Haro~14  do 
not show the most common behavior among BCGs, namely a  brightness excess
at high and intermediate intensity levels and a LSB host that is well described by an exponential function \citep{Papaderos1996a,Marlowe1997,Cairos2001a,Micheva2013a,Micheva2013b,Janowiecki2014}. 
At intermediate radii, the profile displays a "plateau" or flattening, which is probably due to the blue  and extended component that is well visible in the $(V-I)$ color map. 
The LSB galaxy host cannot be fitted using a physically meaningful exponential, but the extrapolation of the fitted function to the central galaxy regions produces a higher surface brightness than that observed. Our SBPs are limited in radius by the MUSE FoV, but the more radially extended profiles derived in the optical (see Figure~1 in \citealp{GildePaz2005}) and in the NIR (Figure~2 in  \citealp{Noeske2003}) exhibit the same behavior out to a larger radius.  \cite{GildePaz2005} indeed fit  an exponential function to the LSB host, but the derived fit has clearly no physical meaning (see their Figure~1).

\subsection{Emission line maps}
\label{Section:ionizedgasmorphology}

The wide FoV and high sensitivity of MUSE allow for the detection of warm ionized gas emission out to large galactocentric distances (about 1.9 to 2.7~kpc) and  
very low surface brightness levels ($\sim$4$\times$10$^{-20}$erg~s$^{-1}$~cm$^{-2}$ in H$\alpha$). 

\smallskip

The intensity maps of Haro\,14 in the brightest emission lines, H$\alpha$ and 
[\ion{O}{iii}]~$\lambda5007$, exhibit an intriguing  morphology (Figure~\ref{Figure:emission}). The  inner galaxy areas are dominated by large \ion{H}{ii} regions,  confirming the VIMOS results at a  better spatial resolution (see Figure~2 in CGP17a) while the current MUSE data reveal an intricate structure of bubbles, arcs, filaments, and faint blobs in the galaxy outskirts. 

\smallskip 

The copious  \ion{H}{ii} regions indicate substantial and extended ongoing SF activity. The largest SF complexes are situated close to the central regions of the galaxy,  forming a linear (chain-like) structure of about 1.8~kpc,  roughly along the north--south direction, and in a horseshoe-like curvilinear feature, with a diameter of about 700~pc, which extends eastwards (see Figure~\ref{Figure:emission}, top-left).
Looking more carefully, we notice that the central SF chain-like  structure is slightly curved, resembling incipient spiral arms. The emission in the galaxy periphery is dominated by LSB filaments and shells  (see Figure~\ref{Figure:emission}, top-right). Particularly impressive are the two curvilinear filaments extending southwest,  highly detectable up to galactocentric distances of 2.0 and 2.3~kpc---such filaments extending far out into the halo are indicative that the galaxy is in an evolved phase of a starburst.  
A shell with a diameter of about 800 pc is clearly distinguished to the north and two smaller ones are visible departing southeast and northwest, respectively. Numerous faint knots appear scattered throughout the field, most of them in the vicinity of LSB structures. Also interesting are two emission blobs  detached from the galaxy main body and highly visible to the west both in H$\alpha$ and [\ion{O}{iii}].

\smallskip 

The H$\alpha$ and [\ion{O}{iii}]~$\lambda5007$ maps are similar but some remarkable differences are present. First, the large filaments expanding southwest and the large bubble extending north, which are both highly visible in  H$\alpha$,  are not detected in [\ion{O}{iii}]~$\lambda5007$---the same applies to other faint extensions. Second, the blobs visible on both maps do not necessarily coincide: several knots are only visible in one of the maps, and many of them are highly visible in one line but only marginally detected in the other. In general, the [\ion{O}{iii}]~$\lambda5007$ emission appears more compact and more knotty.

\begin{figure*}
\centering
\begin{subfigure}{}
\includegraphics[width=8cm]{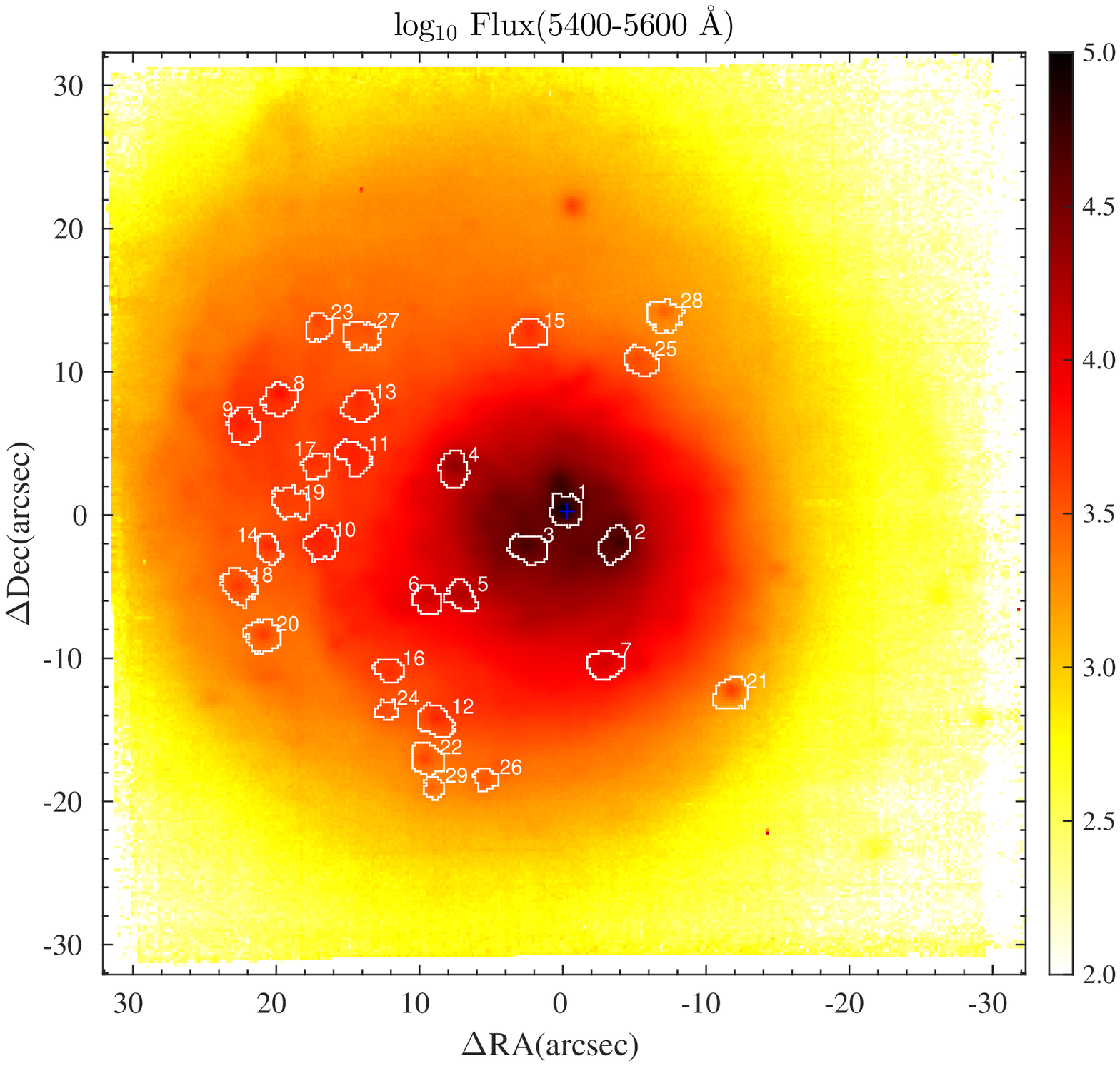}
\end{subfigure}
\begin{subfigure}{}
\includegraphics[width=8cm]{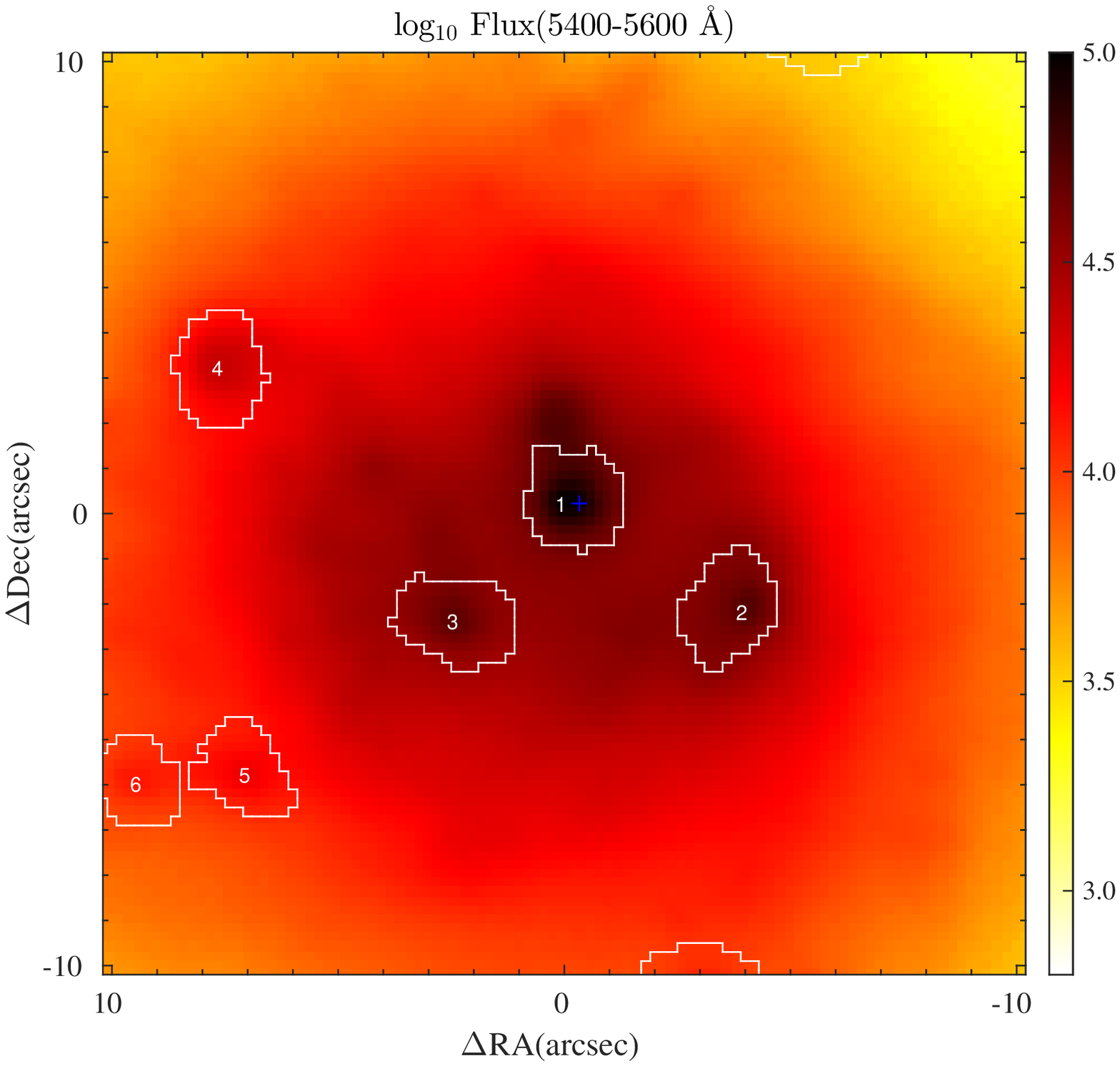}
\end{subfigure}
\begin{subfigure}{}
\includegraphics[width=8cm]{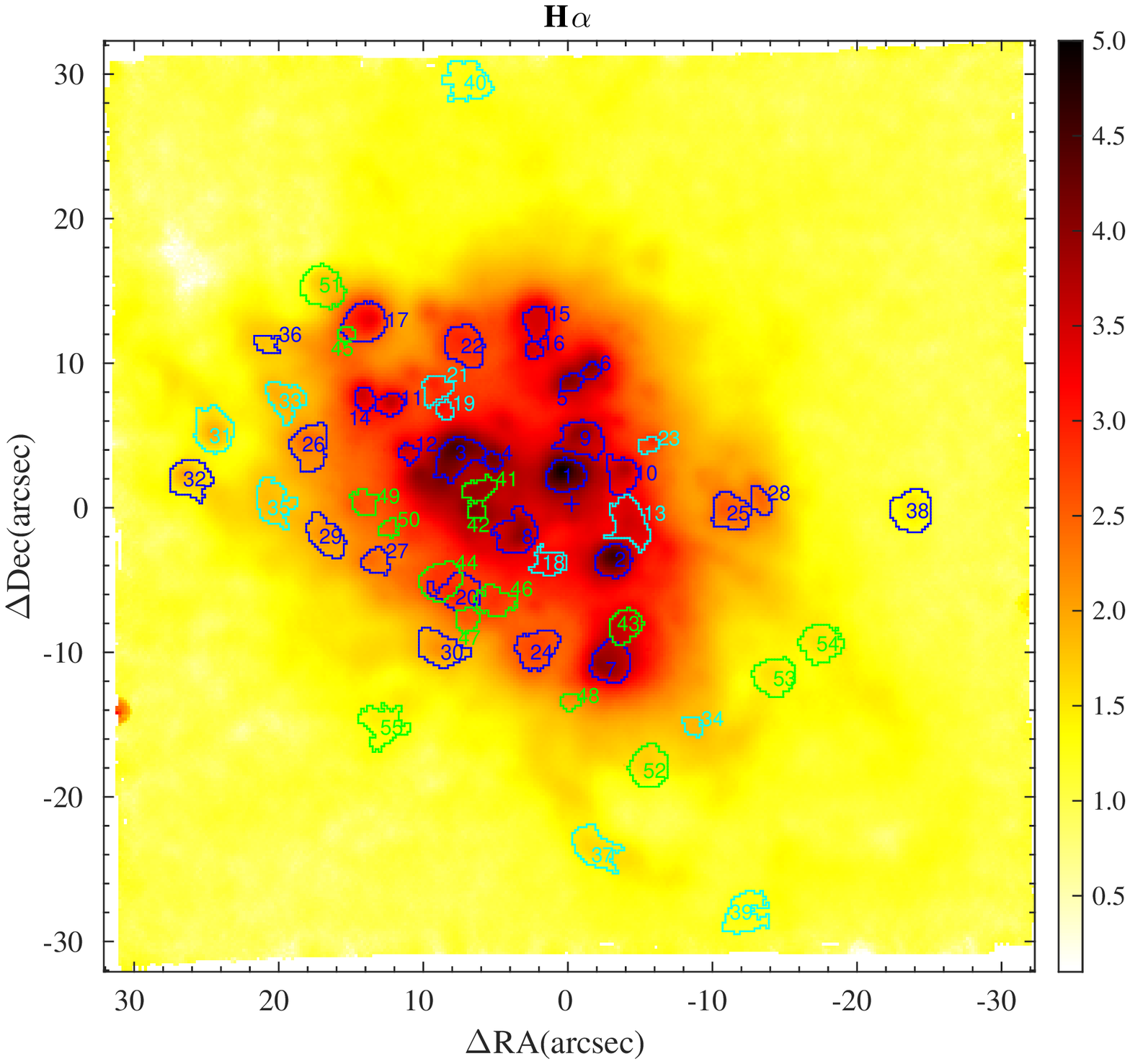}
\end{subfigure}
\begin{subfigure}{}
\includegraphics[width=8cm]{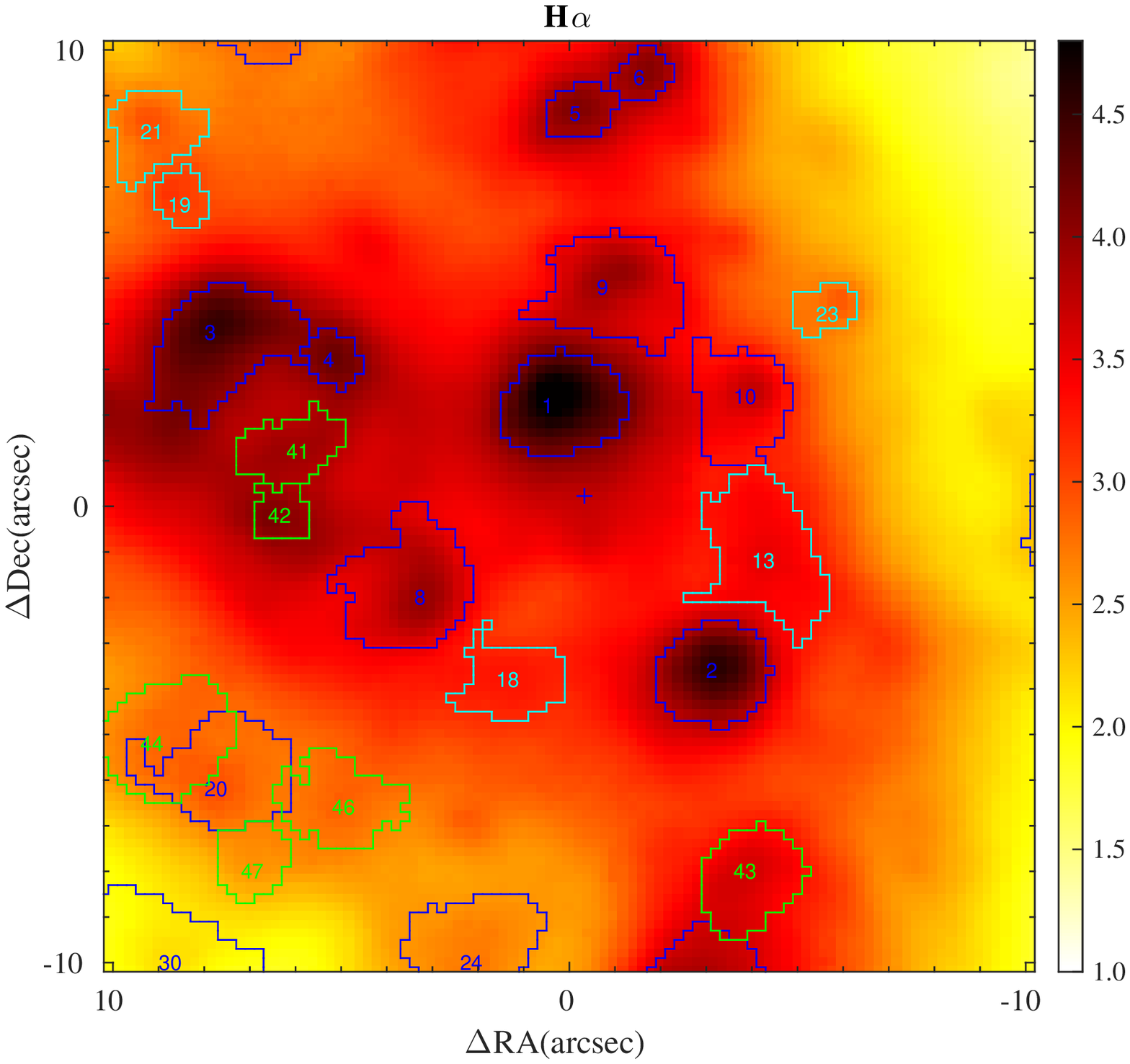}
\end{subfigure}
\caption{Clumps detected in Haro~14. {\em Top-left}: The 29 sources (clumps)  detected in continuum (see also Table~\ref{Table:continuum_sources}) overplotted on the blue (5400-5600~\AA)  continuum map. There is an evident clustering of the clumps at the galaxy eastern regions. {\em Top-right}: Zoom into the central 20$\arcsec \times$ 20$\arcsec$. The flux units are arbitrary. {\em Bottom-left}: The 55 clumps detected in emission lines (see also Table~\ref{Table:line_sources}) overplotted on the H$\alpha$ map; sources detected in H$\alpha$ and [\ion{O}{iii}]~$\lambda5007$ appear in blue, sources detected only in H$\alpha$ in cyan, and those detected only in [\ion{O}{iii}]~$\lambda5007$ in green. {\em Bottom-right}: Zoom into the central galaxy regions. The intensity scale is logarithmic and the cross marks the position of the continuum peak.}
\label{Figure:clumps} 
\end{figure*}

\subsection{Clumps in continuum and emission-line maps}
\label{Section:clumps}

\begin{table*}
\small
\begin{center}
\caption{Properties of the continuum sources detected in Haro~14. \label{Table:continuum_sources}}
\begin{tabular}{lrrccccccc}
\hline
Reg & {$\Delta$X} & {$\Delta$Y} & Area & FWHM& \multicolumn{1}{c}{V} & \multicolumn{1}{c}{V-I} & \multicolumn{1}{c}{V-R} & \multicolumn{1}{c}{M$_V$} & ID  \\
    &  (arcsec)     &  (arcsec)     &  (arcsec$^2)$ & (arcsec)  & (mag) & (mag) &  (mag) & (mag) & \\
    (1) &  (2) & (3) & (4) & (5) & (6) & (7) & (8) & (9) & (10) \\
\hline

C01 &  0.0 &   0.0 &  4.08  &  0.80 &  18.39 (0.02) &   0.86  (0.01) &   0.39  (0.00) & -12.18 & SG\\
C02 &  3.4 &  -2.4 &  4.12  &  1.38 &  19.13 (0.03) &   0.63  (0.01) &   0.19  (0.01) & -11.44 & SG\\
C03 & -2.6 &  -2.7 &  4.32  &  0.94 &  19.61 (0.05) &   0.54  (0.01) &   0.23  (0.00) & -10.96 & SG\\
C04 & -7.8 &   2.9 &  4.12  &  1.67 &  19.68 (0.05) &   0.14  (0.03) &   0.10  (0.01) & -10.89 & HII\\
C05 & -7.3 &  -6.0 &  3.40  &  0.92 &  21.27 (0.14) &   0.51  (0.04) &   0.22  (0.02) &  -9.30 & SG\\
C06 & -9.6 &  -6.3 &  3.16  &  1.16 &  21.32 (0.06) &   0.62  (0.01) &   0.30  (0.00) &  -9.25 & SG\\
C07 &  2.8 & -10.8 &  4.04  &  1.47 &  21.16 (0.10) &  -0.00  (0.11) &   0.27  (0.01) &  -9.41 & HII\\
C08 &-20.0 &   7.8 &  4.16  &  0.83 &  21.79 (0.04) &   0.46  (0.00) &   0.19  (0.00) &  -8.78 & SG\\
C09 &-22.5 &   5.8 &  4.36  &  1.56 &  22.04 (0.05) &   0.61  (0.00) &   0.24  (0.00) &  -8.53 & SG\\
C10 &-17.0 &  -2.3 &  4.16  &  1.95 &  21.92 (0.05) &   0.40  (0.01) &   0.19  (0.01) &  -8.65 & SG\\
C11 &-14.8 &   3.7 &  4.04  &  1.74 &  22.27 (0.03) &   0.08  (0.06) &  -0.03  (0.02) &  -8.30 & HII\\
C12 & -9.1 & -14.7 &  4.32  &  1.05 &  22.28 (0.10) &   0.65  (0.01) &   0.26  (0.01) &  -8.29 & SG\\
C13 &-14.4 &   7.3 &  4.08  &  1.47 &  22.24 (0.07) &  -0.12  (0.07) &   0.24  (0.01) &  -8.33 & HII\\
C14 &-20.7 &  -2.7 &  2.72  &  0.86 &  22.50 (0.04) &   0.54  (0.01) &   0.23  (0.01) &  -8.07 & ?\\
C15 & -2.6 &  12.3 &  4.32  &  1.65 &  21.73 (0.06) &   0.29  (0.06) &   0.32  (0.01) &  -8.84 & HII\\
C16 &-12.3 & -11.2 &  2.68  &  1.25 &  23.27 (0.14) &   0.24  (0.05) &   0.10  (0.02) &  -7.30 & SG\\
C17 &-17.4 &   3.1 &  2.72  &  1.05 &  23.43 (0.09) &   0.36  (0.03) &   0.16  (0.01) &  -7.14 & SG\\
C18 &-22.8 &  -5.3 &  4.52  &  0.77 &  22.21 (0.03) &   0.72  (0.00) &   0.33  (0.00) &  -8.36 & ? \\
C19 &-19.2 &   0.5 &  4.36  &  0.83 &  24.43 (0.35) &   2.34  (0.31) &   0.92  (0.19) &  -6.14 &RSG\\
C20 &-21.1 &  -8.8 &  4.44  &  0.97 &  22.56 (0.03) &   0.44  (0.02) &   0.19  (0.01) &  -8.01 &?\\
C21 & 11.6 & -12.8 &  4.32  &  0.83 &  22.14 (0.08) &   0.98  (0.02) &   0.48  (0.01) &  -8.43 &?\\
C22 & -9.7 & -17.4 &  4.16  &  1.05 &  22.65 (0.10) &   0.51  (0.02) &   0.21  (0.01) &  -7.92 &SG\\
C23 &-17.3 &  12.8 &  2.88  &  0.97 &  23.16 (0.07) &   0.39  (0.02) &   0.10  (0.01) &  -7.41 &?\\
C24 &-12.5 & -14.0 &  1.60  &  1.09 &  23.87 (0.12) &   0.68  (0.01) &   0.26  (0.00) &  -6.70 &SG \\
C25 &  5.2 &  10.4 &  3.72  &  0.89 &  22.96 (0.18) &   0.99  (0.03) &   0.48  (0.02) &  -7.61 &?\\
C26 & -5.6 & -18.8 &  2.00  &  1.18 &  23.04 (0.05) &   0.75  (0.00) &   0.32  (0.00) &  -7.53 & SG\\
C27 &-14.3 &  12.2 &  4.28  &  1.38 &  23.06 (0.06) &   0.18  (0.02) &   0.35  (0.01) &  -7.51 & HII\\
C28 &  6.9 &  13.6 &  4.36  &  1.02 &  22.88 (0.06) &   1.06  (0.01) &   0.48  (0.00) &  -7.69 &SG \\
C29 & -9.2 & -19.4 &  1.76  &  1.18 &  23.31 (0.05) &   0.87  (0.01) &   0.42  (0.01) &  -7.26 &SG \\
\hline
\end{tabular}
\end{center}
\small Notes. Columns (2) and (3) display the offsets in arcsec from the position of the continuum peak. 
Apparent magnitudes and colors and their corresponding uncertainties are shown in Columns 6 to 8.  Photometry  is corrected from Galactic extinction using the reddening coefficients from \cite{Schlafly2011}. In Column (10): SG=stellar group, which includes stellar clusters and associations;  HII= young stellar cluster belonging to an \ion{H}{ii}-region;  red supergiant star (RSG) and undetermined (?).
\end{table*}

Haro~14 exhibits numerous clumps spread out through the galaxy  
both in continuum and in emission-line maps. 
In continuum,  three very bright knots sit roughly at the center of the HSB area and many fainter ones are evident at larger galactocentric distances, mostly clustered in the eastern galaxy regions (Figure~\ref{Figure:Haro14_continuum}). In the emission line maps, the brightest clumps are placed along the central chain-like structure and in the  horseshoe-like curvilinear feature extending  eastwards, whereas fainter knots  are visible in the surroundings  (Figure~\ref{Figure:emission}).

\smallskip 

The clumps in emission lines are  mainly \ion{H}{ii} regions and their ionizing stellar clusters.  However, the nature of the sources detected in continuum is more difficult to asses: 
they could be groups of stars (e.g., stellar associations or clusters) or individual bright stars belonging to Haro~14, background galaxies, or foreground stars in the Milky Way \citep{Leitherer1996,Ostlin1998b,Johnson2000}. Their spatial distribution (the majority of the clumps are gathered in the blue area to the east) suggests that most of them  indeed belong to Haro~14. In order to gain insight into the nature of these clumps, we 
derived their physical parameters---their sizes, magnitudes, and colors.

\subsubsection{Source selection}
\label{Section:sourcedetection}

The first step towards deriving the physical parameters of the sources is their detection and spatial delimitation. To this aim, we wrote a routine that automatically searches for clumps in both the continuum and the emission line maps. The characteristics of these maps are quite distinct: while the line maps have a higher contrast, that is, the local background is dim compared to the sources,  in the continuum frames the local background may be very bright and fluctuating. We developed an algorithm able to work efficiently with both types of images.

\smallskip 

Our routine does not make  {\em a priori} assumptions about the size or shape of the sources, 
it simply detects extended and noncircular features.  The method selects a source by expanding from a local emission maximum 
until it makes contact with a nearby region  or merges into the background. 
The algorithm consists of three steps: detection of the local maxima, iterative growth of the region,  
and stopping criteria.

\smallskip

\begin{enumerate}

\item {\em Detection of the local maxima}. The routine starts smoothing the image with a Gaussian
filter (FWHM=3~pixels). We then search for local maxima  in this smoothed image. 
The flux of the local maxima has to be above the local background by some given threshold which depends on the noise characteristic of the frame.
For this task, we adapted the routine {\tt FastPeakFind.m} from the 
{\sc matlab}  Central File Exchange\footnote{Natan (2020). Fast 2D peak finder (https://www.mathworks.com/matlabcentral/fileexchange/37388-fast-2d-peak-finder), MATLAB Central File Exchange}. We end up with a list with the positions of the local maxima. 

\smallskip 

\item {\em Iterative growth of the sources}. For a given maximum, we started seeding a region with all the pixels around it with a flux 
above 95\% of the peak---typically, between 4 and 15 pixels.

\smallskip 

Now, we proceed iteratively following a strategy
similar to the {\tt HIIphot} routine \citep{Thilker2000}.
We begin defining the threshold, $F_{t}$, as the highest flux of the whole frame, which is then decreased by 1\% in each step of the iteration. 
We incorporate to each source the pixels of its external perimeter (see below) with fluxes above the $F_{t}$, as long as they have not been incorporated
to some other source.

\smallskip 

We distinguish between an {\em external perimeter} and an {\em internal perimeter}. The internal perimeter contains those pixels that belong to the source and are in contact with the current boundary and the external perimeter is composed of those pixels in contact with the boundary but not belonging to the source. In each step of the iteration, we search for pixels in the external perimeter of each region whose flux is above the flux threshold and are not claimed by other sources. 

\smallskip

A problem with this strategy, and particularly relevant for continuum images, 
appears  when there is a gradient on the underlying background, in which case 
the selected region may grow incorrectly along the background gradient. 
To alleviate this, we forced the number of pixels added to the region to be 
at least 30\% of those in the external perimeter and their fluxes 
to be lower than the median of the internal perimeter. 
We only consider pixels whose distance to the center of the region, $d_i$, verifies
\begin{equation*}
\frac{d_i}{D} < f_{e},
\end{equation*}
\noindent where $D$ is the median of the distance to the center of the pixels in the internal perimeter of the region and $f_{e}$ is an input parameter that prevents excessive elongation of the sources. We set $f_{e}=1.25$ for the catalog of features detected in continuum and $f_{e}=1.50$ for those detected in emission lines. This difference arises from the fact that the \ion{H}{ii}  regions are expected to be larger and less round than continuum features.

\smallskip

\item {\em Stopping criteria.}  A region ceases to grow when no more pixels satisfying the above criteria can be added to the region,  the surface brightness profile flattens, or the region comes into contact with another adjacent source.

\smallskip 

The surface brightness profile has flattened when 
\begin{equation*}
\frac{\langle F_{\rm ext}\rangle}{\langle F_{\rm inter}\rangle}> f_{\rm flat},
\end{equation*}
\noindent where $\langle F_{\rm ext}\rangle$ and $\langle F_{\rm inter}\rangle$ are the flux averages of the external and internal perimeters, respectively, and $f_{\rm flat}$ is an input parameter (we adopted $f_{\rm flat}=0.97$ in this work). Also, we require that the flux average of the external and the internal perimeters decrease with each iteration $i$ (i.e., $\langle F_{\rm ext,i}\rangle<\langle F_{\rm ext,i-1}\rangle; \langle F_{\rm int,i}\rangle<\langle F_{\rm int ,i-1}\rangle$).

\smallskip

When two adjacent regions are in contact by more than two pixels, the growth of both is stopped
to avoid the contamination of one source by the other.
In addition, we impose two other conditions to avoid the growth of the sources to unrealistic sizes.  
First, only pixels with a flux above 15\% of the flux of the region peak can be added to the source. Second, we fix an upper bound for a source size: 100 pixels for the continuum sources and 150 pixels for the emission line regions.

\end{enumerate}

The routine provides the list of identified sources,  their spatial position, and a mask with the pixels belonging to each source,   together with a series of parameters helping  to control the operation of the algorithm.

\subsubsection{Background subtraction}
\label{Section:background}

An important albeit difficult step in the characterization of the sources is the subtraction of the background. The clumps we identified lie on top of unresolved stellar and gaseous emission, and therefore to extract their intrinsic flux, we must first subtract the light contribution coming from the underlying components. This is a complicated task because 
the background is highly inhomogeneous, which prevents a global modelling. Also, 
we are interested in extracting the intrinsic spectrum of the region in the whole spectral range and not merely in the photometry of the sources in specific filters, meaning that traditional techniques are not immediately applicable here.

\smallskip

We determine the local background per pixel for each source individually, as the median of the spectra of the pixels in a region around the source. 
We select this region as a strip of two pixels in width, two pixels away from the boundary 
as a compromise in order to get a good signal-to-noise ratio (S/N) spectrum while minimizing contamination from the source. We exclude pixels belonging to nearby regions  from the background
strip. 

\smallskip

\subsubsection{Determination of the parameters of the sources}
\label{Section:parameters}

From the background-subtracted data, we determined the spatial properties of the sources.
The center of the source is located at the centroid of its pixels (i.e., the intensity-weighted means of the marginal profiles in the X and Y directions of all pixels of the source).
The area is simply the area of the pixels belonging to the source. 
However, this value may not be a good characterization of the actual extension of the
region because it could be affected by background inhomogeneities or gradients 
(see discussion above, Sect.~\ref{Section:sourcedetection}). 
A better measure of the size of the source is its FWHM, which we calculate here as the equivalent diameter of the region formed by all the pixels with a flux larger than half the maximum value.

\smallskip

The integrated spectrum of each individual blob is generated by 
adding  the spectra of its corresponding spaxels and removing the local background (see above). 
We then calculate the fluxes in the brighter emission lines (lines are fitted as described in Section~\ref{Section:making_maps}). 

\smallskip

The intrinsic (background-subtracted) photometry in the VRI filters was generated from flux maps. 
The intrinsic inhomogeneity of the underlying galaxy and the systematic errors of the instrument are the dominant factor in the photometric error budget, much larger than the photonic noise (if the background were uniform and the observations photon limited, photometric errors would be well below 0.01 mag). These sources of error are also highly correlated from filter to filter. 
We estimate the uncertainty due to the background subtraction as the standard deviation of the fluxes in the strip around the source divided by the square root of the number of pixels. From this we calculate the corresponding errors in the magnitudes. For most sources, $V$ is determined with a precision of a few hundredths of a magnitude (see Table~\ref{Table:continuum_sources}). 
The few sources with higher values have highly inhomogeneous surroundings for which a background estimate is therefore more uncertain.
The uncertainties on the colors are reduced with respect to a simple quadratic sum
of the broadband errors because of the covariance of the correlated noise.

\subsubsection{Catalogs of continuum and emission line sources}
\label{Section:stellarclusters}

We ran  the routine for source detection in  two continuum maps, one in the blue  spectral region (5400-5600~\AA) and one in the red (7800-8000~\AA), and  found 
42 and 38 objects, respectively. These sources could be 
objects belonging to Haro~14 (stellar groups or individual bright stars), foreground Galactic stars, or  background galaxies. 

\smallskip 

In order to discriminate among the distinct possibilities  we 
used the integrated spectra of the individual sources. 
This way, we discovered two background galaxies (see Figure~\ref{Figure:Haro14_broadband}  and Appendix~1) and eliminated from the catalog several blobs with poor S/N spectrum for which a reliable redshift could not be determined. All remaining clumps are at the same redshift as Haro~14---there are no foreground stars in the field.  

\smallskip 

The final catalog contains 29 sources that undoubtedly belong to Haro~14---all of them were detected by the routine in both the blue and the red map. Their position, area, FWHM, and photometry are presented in Table~\ref{Table:continuum_sources}. 
Figure~\ref{Figure:clumps} (upper left panel) shows that their distribution is not uniform through the galaxy, but most of them gather on the eastern sector. 

\smallskip 

Of the 29 sources, 18 are extended  (we consider extended sources those with FWHM$\geq$1~$\arcsec$) and are therefore definitely  groups of stars (e.g., \ion{H}{ii}-regions complexes or stellar associations). 
Point-like objects may still be unresolved stellar groups or individual stars. Using the 
photometry  we can further constrain their nature: adopting an absolute magnitude 
M$_{V}=-8.5$  as the  cutoff for single stars  \citep{Whitmore1999,Johnson2000},
we conclude that  the four point-like sources brighter that this limit are also not individual stars, but complexes of stars. 
Source {\sc C19}, with $(V-I)=2.34$, is likely a  red supergiant (RSG), as sources with $(V-I)\geq1.50$ can be safely classified as stars (\citealp{Whitmore1995, Miller1997,Harris1996}, 2010 edition), but the nature of the remaining six point-like sources  cannot be unequivocally established using only photometrical criteria.

\medskip

\begin{table*}
\begin{center}
\small
\caption{Properties of the emission-line sources detected in Haro~14. \label{Table:line_sources}}
\begin{tabular}{lrrcccccc}
\hline
Reg & {$\Delta$X} & {$\Delta$Y} & Area & \multicolumn{1}{c}{FWHM}& F(H$\alpha$)  & F([OIII])  & log(L$_{\mathrm{H}\alpha}$) & log(L$_{\mathrm{[OIII]}}$)\\
      &  (arcsec)         & (arcsec)       & (arcsec$^2)$ & \multicolumn{1}{c}{(arcsec)}    & \multicolumn{1}{c}{(erg~s$^{-1}$~cm$^{-2}$)}  & \multicolumn{1}{c}{(erg~s$^{-1}$~cm$^{-2}$)} & (erg~s$^{-1}$) & (erg~s$^{-1}$)\\
      (1) & (2) & (3) & (4) & (5) & (6) & (7) & (8) & (9)  \\
\hline
L01 & -0.4 &   1.9 &  4.96 &  1.07 & 36.720 & 43.350 &   38.9 &   38.9  \\ 
L02 &  3.0 &  -4.0 &  4.56 &  1.18 & 16.396 & 14.470 &   38.5 &   38.5  \\ 
L03 & -7.9 &   3.3 &  6.64 &  1.51 & 21.026 & 23.103 &   38.6 &   38.7  \\ 
L04 & -5.4 &   2.9 &  1.20 &  1.18 &  2.442 &  1.895 &   37.7 &   37.6  \\ 
L05 &  0.0 &   8.3 &  1.52 &  1.31 &  2.700 &  1.797 &   37.7 &   37.6  \\ 
L06 &  1.4 &   9.1 &  1.16 &  1.23 &  2.081 &  1.536 &   37.6 &   37.5  \\ 
L07 &  2.7 & -11.1 &  6.16 &  2.09 &  8.189 &  5.814 &   38.2 &   38.1  \\ 
L08 & -3.7 &  -2.1 &  6.56 &  1.45 &  4.444 &  4.510 &   38.0 &   38.0  \\ 
L09 &  0.7 &   4.4 &  6.16 &  1.62 &  5.578 &  4.446 &   38.1 &   38.0  \\ 
L10 &  3.5 &   2.0 &  4.48 &  1.25 &  1.850 &  1.143 &   37.6 &   37.4  \\ 
L11 &-12.5 &   6.8 &  2.28 &  1.05 &  1.091 &  0.331 &   37.3 &   36.8  \\ 
L12 &-11.2 &   3.4 &  1.32 &  1.16 &  0.597 &  0.347 &   37.1 &   36.8  \\ 
L13 &  4.1 &  -1.4 &  6.68 &  2.09 &  1.772 &  0.436 &   37.6 &   36.9  \\ 
L14 &-14.3 &   7.2 &  1.80 &  1.31 &  0.845 &  0.151 &   37.2 &   36.5  \\ 
L15 & -2.3 &  12.5 &  3.28 &  2.05 &  1.513 &  0.568 &   37.5 &   37.1  \\ 
L16 & -2.6 &  10.6 &  1.16 &  1.23 &  0.317 &  0.172 &   36.8 &   36.5  \\ 
L17 &-14.2 &  12.5 &  6.64 &  1.45 &  1.359 &  0.267 &   37.4 &   36.7  \\ 
L18 & -1.5 &  -4.1 &  3.24 &  1.67 &  0.285 &  0.009 &   36.8 &   35.3  \\ 
L19 & -8.7 &   6.4 &  1.20 &  0.97 &  0.079 &  0.000 &   36.2 &   22.1  \\ 
L20 & -7.8 &  -6.2 &  6.04 &  1.38 &  0.284 &  0.078 &   36.8 &   36.2  \\ 
L21 & -9.3 &   7.9 &  3.04 &  0.97 &  0.116 &  0.016 &   36.4 &   35.5  \\ 
L22 & -7.4 &  10.9 &  6.20 &  2.07 &  0.468 &  0.078 &   37.0 &   36.2  \\ 
L23 &  5.4 &   4.0 &  1.28 &  0.83 &  0.083 &  0.031 &   36.2 &   35.8  \\ 
L24 & -2.3 & -10.1 &  6.08 &  1.89 &  0.237 &  0.115 &   36.7 &   36.4  \\ 
L25 & 11.1 &  -0.6 &  5.44 &  1.97 &  0.181 &  0.417 &   36.6 &   36.9  \\ 
L26 &-18.0 &   3.9 &  6.40 &  1.70 &  0.138 &  0.269 &   36.4 &   36.7  \\ 
L27 &-13.5 &  -4.0 &  2.64 &  1.09 &  0.036 &  0.059 &   35.9 &   36.1  \\ 
L28 & 13.1 &   0.2 &  2.12 &  1.51 &  0.078 &  0.136 &   36.2 &   36.4  \\ 
L29 &-16.9 &  -2.3 &  5.40 &  1.62 &  0.034 &  0.025 &   35.8 &   35.7  \\ 
L30 & -9.0 & -10.1 &  6.32 &  1.97 &  0.073 &  0.014 &   36.2 &   35.5  \\ 
L31 &-24.7 &   5.0 &  6.72 &  1.00 &  0.035 &  0.008 &   35.9 &   35.2  \\ 
L32 &-26.4 &   1.7 &  6.08 &  1.05 &  0.048 &  0.015 &   36.0 &   35.5  \\ 
L33 &-19.8 &   7.1 &  4.96 &  1.80 &  0.021 &  0.003 &   35.6 &   34.8  \\ 
L34 &  8.4 & -15.4 &  1.44 &  1.36 &  0.004 &  0.002 &   34.9 &   34.6  \\ 
L35 &-20.5 &   0.0 &  6.12 &  2.22 &  0.013 &  0.010 &   35.4 &   35.3  \\ 
L36 &-21.0 &  11.0 &  1.72 &  1.49 &  0.006 &  0.004 &   35.0 &   34.9  \\ 
L37 &  1.7 & -23.9 &  6.48 &  2.37 &  0.025 &  0.003 &   35.7 &   34.8  \\ 
L38 & 23.6 &  -0.5 &  6.72 &  2.13 &  0.024 &  0.050 &   35.7 &   36.0  \\ 
L39 & 12.1 & -28.4 &  6.32 &  2.37 &  0.008 &  0.005 &   35.2 &   35.0  \\ 
L40 & -7.2 &  29.2 &  6.80 &  1.59 &  0.002 &  0.004 &   34.6 &   35.0  \\ 
L41 & -6.3 &   1.0 &  2.80 &  1.36 &  0.167 &  0.651 &   36.5 &   37.1  \\ 
L42 & -6.5 &  -0.5 &  1.28 &  1.29 &  1.550 &  0.769 &   37.5 &   37.2  \\ 
L43 &  3.8 &  -8.5 &  4.16 &  1.62 &  2.272 &  2.157 &   37.7 &   37.6  \\ 
L44 & -9.1 &  -5.4 &  6.08 &  1.88 &  0.237 &  0.578 &   36.7 &   37.1  \\ 
L45 &-15.5 &  11.7 &  1.04 &  1.05 &  0.060 &  0.050 &   36.1 &   36.0  \\ 
L46 & -5.2 &  -6.8 &  4.24 &  0.80 &  0.003 &  0.000 &   34.8 &   28.0  \\ 
L47 & -7.2 &  -8.1 &  2.20 &  1.09 &  0.081 &  0.047 &   36.2 &   36.0  \\ 
L48 & -0.1 & -13.8 &  1.24 &  1.20 &  0.008 &  0.029 &   35.2 &   35.8  \\ 
L49 &-14.3 &   0.0 &  2.52 &  1.05 &  0.076 &  0.023 &   36.2 &   35.7  \\ 
L50 &-12.6 &  -1.7 &  1.28 &  0.70 &  0.001 &  0.006 &   34.3 &   35.1  \\ 
L51 &-17.2 &  15.0 &  6.72 &  1.33 &  0.041 &  0.055 &   35.9 &   36.0  \\ 
L52 &  5.4 & -18.3 &  6.20 &  1.70 &  0.015 &  0.039 &   35.5 &   35.9  \\ 
L53 & 14.1 & -12.0 &  6.08 &  1.64 &  0.007 &  0.039 &   35.2 &   35.9  \\ 
L54 & 17.3 &  -9.7 &  6.08 &  1.64 &  0.011 &  0.014 &   35.3 &   35.5  \\ 
L55 &-12.9 & -15.4 &  6.44 &  1.78 &  0.012 &  0.010 &   35.4 &   35.3  \\ 
\hline
\end{tabular}
\end{center}
\small Notes.  Columns (2) and (3) display the offsets in arcsec from the position of the continuum peak. Fluxes are in units of $10^{-15}$erg~s$^{-1}$~cm$^{-2}$. Fluxes and luminosities   (Columns 6 to 9) are corrected from Galactic extinction using the reddening coefficients from \cite{Schlafly2011}.
\end{table*}

\normalsize

We also ran our routine in the emission lines maps. We detected 41 sources in  H$\alpha$ and 41 in [\ion{O}{iii}]~$\lambda5007$; of these, 30 sources appear in both lines, 12 only in H$\alpha$, and 13 only in [\ion{O}{iii}]~$\lambda5007$, which gives a total of 55 individual sources. 
Objects detected in emission lines  are most likely SF regions; although some of them could  also be supernova remnants (SNRs; \citealp{Osterbrock2006,Vucetic2013,Vucetic2015,Vucetic2019}).
Only five of these sources have a clear counterpart in the continuum and they are identified as HII in Table~\ref{Table:continuum_sources}.

\smallskip

The main properties of the emission sources are presented in Table~\ref{Table:line_sources}.  Figure~\ref{Figure:clumps} (lower left panel) shows that, by contrast with the continuum ones, their distribution is more homogeneous through the central part of the galaxy.
Nearly all sources are extended---90\% have FWHM larger than 1~arcsec (63~pc)---,  
hence, they are relatively large \ion{H}{ii}-region complexes, and span a wide range in fluxes and luminosities, with  H$\alpha$ luminosity varying from L$_{\mathrm{H}\alpha}$=10$^{34.3}$-10$^{38.9}$~erg~s$^{-1}$. Most objects present luminosities  typical of classical \ion{H}{ii} regions (L$_{\mathrm{H}\alpha}\leq$10$^{37}$~erg~s$^{-1}$), whereas roughly 30\% can be classified as giant \ion{H}{ii} regions\footnote{As a reference, Orion, with a  diameter of about 5~pc  and L(H$\alpha$)=1.0$\times$10$^{37}$erg~cm$^{-1}$ is a classical \ion{H}{ii}-region, while 30~Dor, which has a diameter of about 200~pc and L(H$\alpha$)=1.5$\times$10$^{40}$erg~s$^{-1}$ is a super giant   \ion{H}{ii}-region (\citealp{Kennicutt1984})} (GHIIR; L$_{\mathrm{H}\alpha}$=10$^{37}$-10$^{39}$~erg~s$^{-1}$). The flux of recombination lines is proportional to the number of photons emitted above the Lyman continuum and therefore to the number of ionizing stars. Regions with luminosities 
L$_{\mathrm{H}\alpha}<10^{37}$~erg~s$^{-1}$ are ionized by one or several stars, while GHIIRs must be ionized by multiple associations or stellar clusters. We note that the values shown in  Table~\ref{Table:line_sources} are not corrected for interstellar extinction: the de-reddened  fluxes are somewhat larger. Applying, for instance, the value of the interstellar extinction coefficient derived in CGP17a for the integrated spectrum of Haro~14, C(H$_{\beta}$)=0.379,  we would derive  fluxes  twice as large as the tabulated ones. However, large spatial  variations of the extinction are expected in Haro~14, meaning that an accurate interstellar extinction correction must be done in two dimensions.

\smallskip

The detailed analysis of the properties of the individual sources detected both in continuum and in emission line maps is out of the scope of this paper and will be presented in a forthcoming publication (Cair\'os et al. in prep.)

\section{Discussion}
\label{Section:discussion}

The huge amount of information contained in the MUSE data enabled us to conduct  a thorough  analysis of the BCG Haro~14.  The unique  combination of high spatial resolution, wide FoV, and extended wavelength coverage mean that MUSE  is a powerful tool with which to disentangle the distinct stellar populations and, in particular, to fully characterize the ongoing starburst. This is the 
first step towards establishing the evolutionary status of the galaxy and deriving its star forming history (SFH). In addition, an accurate depiction of the  properties of the young stars (spatial position, ages and metallicities)  may provide  insights into the mechanism(s) that triggers and controls the SF activity.

\subsection{First view on the stellar populations}
\label{stellarpop}

 Like most BCGs, Haro~14 harbors an inner, mostly line-emitting HSB region embedded in a LSB stellar host  \citep{Marlowe1997,Doublier1999,Noeske2003, GildePaz2003}.  The large FoV of MUSE allowed us    
to perform accurate spectrophotometry for both the HSB and the LSB area:  we were able to  investigate the stellar component in the galaxy outskirts, which are free from the contribution of young stars and gas, as well as to resolve filaments of ionized gas and faint \ion{H}{ii}-regions up to kiloparsec scales. This is a decisive asset with respect to previous IFS BCG studies, which,  due to their limited FoV, could only cover the central HSB area, or in most cases  merely a fraction of it \citep{Vanzi2008,Vanzi2011,GarciaLorenzo2008,Cresci2010,Kehrig2008,Kehrig2013,Cairos2009a,Cairos2009b,Cairos2017a,Cairos2017b,Cairos2020,James2010,James2013a,James2013b,Lagos2018,Kumari2017,Kumari2019}. 

\smallskip

The comparison of Haro~14  continuum and emission line maps 
generated our first interesting results (Figures~\ref{Figure:Haro14_continuum} and \ref{Figure:emission}). Continuum and line maps exhibit very distinct morphologies, the most evident difference being found in the outer galaxy regions: the stellar component shows a regular LSB envelope, while in ionized gas the faint intensity regions appear highly irregular and mostly made of filaments, shells, loops, and arcs. This is indeed what is found in the vast majority of BCGs, a dynamically relaxed stellar host underlying a star-forming population \citep{Loose1987,Marlowe1997,Beck1999,Cairos2001b,GildePaz2003}. 

\smallskip

In Haro~14, however, the spatial patterns of the stars and ionized gas are also markedly different at high and intermediate intensity levels.  
At HSB, the continuum shows a main peak and two subsidiary ones within a radius of $\sim$300~pc (knots~{\sc 1}, {\sc 2} and {\sc 3}, top panels in Figure~\ref{Figure:clumps}), while the gas emission is clumpy and extended (the maxima in gas emission spreads up to kiloparsec scales; see Figure~\ref{Figure:clumps}, bottom panels). 
None of the three major continuum emitters spatially coincides with a peak in emission lines, indicating that they are relatively evolved (nonionizing) stellar clusters or associations.

\smallskip 

At intermediate brightness levels, the morphology of Haro~14 is intriguing: 
in the continuum maps, the intensity decreases unevenly, and we distinguish a diffuse structure extending northeast  (Figure~\ref{Figure:Haro14_continuum}). This structure  is also evident in the SBPs of the galaxy as a flattening between 0.9 and 1.6~kpc (Figure~\ref{Figure:SBP}). In the color maps, it appears as a band with intrinsic blue colors that occupies most of the eastern galaxy region. 

\smallskip

Combining the information from the color and the galaxy maps, we identify three different stellar populations:

\begin{itemize}

\item A very young stellar component, depicted in the emission line maps---only rapidly evolving OB stars (ages$\leq$10~Myr) produce  photons energetic enough  to
ionize hydrogen 
\citep{Leitherer1995,Ekstroem2012,Langer2012}. This current starburst presents a highly irregular morphology, takes place in numerous SF complexes, and is considerably extended:
\ion{H}{ii}-regions are detected  over the whole FoV up to galactocentric distances of $\sim$1.8~kpc. 

\smallskip

\item An intermediate-age component, whose presence is suggested by the pure continuum maps (Figure~\ref{Figure:Haro14_continuum}, right panels) and is confirmed by the color (Figure~\ref{Figure:Haro14_broadband}, left panel) and H$\alpha$ equivalent-width maps (Figure~\ref{Figure:emission} bottom left). The continuum frames exhibit an irregular and diffuse appearance in the eastern galaxy regions,  which points to a  
population  of stars in a nonequilibrium state. In the color index map, this area is visible as a wide band 
with blue colors ($V-I\sim 0.5-0.7$), characteristic of young stars, while the  H$\alpha$  equivalent-width map in this region shows very low values (W($H\alpha)\leq 10 \AA$),  proving that ionizing stars  have already evolved---which clearly distinguishes this stellar population from the ongoing starburst. According to evolutionary synthesis models, such blue colors are consistent with a stellar population with ages of between 10 and 300~Myr  \citep{Fioc1997,Leitherer1999,Bruzual2003}. 

\smallskip 

\item An extended  LSB component, whose red colors and smooth, almost round shape are indicative of an old, relaxed  stellar population.  The comparison of the colors with evolutionary synthesis models \citep{Vazdekis1996,Vazdekis2010,Fioc1997,LeBorgne2004} suggests ages of several Gyr, which is in agreement with the values derived by \cite{Noeske2003} and \cite{Marlowe1997}.

\end{itemize}

\subsection{Stellar complexes in Haro~14}
\label{stellarcomplex}

In addition to the unresolved stellar components   in Haro~14, we detected  many clumps   scattered through the galaxy, most of which we identified as stellar complexes (i.e.,  \ion{H}{ii}-regions, stellar clusters, or stellar associations) in Section~\ref{Section:stellarclusters}. A detailed analysis of the individual 
\ion{H}{ii}-regions, clusters, and associations by means of spectral synthesis techniques to determine their masses, extinction, metallicities, and ages will be presented in a forthcoming publication. Here we advance some preliminary results on the properties of these individual stellar groups based mainly on their photometry.

\smallskip 

The \ion{H}{ii} regions identified in the emission line maps belong  evidently to the ongoing starburst.  Using the H$\alpha$ equivalent width we can further constrain their ages. The most luminous SF regions
all present  W(H$\alpha) \geq 100$~\AA\   (see Figure~\ref{Figure:emission} bottom left); 
according to {\sc starburst~99} synthesis models \citep{Leitherer1999} this translates into ages $\leq 6.3$~Myr
---adopting the models with metallicity $z=0.008$, the closest value to the metallicity derived from the emission-line fluxes (CGP17a). These ages  must be understood, however, as an upper limit, as the measured equivalent widths can decrease as a result of absorption from A-F stars and/or dilution due to the continuum from an older stellar population \citep{Fernandes2003,Levesque2013}.
 
\smallskip  
 
\begin{figure}
\centering
\includegraphics[angle=0, width=\linewidth]{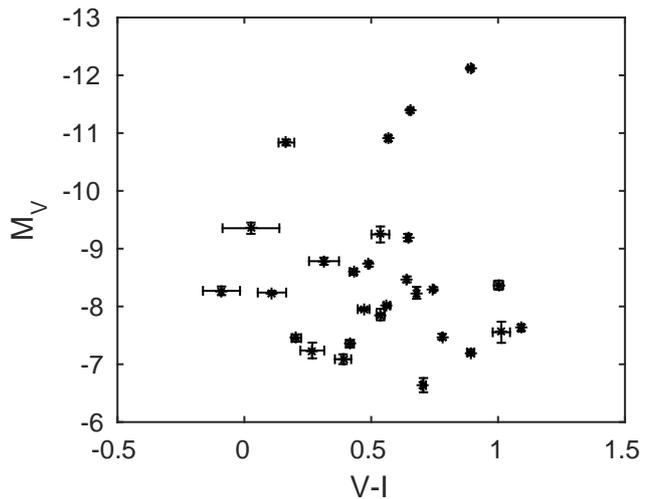}
\caption{M$_{V}$  vs. $(V-I)$  color--magnitude diagram for the sources identified in the continuum maps of Haro~14. 
}
\label{Figure:color_mag}
\end{figure}

By contrast,  the sources detected in continuum are very different regarding their ages and evolutionary paths, ranging from complexes of very young stars still embedded in their natal clouds to ancient globular clusters. The first hint as  to their diversity is  provided by the large variations in  magnitudes and colors they present (Table~\ref{Table:continuum_sources} and Figure~\ref{Figure:color_mag}). Absolute magnitudes vary from $M_{V}=-6.70$ to $M_{V}=-12.18$, while 
the color index $(V-I)$ changes from very blue (-0.12)
to quite red (1.06)---we have excluded here source {\sc c19} identified as a RSG in Section~\ref{Section:stellarclusters}. 
Six stellar complexes show blue colors, are co-spatial with (or very close to) an emission line source,
and have \ion{H}{ii}-like spectra: these are the clusters or associations that are rich in ionizing stars and create the \ion{H}{ii} regions. 
Five clumps are red, with colors---$V-I\geq0.8$---characteristic of intermediate age or old globular clusters.
Most of the sources have intermediate colors in the range $0.2\leq (V-I) \leq 0.8$ and no counterpart in emission lines. These young but nonionizing stellar complexes are gathered in the eastern regions of the galaxy, in the blue zone that is highly visible in the color map, and display similar blue colors. They are probably stellar clusters and/or associations in the process of dissolution.

\smallskip 

The majority of the stellar complexes detected in Haro~14 are extended, that is,  they are  large \ion{H}{ii} regions or stellar associations. Among the point-like objects, the two most luminous,  
{\sc c1} and {\sc c3}, display properties consistent with those of super stellar clusters (SSCs).
In particular, the continuum peak, {\sc c1}, situated  in the inner
galaxy regions and having  $M_{V}=-12.18$, is definitely a very strong SSC candidate.
In CGP17a, we used 
the equivalent widths in absorption of the higher order Balmer lines to constrain the age of the 
continuum peak and found values 
of between 10 and 30~Myr. 
This points to a young age for the SSC candidate c1---
young ages  and red colors  together could be naturally explained by the presence of RSGs.

\smallskip

Super
stellar clusters are extremely compact and luminous objects  thought to be created  in violent episodes of SF
\citep{Arp1985, Holtzman1996,Ho1997}. Their characteristics make them particularly interesting in the context of extragalactic research: because they are bright they can be observed at large distances, and because they are long lived, they retain valuable information  on the SFH of their host galaxies
 \citep{Larsen2010,PortegiesZwart2010,Adamo2020}.

\subsection{The trigger mechanism of SF in Haro~14}
\label{SFtrigger}

An important open question in BCG research is which mechanism triggers such  violent episodes of SF in (apparently) isolated systems \citep{Pustilnik2001,Brosch2004,HunterElmegreen2006, Elmegreen2012}. We find evidence of at least two different SF episodes in Haro~14: the ongoing starburst, with ages of about 6~Myr, and a previous  episode of SF, a precursor to the irregular and  blue population observed in the continuum frames and color maps, which ignited from ten to several hundred million years ago.  We  proposed  in CGP17a that the current starburst was triggered by an earlier (star-induced) SF episode, that is, as the effect of the  collective action of stellar winds and supernova originated in a previous  burst, seemingly ten to thirty million years ago. However, the mechanism responsible for the onset of that previous starburst remains unclear. 

\smallskip 

The findings presented here point to mergers or interactions as a plausible explanation: a pronounced  
asymmetry in the stellar distribution, such as that seen in Haro~14, is a common  feature in galaxy pairs and merger systems  \citep{Conselice2003,Hammer2004,HernandezToledo2005,DePropris2007,Ellison2010,Xu2020}; 
tail structures, such as the one observed in the northeast regions of Haro~14 (see Fig.~\ref{Figure:Haro14_broadband}), also appear as the result of interactions \citep{Toomre1972,Mullan2011,Struck2011,Duc2013};  and the possible 
detection of SSCs suggests at least one episode of violent SF in the past,  as such massive clusters require large  quantities of gas infalling in a small region (high-pressure environments) in a short time interval  \citep{Whitmore2003, Dobbs2020}.

\smallskip 

Our results are in line with an increasing amount of observational evidence suggesting that 
interactions \citep{Brinks1990,Taylor1996,Ostlin2001,Pustilnik2001} or external infalling gas  \citep{LopezSanchez2012,Nidever2013,Ashley2014,Miura2015,Turner2015} do indeed play a major role in the ignition of the starburst in low-mass systems. 
In particular, observational evidence of mergers between dwarf galaxies has grown  considerably  in recent years \citep{MartinezDelgado2012,Paudel2015,Privon2017,Paudel2018,Zhang2020a,Zhang2020b}. 
This opens a very interesting line of research, as mergers and interactions in nearby BCGs provide an excellent chance to probe the hierarchical scenario in conditions very similar to those found in the high-redshift Universe, at the epoch of galaxy formation.

\section{Summary and Conclusions}

This is the first in a series of papers presenting results from a MUSE/VLT-based study of the BCG Haro~14.  MUSE provides integral field data in a large wavelength range (4750-9300~\AA),  with high spatial resolution and unprecedented sensitivity. Moreover, its large FoV  ($1\arcmin\times1\arcmin$=3.8$\times$3.8~kpc$^{2}$)  enables accurate spectroscopy to be obtained not only in the central starburst region but also in the LSB galaxy component. 

\smallskip

We performed an  exhaustive investigation of the morphology, structure, and stellar populations of Haro~14.
We built continuum maps in several spectral regions free of strong emission lines (e.g., 5400-5600 \AA\ and 7800-8000\AA) as well as in the brightest emission lines (H$\alpha$ and [\ion{O}{iii}]~$\lambda5007$). We also generated synthetic broad-band images in the VRI bands of the Johnson-Cousins UBVRI  system, from which we produced color index maps and SBPs. 
We detected numerous discrete sources (clumps) spread out through the galaxy, both in continuum and in emission lines.  We developed a routine that searches for these sources automatically and produced a final catalog with their positions, sizes, and photometry.  

\smallskip

From our analysis we highlight the  following results:

 \begin{itemize}
 
 \item The stellar distribution of Haro~14 is markedly asymmetric.  In continuum maps free from line emission, the intensity peak  is not centered with respect to the LSB host, but displaced by  $\sim$500~pc southwest; from there, the intensity decreases unevenly, with galaxy isophotes clearly elongated in the northeast direction. We identify stellar structure resembling a tail, extending about 1.9~kpc northeast, which likely traces a nonrelaxed stellar component.

\item The galaxy color maps  reveal a blue, but nonionizing stellar component, which covers almost the whole eastern part of the galaxy. This blue component extends to kiloparsec scales and largely overlaps with the tail structure detected in the continuum maps.

\item  We identified (at least) three different stellar populations in Haro~14: an ongoing starburst, a young but nonionizing stellar population, and the extended LSB host. The recent starburst is composed of a large number of  \ion{H}{ii}-region complexes distributed all over the galaxy main body; the brightest  \ion{H}{ii} regions have ages of about 6~Myr. The highly irregular, blue but nonionizing stellar component dominates the emission in the eastern part of the galaxy, and shows observables consistent with ages of between ten and several hundred million years. Finally, the LSB host is a red and regular-shaped component of several gigayears.

\item In addition to the unresolved stellar components, we also identify a large number of discrete sources in the galaxy: 55 in emission line maps and 29 in the continuum---only 6 belonging to both groups. 
The gas emission sources are most likely H~{\sc ii} regions.
The sources in the continuum include, along with the six ionizing groups associated with H~{\sc ii} regions, at least 16 other stellar complexes in different evolutive stages. 
The two most luminous ones have properties consistent with an SSC, in particular 
the continuum peak (M$_{v}$=-12.18)  is a strong SSC candidate.


\end{itemize}

\smallskip 

The striking asymmetry of the stellar distribution in Haro~14 in continuum maps, the highly irregular blue stellar population detected in the color map, the anomalous behavior of its SBP  which  cannot be described by a simple law even at large radii, and the presence of numerous stellar complexes, some of them young massive and super stellar cluster candidates, speak in favor of the possibility that interactions or mergers have taken place in Haro~14.  
This is indeed an appealing hypothesis, as interactions or mergers in nearby, low-mass, metal-poor starburst systems would provide an excellent opportunity to probe the hierarchical scenario in conditions very similar to those found in the early Universe. 
In a forthcoming publication, we shall explore the interaction or merger scenario with results from a kinematical analysis of the MUSE data.

\smallskip 

 Studies focused on the stellar content and SFH of low-mass metal-poor starbursts are crucial for  understanding galaxy formation and evolution. It is particularly important to determine the recent SFH in these objects, as this contains valuable information on the impact of massive stellar feedback within a low gravitational potential, and on the mechanism that ignites the burst. Detailed studies on the SFH of BCGs are unfortunately scarce: the color--magnitude diagram (CMD) synthesis method can be applied only to nearby systems \citep{Tolstoy2009} and BCGs are rare objects; only a handful are close enough to be resolved into stars \citep{Lee2009,Karachentsev2019}. 
The few BCGs analyzed by means of CMD methods indeed show a complex and varied recent SFH---e.g., VII~Zw~403 \citep{SchulteLadbeck1999a,SchulteLadbeck1999b}; Mrk~178 \citep{SchulteLadbeck2000}; NGC~1705 \citep{Annibali2003,Annibali2009}; NGC~5253 \citep{Davidge2007,McQuinn2010}; UGC~4483 \citep{Izotov2002,McQuinn2010,Sacchi2021}. This emphasizes the importance of determining the SFH in a large and representative sample of low-mass starbursts.

\smallskip

The present work shows that integral field spectroscopy 
provides an excellent alternative technique with which to investigate the stellar populations and SFH of galaxies that cannot be resolved into stars. 
In particular, the presently unique capabilities of MUSE, namely its wide FoV, large spectral range, and high spatial resolution, play a decisive role in investigations of the stellar populations and SFHs of dwarf star-forming galaxies.

\begin{acknowledgements} LMC acknowledges support from the Deutsche Forschungsgemeinschaft
(CA~1243/1-1 and CA~1234/1-2). PMW received support from BMBF Verbundforschung (project MUSE-NFM, grant 05A17BAA).  Based on observations collected at the European Southern Observatory under ESO programme ID 60.A-9186(A). 
This research has made use of the NASA/IPAC Extragalactic
Database (NED), which is operated by the Jet Propulsion Laboratory, Caltech,
under contract with the National Aeronautics and Space Administration. We acknowledge the usage of the HyperLeda database (http://leda.univ-lyon1.fr).  This work also used IRAF package, which are distributed by the National Optical Astronomy Observatory, which is operated by the Association of Universities for Research in Astronomy, Inc., under contract with the National Science Foundation.
\end{acknowledgements}

\bibliographystyle{aa}
\bibliography{muse}

\begin{thebibliography}{193}
\expandafter\ifx\csname natexlab\endcsname\relax\def\natexlab#1{#1}\fi

\bibitem[{{Adamo} {et~al.}(2020){Adamo}, {Zeidler}, {Kruijssen}, {Chevance},
  {Gieles}, {Calzetti}, {Charbonnel}, {Zinnecker}, \& {Krause}}]{Adamo2020}
{Adamo}, A., {Zeidler}, P., {Kruijssen}, J.~M.~D., {et~al.} 2020, \ssr, 216, 69

\bibitem[{{Aller}(1984)}]{Aller1984}
{Aller}, L.~H. 1984, {Physics of thermal gaseous nebulae}

\bibitem[{{Anders} \& {Fritze-v. Alvensleben}(2003)}]{Anders2003}
{Anders}, P. \& {Fritze-v. Alvensleben}, U. 2003, \aap, 401, 1063

\bibitem[{{Annibali} {et~al.}(2003){Annibali}, {Greggio}, {Tosi}, {Aloisi}, \&
  {Leitherer}}]{Annibali2003}
{Annibali}, F., {Greggio}, L., {Tosi}, M., {Aloisi}, A., \& {Leitherer}, C.
  2003, \aj, 126, 2752

\bibitem[{{Annibali} {et~al.}(2009){Annibali}, {Tosi}, {Monelli}, {Sirianni},
  {Montegriffo}, {Aloisi}, \& {Greggio}}]{Annibali2009}
{Annibali}, F., {Tosi}, M., {Monelli}, M., {et~al.} 2009, \aj, 138, 169

\bibitem[{{Arp} \& {Sandage}(1985)}]{Arp1985}
{Arp}, H. \& {Sandage}, A. 1985, \aj, 90, 1163

\bibitem[{{Ashley} {et~al.}(2014){Ashley}, {Elmegreen}, {Johnson}, {Nidever},
  {Simpson}, \& {Pokhrel}}]{Ashley2014}
{Ashley}, T., {Elmegreen}, B.~G., {Johnson}, M., {et~al.} 2014, \aj, 148, 130

\bibitem[{{Bacon} {et~al.}(2010){Bacon}, {Accardo}, {Adjali}, {Anwand},
  {Bauer}, {Biswas}, {Blaizot}, {Boudon}, {Brau-Nogue}, {Brinchmann},
  {Caillier}, {Capoani}, {Carollo}, {Contini}, {Couderc}, {Daguis{\'e}},
  {Deiries}, {Delabre}, {Dreizler}, {Dubois}, {Dupieux}, {Dupuy}, {Emsellem},
  {Fechner}, {Fleischmann}, {Fran{\c c}ois}, {Gallou}, {Gharsa}, {Glindemann},
  {Gojak}, {Guiderdoni}, {Hansali}, {Hahn}, {Jarno}, {Kelz}, {Koehler},
  {Kosmalski}, {Laurent}, {Le Floch}, {Lilly}, {Lizon}, {Loupias}, {Manescau},
  {Monstein}, {Nicklas}, {Olaya}, {Pares}, {Pasquini}, {P{\'e}contal-Rousset},
  {Pell{\'o}}, {Petit}, {Popow}, {Reiss}, {Remillieux}, {Renault}, {Roth},
  {Rupprecht}, {Serre}, {Schaye}, {Soucail}, {Steinmetz}, {Streicher}, {Stuik},
  {Valentin}, {Vernet}, {Weilbacher}, {Wisotzki}, \& {Yerle}}]{Bacon2010}
{Bacon}, R., {Accardo}, M., {Adjali}, L., {et~al.} 2010, in \procspie, Vol.
  7735, Ground-based and Airborne Instrumentation for Astronomy III, 773508

\bibitem[{{Bassino} \& {Caso}(2017)}]{Bassino2017}
{Bassino}, L.~P. \& {Caso}, J.~P. 2017, \mnras, 466, 4259

\bibitem[{{Beck} \& {Kovo}(1999)}]{Beck1999}
{Beck}, S.~C. \& {Kovo}, O. 1999, \aj, 117, 190

\bibitem[{{Bessell}(1990)}]{Bessell1990}
{Bessell}, M.~S. 1990, \pasp, 102, 1181

\bibitem[{{Bessell} {et~al.}(1998){Bessell}, {Castelli}, \&
  {Plez}}]{Bessell1998}
{Bessell}, M.~S., {Castelli}, F., \& {Plez}, B. 1998, \aap, 333, 231

\bibitem[{{Bik} {et~al.}(2015){Bik}, {{\"O}stlin}, {Hayes}, {Adamo},
  {Melinder}, \& {Amram}}]{Bik2015}
{Bik}, A., {{\"O}stlin}, G., {Hayes}, M., {et~al.} 2015, \aap, 576, L13

\bibitem[{{Bik} {et~al.}(2018){Bik}, {{\"O}stlin}, {Menacho}, {Adamo}, {Hayes},
  {Herenz}, \& {Melinder}}]{Bik2018}
{Bik}, A., {{\"O}stlin}, G., {Menacho}, V., {et~al.} 2018, \aap, 619, A131

\bibitem[{{Brinks}(1990)}]{Brinks1990}
{Brinks}, E. 1990, {II Zwicky 33: star formation induced by a recent
  interaction.}, ed. {Wielen, R.}, 146--149

\bibitem[{{Brosch} {et~al.}(2004){Brosch}, {Almoznino}, \&
  {Heller}}]{Brosch2004}
{Brosch}, N., {Almoznino}, E., \& {Heller}, A.~B. 2004, \mnras, 349, 357

\bibitem[{{Bruzual} \& {Charlot}(2003)}]{Bruzual2003}
{Bruzual}, G. \& {Charlot}, S. 2003, \mnras, 344, 1000

\bibitem[{{Cair{\'o}s} {et~al.}(2012){Cair{\'o}s}, {Caon}, {Garc{\'{\i}}a
  Lorenzo}, {Kelz}, {Roth}, {Papaderos}, \& {Streicher}}]{Cairos2012}
{Cair{\'o}s}, L.~M., {Caon}, N., {Garc{\'{\i}}a Lorenzo}, B., {et~al.} 2012,
  \aap, 547, A24

\bibitem[{{Cair{\'o}s} {et~al.}(2007){Cair{\'o}s}, {Caon},
  {Garc{\'{\i}}a-Lorenzo}, {Monreal-Ibero}, {Amor{\'{\i}}n}, {Weilbacher}, \&
  {Papaderos}}]{Cairos2007}
{Cair{\'o}s}, L.~M., {Caon}, N., {Garc{\'{\i}}a-Lorenzo}, B., {et~al.} 2007,
  \apj, 669, 251

\bibitem[{{Cair{\'o}s} {et~al.}(2002){Cair{\'o}s}, {Caon},
  {Garc{\'{\i}}a-Lorenzo}, {V{\'{\i}}lchez}, \&
  {Mu{\~n}oz-Tu{\~n}{\'o}n}}]{Cairos2002}
{Cair{\'o}s}, L.~M., {Caon}, N., {Garc{\'{\i}}a-Lorenzo}, B., {V{\'{\i}}lchez},
  J.~M., \& {Mu{\~n}oz-Tu{\~n}{\'o}n}, C. 2002, \apj, 577, 164

\bibitem[{{Cair{\'o}s} {et~al.}(2009{\natexlab{a}}){Cair{\'o}s}, {Caon},
  {Papaderos}, {Kehrig}, {Weilbacher}, {Roth}, \& {Zurita}}]{Cairos2009b}
{Cair{\'o}s}, L.~M., {Caon}, N., {Papaderos}, P., {et~al.} 2009{\natexlab{a}},
  \apj, 707, 1676

\bibitem[{{Cair{\'o}s} {et~al.}(2003){Cair{\'o}s}, {Caon}, {Papaderos},
  {Noeske}, {V{\'{\i}}lchez}, {Lorenzo}, \&
  {Mu{\~n}oz-Tu{\~n}{\'o}n}}]{Cairos2003}
{Cair{\'o}s}, L.~M., {Caon}, N., {Papaderos}, P., {et~al.} 2003, \apj, 593, 312

\bibitem[{{Cair{\'o}s} {et~al.}(2001{\natexlab{a}}){Cair{\'o}s}, {Caon},
  {V{\'{\i}}lchez}, {Gonz{\'a}lez-P{\'e}rez}, \&
  {Mu{\~n}oz-Tu{\~n}{\'o}n}}]{Cairos2001b}
{Cair{\'o}s}, L.~M., {Caon}, N., {V{\'{\i}}lchez}, J.~M.,
  {Gonz{\'a}lez-P{\'e}rez}, J.~N., \& {Mu{\~n}oz-Tu{\~n}{\'o}n}, C.
  2001{\natexlab{a}}, \apjs, 136, 393

\bibitem[{{Cair{\'o}s} {et~al.}(2015){Cair{\'o}s}, {Caon}, \&
  {Weilbacher}}]{Cairos2015}
{Cair{\'o}s}, L.~M., {Caon}, N., \& {Weilbacher}, P.~M. 2015, \aap, 577, A21

\bibitem[{{Cair{\'o}s} {et~al.}(2010){Cair{\'o}s}, {Caon}, {Zurita}, {Kehrig},
  {Roth}, \& {Weilbacher}}]{Cairos2010}
{Cair{\'o}s}, L.~M., {Caon}, N., {Zurita}, C., {et~al.} 2010, \aap, 520, A90+

\bibitem[{{Cair{\'o}s} {et~al.}(2009{\natexlab{b}}){Cair{\'o}s}, {Caon},
  {Zurita}, {Kehrig}, {Weilbacher}, \& {Roth}}]{Cairos2009a}
{Cair{\'o}s}, L.~M., {Caon}, N., {Zurita}, C., {et~al.} 2009{\natexlab{b}},
  \aap, 507, 1291

\bibitem[{{Cair{\'o}s} \&
  {Gonz{\'a}lez-P{\'e}rez}(2017{\natexlab{a}})}]{Cairos2017a}
{Cair{\'o}s}, L.~M. \& {Gonz{\'a}lez-P{\'e}rez}, J.~N. 2017{\natexlab{a}},
  \aap, 600, A125

\bibitem[{{Cair{\'o}s} \&
  {Gonz{\'a}lez-P{\'e}rez}(2017{\natexlab{b}})}]{Cairos2017b}
{Cair{\'o}s}, L.~M. \& {Gonz{\'a}lez-P{\'e}rez}, J.~N. 2017{\natexlab{b}},
  \aap, 608, A119

\bibitem[{{Cair{\'o}s} \& {Gonz{\'a}lez-P{\'e}rez}(2020)}]{Cairos2020}
{Cair{\'o}s}, L.~M. \& {Gonz{\'a}lez-P{\'e}rez}, J.~N. 2020, \aap, 634, A95

\bibitem[{{Cair{\'o}s} {et~al.}(2001{\natexlab{b}}){Cair{\'o}s},
  {V{\'{\i}}lchez}, {Gonz{\'a}lez P{\'e}rez}, {Iglesias-P{\'a}ramo}, \&
  {Caon}}]{Cairos2001a}
{Cair{\'o}s}, L.~M., {V{\'{\i}}lchez}, J.~M., {Gonz{\'a}lez P{\'e}rez}, J.~N.,
  {Iglesias-P{\'a}ramo}, J., \& {Caon}, N. 2001{\natexlab{b}}, \apjs, 133, 321

\bibitem[{{Calzetti} {et~al.}(1999){Calzetti}, {Conselice}, {Gallagher}, \&
  {Kinney}}]{Calzetti1999}
{Calzetti}, D., {Conselice}, C.~J., {Gallagher}, John~S., I., \& {Kinney},
  A.~L. 1999, \aj, 118, 797

\bibitem[{{Calzetti} {et~al.}(2004){Calzetti}, {Harris}, {Gallagher}, {Smith},
  {Conselice}, {Homeier}, \& {Kewley}}]{Calzetti2004}
{Calzetti}, D., {Harris}, J., {Gallagher}, John~S., I., {et~al.} 2004, \aj,
  127, 1405

\bibitem[{{Chisholm} {et~al.}(2018){Chisholm}, {Tremonti}, \&
  {Leitherer}}]{Chisholm2018}
{Chisholm}, J., {Tremonti}, C., \& {Leitherer}, C. 2018, \mnras, 481, 1690

\bibitem[{{Conselice}(2003)}]{Conselice2003}
{Conselice}, C.~J. 2003, \apjs, 147, 1

\bibitem[{{Cresci} {et~al.}(2010){Cresci}, {Vanzi}, {Sauvage}, {Santangelo}, \&
  {van der Werf}}]{Cresci2010}
{Cresci}, G., {Vanzi}, L., {Sauvage}, M., {Santangelo}, G., \& {van der Werf},
  P. 2010, \aap, 520, A82

\bibitem[{{Cresci} {et~al.}(2017){Cresci}, {Vanzi}, {Telles}, {Lanzuisi},
  {Brusa}, {Mingozzi}, {Sauvage}, \& {Johnson}}]{Cresci2017}
{Cresci}, G., {Vanzi}, L., {Telles}, E., {et~al.} 2017, \aap, 604, A101

\bibitem[{{Davidge}(2007)}]{Davidge2007}
{Davidge}, T.~J. 2007, \aj, 134, 1799

\bibitem[{{Davies}(2007)}]{Davies2007}
{Davies}, R.~I. 2007, \mnras, 375, 1099

\bibitem[{{De Propris} {et~al.}(2007){De Propris}, {Conselice}, {Liske},
  {Driver}, {Patton}, {Graham}, \& {Allen}}]{DePropris2007}
{De Propris}, R., {Conselice}, C.~J., {Liske}, J., {et~al.} 2007, \apj, 666,
  212

\bibitem[{{de Vaucouleurs} {et~al.}(1991){de Vaucouleurs}, {de Vaucouleurs},
  {Corwin}, {Buta}, {Paturel}, \& {Fouque}}]{deVaucouleurs1991}
{de Vaucouleurs}, G., {de Vaucouleurs}, A., {Corwin}, Jr., H.~G., {et~al.}
  1991, {Third Reference Catalogue of Bright Galaxies}, ed. {de Vaucouleurs,
  G., de Vaucouleurs, A., Corwin, H.~G., Jr., Buta, R.~J., Paturel, G., \&
  Fouque, P.}

\bibitem[{{De Young} \& {Heckman}(1994)}]{DeYoung1994}
{De Young}, D.~S. \& {Heckman}, T.~M. 1994, \apj, 431, 598

\bibitem[{{Dekel} \& {Silk}(1986)}]{Dekel1986}
{Dekel}, A. \& {Silk}, J. 1986, \apj, 303, 39

\bibitem[{{Dobbs} {et~al.}(2020){Dobbs}, {Liow}, \& {Rieder}}]{Dobbs2020}
{Dobbs}, C.~L., {Liow}, K.~Y., \& {Rieder}, S. 2020, \mnras, 496, L1

\bibitem[{{Doublier} {et~al.}(1999){Doublier}, {Caulet}, \&
  {Comte}}]{Doublier1999}
{Doublier}, V., {Caulet}, A., \& {Comte}, G. 1999, \aaps, 138, 213

\bibitem[{{Doublier} {et~al.}(2001){Doublier}, {Caulet}, \&
  {Comte}}]{Doublier2001}
{Doublier}, V., {Caulet}, A., \& {Comte}, G. 2001, \aap, 367, 33

\bibitem[{{Doublier} {et~al.}(1997){Doublier}, {Comte}, {Petrosian}, {Surace},
  \& {Turatto}}]{Doublier1997}
{Doublier}, V., {Comte}, G., {Petrosian}, A., {Surace}, C., \& {Turatto}, M.
  1997, \aaps, 124, 405

\bibitem[{{Duc} \& {Renaud}(2013)}]{Duc2013}
{Duc}, P.-A. \& {Renaud}, F. 2013, {Tides in Colliding Galaxies}, ed.
  J.~{Souchay}, S.~{Mathis}, \& T.~{Tokieda}, Vol. 861, 327

\bibitem[{{Ekstr{\"o}m} {et~al.}(2012){Ekstr{\"o}m}, {Georgy}, {Eggenberger},
  {Meynet}, {Mowlavi}, {Wyttenbach}, {Granada}, {Decressin}, {Hirschi},
  {Frischknecht}, {Charbonnel}, \& {Maeder}}]{Ekstroem2012}
{Ekstr{\"o}m}, S., {Georgy}, C., {Eggenberger}, P., {et~al.} 2012, \aap, 537,
  A146

\bibitem[{{Ellison} {et~al.}(2010){Ellison}, {Patton}, {Simard}, {McConnachie},
  {Baldry}, \& {Mendel}}]{Ellison2010}
{Ellison}, S.~L., {Patton}, D.~R., {Simard}, L., {et~al.} 2010, \mnras, 407,
  1514

\bibitem[{{Elmegreen} {et~al.}(2012){Elmegreen}, {Zhang}, \&
  {Hunter}}]{Elmegreen2012}
{Elmegreen}, B.~G., {Zhang}, H.-X., \& {Hunter}, D.~A. 2012, \apj, 747, 105

\bibitem[{{Fernandes} {et~al.}(2003){Fernandes}, {Le{\~a}o}, \&
  {Lacerda}}]{Fernandes2003}
{Fernandes}, R.~C., {Le{\~a}o}, J.~R.~S., \& {Lacerda}, R.~R. 2003, \mnras,
  340, 29

\bibitem[{{Fioc} \& {Rocca-Volmerange}(1997)}]{Fioc1997}
{Fioc}, M. \& {Rocca-Volmerange}, B. 1997, \aap, 500, 507

\bibitem[{{Fukugita} {et~al.}(1996){Fukugita}, {Ichikawa}, {Gunn}, {Doi},
  {Shimasaku}, \& {Schneider}}]{Fukugita1996}
{Fukugita}, M., {Ichikawa}, T., {Gunn}, J.~E., {et~al.} 1996, \aj, 111, 1748

\bibitem[{{Garc{\'{\i}}a-Lorenzo} {et~al.}(2008){Garc{\'{\i}}a-Lorenzo},
  {Cair{\'o}s}, {Caon}, {Monreal-Ibero}, \& {Kehrig}}]{GarciaLorenzo2008}
{Garc{\'{\i}}a-Lorenzo}, B., {Cair{\'o}s}, L.~M., {Caon}, N., {Monreal-Ibero},
  A., \& {Kehrig}, C. 2008, \apj, 677, 201

\bibitem[{{Garnett}(2002)}]{Garnett2002}
{Garnett}, D.~R. 2002, \apj, 581, 1019

\bibitem[{{Gil de Paz} \& {Madore}(2005)}]{GildePaz2005}
{Gil de Paz}, A. \& {Madore}, B.~F. 2005, \apjs, 156, 345

\bibitem[{{Gil de Paz} {et~al.}(2003){Gil de Paz}, {Madore}, \&
  {Pevunova}}]{GildePaz2003}
{Gil de Paz}, A., {Madore}, B.~F., \& {Pevunova}, O. 2003, \apjs, 147, 29

\bibitem[{{Gonz{\'a}lez Delgado} \& {Leitherer}(1999)}]{GonzalezDelgado1999a}
{Gonz{\'a}lez Delgado}, R.~M. \& {Leitherer}, C. 1999, \apjs, 125, 479

\bibitem[{{Gonz{\'a}lez Delgado} {et~al.}(1999){Gonz{\'a}lez Delgado},
  {Leitherer}, \& {Heckman}}]{GonzalezDelgado1999b}
{Gonz{\'a}lez Delgado}, R.~M., {Leitherer}, C., \& {Heckman}, T.~M. 1999,
  \apjs, 125, 489

\bibitem[{{Gordon} \& {Gottesman}(1981)}]{Gordon1981}
{Gordon}, D. \& {Gottesman}, S.~T. 1981, \aj, 86, 161

\bibitem[{{Guseva} {et~al.}(2001){Guseva}, {Izotov}, {Papaderos}, {Chaffee},
  {Foltz}, {Green}, {Thuan}, {Fricke}, \& {Noeske}}]{Guseva2001}
{Guseva}, N.~G., {Izotov}, Y.~I., {Papaderos}, P., {et~al.} 2001, \aap, 378,
  756

\bibitem[{{Guseva} {et~al.}(2003{\natexlab{a}}){Guseva}, {Papaderos}, {Izotov},
  {Green}, {Fricke}, {Thuan}, \& {Noeske}}]{Guseva2003c}
{Guseva}, N.~G., {Papaderos}, P., {Izotov}, Y.~I., {et~al.} 2003{\natexlab{a}},
  \aap, 407, 75

\bibitem[{{Guseva} {et~al.}(2003{\natexlab{b}}){Guseva}, {Papaderos}, {Izotov},
  {Green}, {Fricke}, {Thuan}, \& {Noeske}}]{Guseva2003b}
{Guseva}, N.~G., {Papaderos}, P., {Izotov}, Y.~I., {et~al.} 2003{\natexlab{b}},
  \aap, 407, 91

\bibitem[{{Guseva} {et~al.}(2003{\natexlab{c}}){Guseva}, {Papaderos}, {Izotov},
  {Green}, {Fricke}, {Thuan}, \& {Noeske}}]{Guseva2003a}
{Guseva}, N.~G., {Papaderos}, P., {Izotov}, Y.~I., {et~al.} 2003{\natexlab{c}},
  \aap, 407, 105

\bibitem[{{Guseva} {et~al.}(2004){Guseva}, {Papaderos}, {Izotov}, {Noeske}, \&
  {Fricke}}]{Guseva2004}
{Guseva}, N.~G., {Papaderos}, P., {Izotov}, Y.~I., {Noeske}, K.~G., \&
  {Fricke}, K.~J. 2004, \aap, 421, 519

\bibitem[{{Harris}(1996)}]{Harris1996}
{Harris}, W.~E. 1996, \aj, 112, 1487

\bibitem[{{Hart}(2019)}]{Hart2019}
{Hart}, M. 2019, \aj, 157, 221

\bibitem[{{Heckman} {et~al.}(1990){Heckman}, {Armus}, \& {Miley}}]{Heckman1990}
{Heckman}, T.~M., {Armus}, L., \& {Miley}, G.~K. 1990, \apjs, 74, 833

\bibitem[{{Hern{\'a}ndez-Toledo} {et~al.}(2005){Hern{\'a}ndez-Toledo},
  {Avila-Reese}, {Conselice}, \& {Puerari}}]{HernandezToledo2005}
{Hern{\'a}ndez-Toledo}, H.~M., {Avila-Reese}, V., {Conselice}, C.~J., \&
  {Puerari}, I. 2005, \aj, 129, 682

\bibitem[{{Ho}(1997)}]{Ho1997}
{Ho}, L.~C. 1997, in Revista Mexicana de Astronomia y Astrofisica Conference
  Series, Vol.~6, Revista Mexicana de Astronomia y Astrofisica Conference
  Series, ed. J.~{Franco}, R.~{Terlevich}, \& A.~{Serrano}, 5

\bibitem[{{Holtzman} {et~al.}(1996){Holtzman}, {Watson}, {Mould}, {Gallagher},
  {Ballester}, {Burrows}, {Clarke}, {Crisp}, {Evans}, {Griffiths}, {Hester},
  {Hoessel}, {Scowen}, {Stapelfeldt}, {Trauger}, \& {Westphal}}]{Holtzman1996}
{Holtzman}, J.~A., {Watson}, A.~M., {Mould}, J.~R., {et~al.} 1996, \aj, 112,
  416

\bibitem[{{Hunter}(1997)}]{Hunter1997}
{Hunter}, D. 1997, \pasp, 109, 937

\bibitem[{{Hunter} \& {Elmegreen}(2004)}]{HunterElmegreen2004}
{Hunter}, D.~A. \& {Elmegreen}, B.~G. 2004, \aj, 128, 2170

\bibitem[{{Hunter} \& {Elmegreen}(2006)}]{HunterElmegreen2006}
{Hunter}, D.~A. \& {Elmegreen}, B.~G. 2006, \apjs, 162, 49

\bibitem[{{Hunter} \& {Hoffman}(1999)}]{Hunter1999}
{Hunter}, D.~A. \& {Hoffman}, L. 1999, \aj, 117, 2789

\bibitem[{{Izotov} {et~al.}(2011){Izotov}, {Guseva}, \& {Thuan}}]{Izotov2011}
{Izotov}, Y.~I., {Guseva}, N.~G., \& {Thuan}, T.~X. 2011, \apj, 728, 161

\bibitem[{{Izotov} \& {Thuan}(1999)}]{Izotov1999}
{Izotov}, Y.~I. \& {Thuan}, T.~X. 1999, \apj, 511, 639

\bibitem[{{Izotov} \& {Thuan}(2002)}]{Izotov2002}
{Izotov}, Y.~I. \& {Thuan}, T.~X. 2002, \apj, 567, 875

\bibitem[{{Izotov} {et~al.}(1997){Izotov}, {Thuan}, \&
  {Lipovetsky}}]{Izotov1997}
{Izotov}, Y.~I., {Thuan}, T.~X., \& {Lipovetsky}, V.~A. 1997, \apjs, 108, 1

\bibitem[{{James} {et~al.}(2020){James}, {Kumari}, {Emerick}, {Koposov},
  {McQuinn}, {Stark}, {Belokurov}, \& {Maiolino}}]{James2020}
{James}, B.~L., {Kumari}, N., {Emerick}, A., {et~al.} 2020, \mnras, 495, 2564

\bibitem[{{James} {et~al.}(2010){James}, {Tsamis}, \& {Barlow}}]{James2010}
{James}, B.~L., {Tsamis}, Y.~G., \& {Barlow}, M.~J. 2010, \mnras, 401, 759

\bibitem[{{James} {et~al.}(2013{\natexlab{a}}){James}, {Tsamis}, {Barlow},
  {Walsh}, \& {Westmoquette}}]{James2013b}
{James}, B.~L., {Tsamis}, Y.~G., {Barlow}, M.~J., {Walsh}, J.~R., \&
  {Westmoquette}, M.~S. 2013{\natexlab{a}}, \mnras, 428, 86

\bibitem[{{James} {et~al.}(2013{\natexlab{b}}){James}, {Tsamis}, {Walsh},
  {Barlow}, \& {Westmoquette}}]{James2013a}
{James}, B.~L., {Tsamis}, Y.~G., {Walsh}, J.~R., {Barlow}, M.~J., \&
  {Westmoquette}, M.~S. 2013{\natexlab{b}}, \mnras, 430, 2097

\bibitem[{{Janowiecki} \& {Salzer}(2014)}]{Janowiecki2014}
{Janowiecki}, S. \& {Salzer}, J.~J. 2014, \apj, 793, 109

\bibitem[{{Jedrzejewski}(1987)}]{Jedrzejewski1987}
{Jedrzejewski}, R.~I. 1987, \mnras, 226, 747

\bibitem[{{Johnson} {et~al.}(2000){Johnson}, {Leitherer}, {Vacca}, \&
  {Conti}}]{Johnson2000}
{Johnson}, K.~E., {Leitherer}, C., {Vacca}, W.~D., \& {Conti}, P.~S. 2000, \aj,
  120, 1273

\bibitem[{{Karachentsev} \& {Kaisina}(2019)}]{Karachentsev2019}
{Karachentsev}, I.~D. \& {Kaisina}, E.~I. 2019, Astrophysical Bulletin, 74, 111

\bibitem[{{Kauffmann} {et~al.}(1994){Kauffmann}, {Guiderdoni}, \&
  {White}}]{Kauffmann1994}
{Kauffmann}, G., {Guiderdoni}, B., \& {White}, S.~D.~M. 1994, \mnras, 267, 981

\bibitem[{{Kauffmann} {et~al.}(1993){Kauffmann}, {White}, \&
  {Guiderdoni}}]{Kauffmann1993}
{Kauffmann}, G., {White}, S.~D.~M., \& {Guiderdoni}, B. 1993, \mnras, 264, 201

\bibitem[{{Kehrig} {et~al.}(2013){Kehrig}, {P{\'e}rez-Montero}, {V{\'\i}lchez},
  {Brinchmann}, {Kunth}, {Garc{\'\i}a-Benito}, {Crowther},
  {Hern{\'a}ndez-Fern{\'a}ndez}, {Durret}, {Contini},
  {Fern{\'a}ndez-Mart{\'\i}n}, \& {James}}]{Kehrig2013}
{Kehrig}, C., {P{\'e}rez-Montero}, E., {V{\'\i}lchez}, J.~M., {et~al.} 2013,
  \mnras, 432, 2731

\bibitem[{{Kehrig} {et~al.}(2018){Kehrig}, {V{\'\i}lchez}, {Guerrero},
  {Iglesias-P{\'a}ramo}, {Hunt}, {Duarte-Puertas}, \&
  {Ramos-Larios}}]{Kehrig2018}
{Kehrig}, C., {V{\'\i}lchez}, J.~M., {Guerrero}, M.~A., {et~al.} 2018, \mnras,
  480, 1081

\bibitem[{{Kehrig} {et~al.}(2008){Kehrig}, {V{\'{\i}}lchez}, {S{\'a}nchez},
  {Telles}, {P{\'e}rez-Montero}, \& {Mart{\'{\i}}n-Gord{\'o}n}}]{Kehrig2008}
{Kehrig}, C., {V{\'{\i}}lchez}, J.~M., {S{\'a}nchez}, S.~F., {et~al.} 2008,
  \aap, 477, 813

\bibitem[{{Kennicutt}(1984)}]{Kennicutt1984}
{Kennicutt}, R.~C., J. 1984, \apj, 287, 116

\bibitem[{{Koleva} {et~al.}(2014){Koleva}, {De Rijcke}, {Zeilinger}, {Verbeke},
  {Schroyen}, \& {Vermeylen}}]{Koleva2014}
{Koleva}, M., {De Rijcke}, S., {Zeilinger}, W.~W., {et~al.} 2014, \mnras, 441,
  452

\bibitem[{{Krueger} {et~al.}(1995){Krueger}, {Fritze-v. Alvensleben}, \&
  {Loose}}]{Krueger1995}
{Krueger}, H., {Fritze-v. Alvensleben}, U., \& {Loose}, H.~H. 1995, \aap, 303,
  41

\bibitem[{{Kumari} {et~al.}(2017){Kumari}, {James}, \& {Irwin}}]{Kumari2017}
{Kumari}, N., {James}, B.~L., \& {Irwin}, M.~J. 2017, \mnras, 470, 4618

\bibitem[{{Kumari} {et~al.}(2019){Kumari}, {James}, {Irwin}, \&
  {Aloisi}}]{Kumari2019}
{Kumari}, N., {James}, B.~L., {Irwin}, M.~J., \& {Aloisi}, A. 2019, \mnras,
  485, 1103

\bibitem[{{Kunth} \& {{\"O}stlin}(2000)}]{Kunth2000}
{Kunth}, D. \& {{\"O}stlin}, G. 2000, \aapr, 10, 1

\bibitem[{{Kurtz} \& {Mink}(2000)}]{Kurtz2000}
{Kurtz}, M.~J. \& {Mink}, D.~J. 2000, \apjl, 533, L183

\bibitem[{{Lagos} {et~al.}(2018){Lagos}, {Scott}, {Nigoche-Netro}, {Demarco},
  {Humphrey}, \& {Papaderos}}]{Lagos2018}
{Lagos}, P., {Scott}, T.~C., {Nigoche-Netro}, A., {et~al.} 2018, \mnras, 477,
  392

\bibitem[{{Langer}(2012)}]{Langer2012}
{Langer}, N. 2012, \araa, 50, 107

\bibitem[{{Larsen}(2010)}]{Larsen2010}
{Larsen}, S.~S. 2010, Philosophical Transactions of the Royal Society of London
  Series A, 368, 867

\bibitem[{{Larson}(1974)}]{Larson1974}
{Larson}, R.~B. 1974, \mnras, 169, 229

\bibitem[{{Le Borgne} {et~al.}(2004){Le Borgne}, {Rocca-Volmerange},
  {Prugniel}, {Lan{\c{c}}on}, {Fioc}, \& {Soubiran}}]{LeBorgne2004}
{Le Borgne}, D., {Rocca-Volmerange}, B., {Prugniel}, P., {et~al.} 2004, \aap,
  425, 881

\bibitem[{{Lee} {et~al.}(2009){Lee}, {Kennicutt}, {Funes}, {Sakai}, \&
  {Akiyama}}]{Lee2009}
{Lee}, J.~C., {Kennicutt}, Robert~C., J., {Funes}, S.~J. J.~G., {Sakai}, S., \&
  {Akiyama}, S. 2009, \apj, 692, 1305

\bibitem[{{Leibundgut} {et~al.}(2017){Leibundgut}, {Bacon}, {Jaff{\'e}},
  {Johnston}, {Kuntschner}, {Selman}, {Valenti}, {Vernet}, \&
  {Vogt}}]{Leibundgut2017}
{Leibundgut}, B., {Bacon}, R., {Jaff{\'e}}, Y.~L., {et~al.} 2017, The
  Messenger, 170, 20

\bibitem[{{Leitherer} \& {Heckman}(1995)}]{Leitherer1995}
{Leitherer}, C. \& {Heckman}, T.~M. 1995, \apjs, 96, 9

\bibitem[{{Leitherer} {et~al.}(1999){Leitherer}, {Schaerer}, {Goldader},
  {Delgado}, {Robert}, {Kune}, {de Mello}, {Devost}, \&
  {Heckman}}]{Leitherer1999}
{Leitherer}, C., {Schaerer}, D., {Goldader}, J.~D., {et~al.} 1999, \apjs, 123,
  3

\bibitem[{{Leitherer} {et~al.}(1996){Leitherer}, {Vacca}, {Conti},
  {Filippenko}, {Robert}, \& {Sargent}}]{Leitherer1996}
{Leitherer}, C., {Vacca}, W.~D., {Conti}, P.~S., {et~al.} 1996, \apj, 465, 717

\bibitem[{{Levesque} \& {Leitherer}(2013)}]{Levesque2013}
{Levesque}, E.~M. \& {Leitherer}, C. 2013, \apj, 779, 170

\bibitem[{{Lindegren} {et~al.}(2016){Lindegren}, {Lammers}, {Bastian},
  {Hern{\'a}ndez}, {Klioner}, {Hobbs}, {Bombrun}, {Michalik}, {Ramos-Lerate},
  {Butkevich}, {Comoretto}, {Joliet}, {Holl}, {Hutton}, {Parsons},
  {Steidelm{\"u}ller}, {Abbas}, {Altmann}, {Andrei}, {Anton}, {Bach},
  {Barache}, {Becciani}, {Berthier}, {Bianchi}, {Biermann}, {Bouquillon},
  {Bourda}, {Br{\"u}semeister}, {Bucciarelli}, {Busonero}, {Carlucci},
  {Casta{\~n}eda}, {Charlot}, {Clotet}, {Crosta}, {Davidson}, {de Felice},
  {Drimmel}, {Fabricius}, {Fienga}, {Figueras}, {Fraile}, {Gai}, {Garralda},
  {Geyer}, {Gonz{\'a}lez-Vidal}, {Guerra}, {Hambly}, {Hauser}, {Jordan},
  {Lattanzi}, {Lenhardt}, {Liao}, {L{\"o}ffler}, {McMillan}, {Mignard}, {Mora},
  {Morbidelli}, {Portell}, {Riva}, {Sarasso}, {Serraller}, {Siddiqui}, {Smart},
  {Spagna}, {Stampa}, {Steele}, {Taris}, {Torra}, {van Reeven}, {Vecchiato},
  {Zschocke}, {de Bruijne}, {Gracia}, {Raison}, {Lister}, {Marchant},
  {Messineo}, {Soffel}, {Osorio}, {de Torres}, \& {O'Mullane}}]{Lindegren2016}
{Lindegren}, L., {Lammers}, U., {Bastian}, U., {et~al.} 2016, \aap, 595, A4

\bibitem[{{Loeb} \& {Furlanetto}(2013)}]{Loeb2013}
{Loeb}, A. \& {Furlanetto}, S.~R. 2013, {The First Galaxies in the Universe}

\bibitem[{{Loose} \& {Thuan}(1986)}]{Loose1987}
{Loose}, H.-H. \& {Thuan}, T.~X. 1986, in Star-forming Dwarf Galaxies and
  Related Objects, 73--88

\bibitem[{{L{\'o}pez-S{\'a}nchez} {et~al.}(2012){L{\'o}pez-S{\'a}nchez},
  {Koribalski}, {van Eymeren}, {Esteban}, {Kirby}, {Jerjen}, \&
  {Lonsdale}}]{LopezSanchez2012}
{L{\'o}pez-S{\'a}nchez}, {\'A}.~R., {Koribalski}, B.~S., {van Eymeren}, J.,
  {et~al.} 2012, \mnras, 419, 1051

\bibitem[{{Mac Low} \& {Ferrara}(1999)}]{MacLow1999}
{Mac Low}, M.-M. \& {Ferrara}, A. 1999, \apj, 513, 142

\bibitem[{{Marlowe} {et~al.}(1999){Marlowe}, {Meurer}, \&
  {Heckman}}]{Marlowe1999}
{Marlowe}, A.~T., {Meurer}, G.~R., \& {Heckman}, T.~M. 1999, \apj, 522, 183

\bibitem[{{Marlowe} {et~al.}(1997){Marlowe}, {Meurer}, {Heckman}, \&
  {Schommer}}]{Marlowe1997}
{Marlowe}, A.~T., {Meurer}, G.~R., {Heckman}, T.~M., \& {Schommer}, R. 1997,
  \apjs, 112, 285

\bibitem[{{Martin}(1997)}]{Martin1997}
{Martin}, C.~L. 1997, \apj, 491, 561

\bibitem[{{Martin}(1998)}]{Martin1998}
{Martin}, C.~L. 1998, \apj, 506, 222

\bibitem[{{Mart{\'\i}nez-Delgado} {et~al.}(2012){Mart{\'\i}nez-Delgado},
  {Romanowsky}, {Gabany}, {Annibali}, {Arnold}, {Fliri}, {Zibetti}, {van der
  Marel}, {Rix}, {Chonis}, {Carballo-Bello}, {Aloisi}, {Macci{\`o}},
  {Gallego-Laborda}, {Brodie}, \& {Merrifield}}]{MartinezDelgado2012}
{Mart{\'\i}nez-Delgado}, D., {Romanowsky}, A.~J., {Gabany}, R.~J., {et~al.}
  2012, \apjl, 748, L24

\bibitem[{{McCall} {et~al.}(1985){McCall}, {Rybski}, \& {Shields}}]{McCall1985}
{McCall}, M.~L., {Rybski}, P.~M., \& {Shields}, G.~A. 1985, \apjs, 57, 1

\bibitem[{{McCray} \& {Kafatos}(1987)}]{McCray1987}
{McCray}, R. \& {Kafatos}, M. 1987, \apj, 317, 190

\bibitem[{{McKee} \& {Ostriker}(2007)}]{McKee2007}
{McKee}, C.~F. \& {Ostriker}, E.~C. 2007, \araa, 45, 565

\bibitem[{{McLeod} {et~al.}(2015){McLeod}, {Dale}, {Ginsburg}, {Ercolano},
  {Gritschneder}, {Ramsay}, \& {Testi}}]{McLeod2015}
{McLeod}, A.~F., {Dale}, J.~E., {Ginsburg}, A., {et~al.} 2015, \mnras, 450,
  1057

\bibitem[{{McQuinn} {et~al.}(2010){McQuinn}, {Skillman}, {Cannon}, {Dalcanton},
  {Dolphin}, {Hidalgo-Rodr{\'\i}guez}, {Holtzman}, {Stark}, {Weisz}, \&
  {Williams}}]{McQuinn2010}
{McQuinn}, K. B.~W., {Skillman}, E.~D., {Cannon}, J.~M., {et~al.} 2010, \apj,
  721, 297

\bibitem[{{Menacho} {et~al.}(2021){Menacho}, {Bik}, {Adamo}, {Bergvall}, {Della
  Bruna}, {Hayes}, {Melinder}, \& {Rivera-Thorsen}}]{Menacho2021}
{Menacho}, V., {Bik}, G. {\"O}.~A., {Adamo}, A., {et~al.} 2021, arXiv e-prints,
  arXiv:2105.11017

\bibitem[{{Menacho} {et~al.}(2019){Menacho}, {{\"O}stlin}, {Bik}, {Della
  Bruna}, {Melinder}, {Adamo}, {Hayes}, {Herenz}, \& {Bergvall}}]{Menacho2019}
{Menacho}, V., {{\"O}stlin}, G., {Bik}, A., {et~al.} 2019, \mnras, 487, 3183

\bibitem[{{Micheva} {et~al.}(2013{\natexlab{a}}){Micheva}, {{\"O}stlin},
  {Bergvall}, {Zackrisson}, {Masegosa}, {Marquez}, {Marquart}, \&
  {Durret}}]{Micheva2013a}
{Micheva}, G., {{\"O}stlin}, G., {Bergvall}, N., {et~al.} 2013{\natexlab{a}},
  \mnras, 431, 102

\bibitem[{{Micheva} {et~al.}(2013{\natexlab{b}}){Micheva}, {{\"O}stlin},
  {Zackrisson}, {Bergvall}, {Marquart}, {Masegosa}, {Marquez}, {Cumming}, \&
  {Durret}}]{Micheva2013b}
{Micheva}, G., {{\"O}stlin}, G., {Zackrisson}, E., {et~al.} 2013{\natexlab{b}},
  \aap, 556, A10

\bibitem[{{Miller} {et~al.}(1997){Miller}, {Whitmore}, {Schweizer}, \&
  {Fall}}]{Miller1997}
{Miller}, B.~W., {Whitmore}, B.~C., {Schweizer}, F., \& {Fall}, S.~M. 1997,
  \aj, 114, 2381

\bibitem[{{Miura} {et~al.}(2015){Miura}, {Espada}, {Sugai}, {Nakanishi}, \&
  {Hirota}}]{Miura2015}
{Miura}, R.~E., {Espada}, D., {Sugai}, H., {Nakanishi}, K., \& {Hirota}, A.
  2015, \pasj, 67, L1

\bibitem[{{Moustakas} \& {Kennicutt}(2006)}]{Moustakas2006}
{Moustakas}, J. \& {Kennicutt}, Jr., R.~C. 2006, \apjs, 164, 81

\bibitem[{{Mullan} {et~al.}(2011){Mullan}, {Konstantopoulos}, {Kepley}, {Lee},
  {Charlton}, {Knierman}, {Bastian}, {Chandar}, {Durrell}, {Elmegreen},
  {English}, {Gallagher}, {Gronwall}, {Hibbard}, {Hunsberger}, {Johnson},
  {Maybhate}, {Palma}, {Trancho}, \& {Vacca}}]{Mullan2011}
{Mullan}, B., {Konstantopoulos}, I.~S., {Kepley}, A.~A., {et~al.} 2011, \apj,
  731, 93

\bibitem[{{Naab} \& {Ostriker}(2017)}]{Naab2017}
{Naab}, T. \& {Ostriker}, J.~P. 2017, \araa, 55, 59

\bibitem[{{Nidever} {et~al.}(2013){Nidever}, {Ashley}, {Slater}, {Ott},
  {Johnson}, {Bell}, {Stanimirovi{\'c}}, {Putman}, {Majewski}, {Simpson},
  {J{\"u}tte}, {Oosterloo}, \& {Butler Burton}}]{Nidever2013}
{Nidever}, D.~L., {Ashley}, T., {Slater}, C.~T., {et~al.} 2013, \apjl, 779, L15

\bibitem[{{Noeske} {et~al.}(2003){Noeske}, {Papaderos}, {Cair{\'o}s}, \&
  {Fricke}}]{Noeske2003}
{Noeske}, K.~G., {Papaderos}, P., {Cair{\'o}s}, L.~M., \& {Fricke}, K.~J. 2003,
  \aap, 410, 481

\bibitem[{{Olofsson}(1995)}]{Olofsson1995}
{Olofsson}, K. 1995, \aaps, 111, 57

\bibitem[{{Osterbrock} \& {Ferland}(2006)}]{Osterbrock2006}
{Osterbrock}, D.~E. \& {Ferland}, G.~J. 2006, {Astrophysics of gaseous nebulae
  and active galactic nuclei} (2nd.~ed.~by D.E.~Osterbrock and
  G.J.~Ferland.~Sausalito, CA: University Science Books, 2006)

\bibitem[{{{\"O}stlin} {et~al.}(2001){{\"O}stlin}, {Amram}, {Bergvall},
  {Masegosa}, {Boulesteix}, \& {M{\'a}rquez}}]{Ostlin2001}
{{\"O}stlin}, G., {Amram}, P., {Bergvall}, N., {et~al.} 2001, \aap, 374, 800

\bibitem[{{\"Ostlin} {et~al.}(1998){\"Ostlin}, {Bergvall}, \&
  {Roennback}}]{Ostlin1998b}
{\"Ostlin}, G., {Bergvall}, N., \& {Roennback}, J. 1998, \aap, 335, 85

\bibitem[{{Papaderos} {et~al.}(1998){Papaderos}, {Izotov}, {Fricke}, {Thuan},
  \& {Guseva}}]{Papaderos1998}
{Papaderos}, P., {Izotov}, Y.~I., {Fricke}, K.~J., {Thuan}, T.~X., \& {Guseva},
  N.~G. 1998, \aap, 338, 43

\bibitem[{{Papaderos} {et~al.}(2002){Papaderos}, {Izotov}, {Thuan}, {Noeske},
  {Fricke}, {Guseva}, \& {Green}}]{Papaderos2002}
{Papaderos}, P., {Izotov}, Y.~I., {Thuan}, T.~X., {et~al.} 2002, \aap, 393, 461

\bibitem[{{Papaderos} {et~al.}(1996){Papaderos}, {Loose}, {Thuan}, \&
  {Fricke}}]{Papaderos1996a}
{Papaderos}, P., {Loose}, H.-H., {Thuan}, T.~X., \& {Fricke}, K.~J. 1996,
  \aaps, 120, 207

\bibitem[{{Paudel} {et~al.}(2015){Paudel}, {Duc}, \& {Ree}}]{Paudel2015}
{Paudel}, S., {Duc}, P.~A., \& {Ree}, C.~H. 2015, \aj, 149, 114

\bibitem[{{Paudel} {et~al.}(2018){Paudel}, {Smith}, {Yoon},
  {Calder{\'o}n-Castillo}, \& {Duc}}]{Paudel2018}
{Paudel}, S., {Smith}, R., {Yoon}, S.~J., {Calder{\'o}n-Castillo}, P., \&
  {Duc}, P.-A. 2018, \apjs, 237, 36

\bibitem[{{Pearson} {et~al.}(2018){Pearson}, {Privon}, {Besla}, {Putman},
  {Mart{\'\i}nez-Delgado}, {Johnston}, {Gabany}, {Patton}, \&
  {Kallivayalil}}]{Pearson2018}
{Pearson}, S., {Privon}, G.~C., {Besla}, G., {et~al.} 2018, \mnras, 480, 3069

\bibitem[{{Portegies Zwart} {et~al.}(2010){Portegies Zwart}, {McMillan}, \&
  {Gieles}}]{PortegiesZwart2010}
{Portegies Zwart}, S.~F., {McMillan}, S. L.~W., \& {Gieles}, M. 2010, \araa,
  48, 431

\bibitem[{{Privon} {et~al.}(2017){Privon}, {Stierwalt}, {Patton}, {Besla},
  {Pearson}, {Putman}, {Johnson}, {Kallivayalil}, {Liss}, \&
  {Titans}}]{Privon2017}
{Privon}, G.~C., {Stierwalt}, S., {Patton}, D.~R., {et~al.} 2017, \apj, 846, 74

\bibitem[{{Pustilnik} {et~al.}(2001){Pustilnik}, {Kniazev}, {Lipovetsky}, \&
  {Ugryumov}}]{Pustilnik2001}
{Pustilnik}, S.~A., {Kniazev}, A.~Y., {Lipovetsky}, V.~A., \& {Ugryumov}, A.~V.
  2001, \aap, 373, 24

\bibitem[{{Sacchi} {et~al.}(2021){Sacchi}, {Aloisi}, {Correnti}, {Annibali},
  {Tosi}, {Garofalo}, {Clementini}, {Cignoni}, {James}, {Marconi}, {Muraveva},
  \& {van der Marel}}]{Sacchi2021}
{Sacchi}, E., {Aloisi}, A., {Correnti}, M., {et~al.} 2021, \apj, 911, 62

\bibitem[{{Salzer} {et~al.}(2002){Salzer}, {Rosenberg}, {Weisstein},
  {Mazzarella}, \& {Bothun}}]{Salzer2002}
{Salzer}, J.~J., {Rosenberg}, J.~L., {Weisstein}, E.~W., {Mazzarella}, J.~M.,
  \& {Bothun}, G.~D. 2002, \aj, 124, 191

\bibitem[{{Schlafly} \& {Finkbeiner}(2011)}]{Schlafly2011}
{Schlafly}, E.~F. \& {Finkbeiner}, D.~P. 2011, \apj, 737, 103

\bibitem[{{Schulte-Ladbeck} {et~al.}(1999{\natexlab{a}}){Schulte-Ladbeck},
  {Hopp}, {Crone}, \& {Greggio}}]{SchulteLadbeck1999a}
{Schulte-Ladbeck}, R.~E., {Hopp}, U., {Crone}, M.~M., \& {Greggio}, L.
  1999{\natexlab{a}}, \apj, 525, 709

\bibitem[{{Schulte-Ladbeck} {et~al.}(1999{\natexlab{b}}){Schulte-Ladbeck},
  {Hopp}, {Greggio}, \& {Crone}}]{SchulteLadbeck1999b}
{Schulte-Ladbeck}, R.~E., {Hopp}, U., {Greggio}, L., \& {Crone}, M.~M.
  1999{\natexlab{b}}, \aj, 118, 2705

\bibitem[{{Schulte-Ladbeck} {et~al.}(2000){Schulte-Ladbeck}, {Hopp}, {Greggio},
  \& {Crone}}]{SchulteLadbeck2000}
{Schulte-Ladbeck}, R.~E., {Hopp}, U., {Greggio}, L., \& {Crone}, M.~M. 2000,
  \aj, 120, 1713

\bibitem[{{Sharp} \& {Parkinson}(2010)}]{SharpParkinson2010}
{Sharp}, R. \& {Parkinson}, H. 2010, \mnras, 408, 2495

\bibitem[{{Somerville} \& {Dav{\'e}}(2015)}]{Somerville2015}
{Somerville}, R.~S. \& {Dav{\'e}}, R. 2015, \araa, 53, 51

\bibitem[{{Soto} {et~al.}(2016){Soto}, {Lilly}, {Bacon}, {Richard}, \&
  {Conseil}}]{Soto2016}
{Soto}, K.~T., {Lilly}, S.~J., {Bacon}, R., {Richard}, J., \& {Conseil}, S.
  2016, \mnras, 458, 3210

\bibitem[{{Springel} {et~al.}(2006){Springel}, {Frenk}, \&
  {White}}]{Springel2006}
{Springel}, V., {Frenk}, C.~S., \& {White}, S.~D.~M. 2006, \nat, 440, 1137

\bibitem[{{Springel} \& {Hernquist}(2003)}]{Springel2003}
{Springel}, V. \& {Hernquist}, L. 2003, \mnras, 339, 312

\bibitem[{{Stierwalt} {et~al.}(2015){Stierwalt}, {Besla}, {Patton}, {Johnson},
  {Kallivayalil}, {Putman}, {Privon}, \& {Ross}}]{Stierwalt2015}
{Stierwalt}, S., {Besla}, G., {Patton}, D., {et~al.} 2015, \apj, 805, 2

\bibitem[{{Struck}(2011)}]{Struck2011}
{Struck}, C. 2011, {Galaxy Collisions}

\bibitem[{{Taylor}(1997)}]{Taylor1997}
{Taylor}, C.~L. 1997, \apj, 480, 524

\bibitem[{{Taylor} {et~al.}(1996){Taylor}, {Thomas}, {Brinks}, \&
  {Skillman}}]{Taylor1996}
{Taylor}, C.~L., {Thomas}, D.~L., {Brinks}, E., \& {Skillman}, E.~D. 1996,
  \apjs, 107, 143

\bibitem[{{Theuns} {et~al.}(2002){Theuns}, {Viel}, {Kay}, {Schaye}, {Carswell},
  \& {Tzanavaris}}]{Theuns2002}
{Theuns}, T., {Viel}, M., {Kay}, S., {et~al.} 2002, \apjl, 578, L5

\bibitem[{{Thilker} {et~al.}(2000){Thilker}, {Braun}, \&
  {Walterbos}}]{Thilker2000}
{Thilker}, D.~A., {Braun}, R., \& {Walterbos}, R. A.~M. 2000, \aj, 120, 3070

\bibitem[{{Thuan} \& {Martin}(1981)}]{ThuanMartin1981}
{Thuan}, T.~X. \& {Martin}, G.~E. 1981, \apj, 247, 823

\bibitem[{{Tolstoy} {et~al.}(2009){Tolstoy}, {Hill}, \& {Tosi}}]{Tolstoy2009}
{Tolstoy}, E., {Hill}, V., \& {Tosi}, M. 2009, \araa, 47, 371

\bibitem[{{Tomisaka} {et~al.}(1981){Tomisaka}, {Habe}, \&
  {Ikeuchi}}]{Tomisaka1981}
{Tomisaka}, K., {Habe}, A., \& {Ikeuchi}, S. 1981, \apss, 78, 273

\bibitem[{{Toomre} \& {Toomre}(1972)}]{Toomre1972}
{Toomre}, A. \& {Toomre}, J. 1972, \apj, 178, 623

\bibitem[{{Tremonti} {et~al.}(2004){Tremonti}, {Heckman}, {Kauffmann},
  {Brinchmann}, {Charlot}, {White}, {Seibert}, {Peng}, {Schlegel}, {Uomoto},
  {Fukugita}, \& {Brinkmann}}]{Tremonti2004}
{Tremonti}, C.~A., {Heckman}, T.~M., {Kauffmann}, G., {et~al.} 2004, \apj, 613,
  898

\bibitem[{{Turner} {et~al.}(2015){Turner}, {Beck}, {Benford}, {Consiglio},
  {Ho}, {Kov{\'a}cs}, {Meier}, \& {Zhao}}]{Turner2015}
{Turner}, J.~L., {Beck}, S.~C., {Benford}, D.~J., {et~al.} 2015, \nat, 519, 331

\bibitem[{{Vanzi} {et~al.}(2011){Vanzi}, {Cresci}, {Sauvage}, \&
  {Thompson}}]{Vanzi2011}
{Vanzi}, L., {Cresci}, G., {Sauvage}, M., \& {Thompson}, R. 2011, \aap, 534,
  A70

\bibitem[{{Vanzi} {et~al.}(2008){Vanzi}, {Cresci}, {Telles}, \&
  {Melnick}}]{Vanzi2008}
{Vanzi}, L., {Cresci}, G., {Telles}, E., \& {Melnick}, J. 2008, \aap, 486, 393

\bibitem[{{Vazdekis} {et~al.}(1996){Vazdekis}, {Casuso}, {Peletier}, \&
  {Beckman}}]{Vazdekis1996}
{Vazdekis}, A., {Casuso}, E., {Peletier}, R.~F., \& {Beckman}, J.~E. 1996,
  \apjs, 106, 307

\bibitem[{{Vazdekis} {et~al.}(2010){Vazdekis}, {S{\'a}nchez-Bl{\'a}zquez},
  {Falc{\'o}n-Barroso}, {Cenarro}, {Beasley}, {Cardiel}, {Gorgas}, \&
  {Peletier}}]{Vazdekis2010}
{Vazdekis}, A., {S{\'a}nchez-Bl{\'a}zquez}, P., {Falc{\'o}n-Barroso}, J.,
  {et~al.} 2010, \mnras, 404, 1639

\bibitem[{{Veilleux} {et~al.}(2005){Veilleux}, {Cecil}, \&
  {Bland-Hawthorn}}]{Veilleux2005}
{Veilleux}, S., {Cecil}, G., \& {Bland-Hawthorn}, J. 2005, \araa, 43, 769

\bibitem[{{Vu{\v{c}}eti{\'c}} {et~al.}(2013){Vu{\v{c}}eti{\'c}}, {Arbutina},
  {Urosevic}, {Dobardzic}, {Pavlovic}, {Pannuti}, \& {Petrov}}]{Vucetic2013}
{Vu{\v{c}}eti{\'c}}, M.~M., {Arbutina}, B., {Urosevic}, D., {et~al.} 2013,
  Serbian Astronomical Journal, 187, 11

\bibitem[{{Vu{\v{c}}eti{\'c}} {et~al.}(2015){Vu{\v{c}}eti{\'c}},
  {{\'C}iprijanovi{\'c}}, {Pavlovi{\'c}}, {Pannuti}, {Petrov}, {G{\"o}ker}, \&
  {Ercan}}]{Vucetic2015}
{Vu{\v{c}}eti{\'c}}, M.~M., {{\'C}iprijanovi{\'c}}, A., {Pavlovi{\'c}}, M.~Z.,
  {et~al.} 2015, Serbian Astronomical Journal, 191, 67

\bibitem[{{Vu{\v{c}}eti{\'c}} {et~al.}(2019){Vu{\v{c}}eti{\'c}}, {Oni{\'c}},
  {Petrov}, {{\'C}iprijanovi{\'c}}, \& {Pavlovi{\'c}}}]{Vucetic2019}
{Vu{\v{c}}eti{\'c}}, M.~M., {Oni{\'c}}, D., {Petrov}, N.,
  {{\'C}iprijanovi{\'c}}, A., \& {Pavlovi{\'c}}, M.~Z. 2019, Serbian
  Astronomical Journal, 198, 13

\bibitem[{{Walter}(1999)}]{Walter1999}
{Walter}, F. 1999, \pasa, 16, 106

\bibitem[{{Weilbacher} {et~al.}(2020){Weilbacher}, {Palsa}, {Streicher},
  {Bacon}, {Urrutia}, {Wisotzki}, {Conseil}, {Husemann}, {Jarno}, {Kelz},
  {P{\'e}contal-Rousset}, {Richard}, {Roth}, {Selman}, \&
  {Vernet}}]{Weilbacher2020}
{Weilbacher}, P.~M., {Palsa}, R., {Streicher}, O., {et~al.} 2020, \aap, 641,
  A28

\bibitem[{{Weilbacher} {et~al.}(2016){Weilbacher}, {Streicher}, \&
  {Palsa}}]{Weilbacher2016}
{Weilbacher}, P.~M., {Streicher}, O., \& {Palsa}, R. 2016, {MUSE-DRP: MUSE Data
  Reduction Pipeline}, Astrophysics Source Code Library

\bibitem[{{Whitmore}(2003)}]{Whitmore2003}
{Whitmore}, B. 2003, in Extragalactic Globular Cluster Systems, ed.
  M.~{Kissler-Patig}, 336

\bibitem[{{Whitmore} {et~al.}(1995){Whitmore}, {Sparks}, {Lucas}, {Macchetto},
  \& {Biretta}}]{Whitmore1995}
{Whitmore}, B.~C., {Sparks}, W.~B., {Lucas}, R.~A., {Macchetto}, F.~D., \&
  {Biretta}, J.~A. 1995, \apjl, 454, L73

\bibitem[{{Whitmore} {et~al.}(1999){Whitmore}, {Zhang}, {Leitherer}, {Fall},
  {Schweizer}, \& {Miller}}]{Whitmore1999}
{Whitmore}, B.~C., {Zhang}, Q., {Leitherer}, C., {et~al.} 1999, \aj, 118, 1551

\bibitem[{{Wild} \& {Hewett}(2005)}]{Wild2005}
{Wild}, V. \& {Hewett}, P.~C. 2005, \mnras, 358, 1083

\bibitem[{{Xu} {et~al.}(2020){Xu}, {Liu}, {Jing}, {Wang}, \& {Lu}}]{Xu2020}
{Xu}, K., {Liu}, C., {Jing}, Y., {Wang}, Y., \& {Lu}, S. 2020, \apj, 895, 100

\bibitem[{{Zackrisson} {et~al.}(2001){Zackrisson}, {Bergvall}, {Olofsson}, \&
  {Siebert}}]{Zackrisson2001}
{Zackrisson}, E., {Bergvall}, N., {Olofsson}, K., \& {Siebert}, A. 2001, \aap,
  375, 814

\bibitem[{{Zhang} {et~al.}(2020{\natexlab{a}}){Zhang}, {Paudel}, {Smith},
  {Duc}, {Puzia}, {Peng}, {C{\^o}te}, {Ferrarese}, {Boselli}, {Wang}, \&
  {Oh}}]{Zhang2020a}
{Zhang}, H.-X., {Paudel}, S., {Smith}, R., {et~al.} 2020{\natexlab{a}}, \apjl,
  891, L23

\bibitem[{{Zhang} {et~al.}(2020{\natexlab{b}}){Zhang}, {Smith}, {Oh}, {Paudel},
  {Duc}, {Boselli}, {C{\^o}t{\'e}}, {Ferrarese}, {Gao}, {Hunter}, {Puzia},
  {Peng}, {Rong}, {Shin}, \& {Zhao}}]{Zhang2020b}
{Zhang}, H.-X., {Smith}, R., {Oh}, S.-H., {et~al.} 2020{\natexlab{b}}, \apj,
  900, 152

\bibitem[{{Zheng} {et~al.}(2004{\natexlab{a}}){Zheng}, {Hammer}, {Flores},
  {Ass{\'e}mat}, \& {Pelat}}]{Zheng2004}
{Zheng}, X.~Z., {Hammer}, F., {Flores}, H., {Ass{\'e}mat}, F., \& {Pelat}, D.
  2004{\natexlab{a}}, \aap, 421, 847

\bibitem[{{Zheng} {et~al.}(2004{\natexlab{b}}){Zheng}, {Hammer}, {Flores},
  {Ass{\'e}mat}, \& {Pelat}}]{Hammer2004}
{Zheng}, X.~Z., {Hammer}, F., {Flores}, H., {Ass{\'e}mat}, F., \& {Pelat}, D.
  2004{\natexlab{b}}, \aap, 421, 847

\end{thebibliography}


\begin{appendix}
\section{Background galaxies}
\label{backgroundgalaxies}

\begin{figure*}
\centering
\includegraphics[angle=0, width=\linewidth]{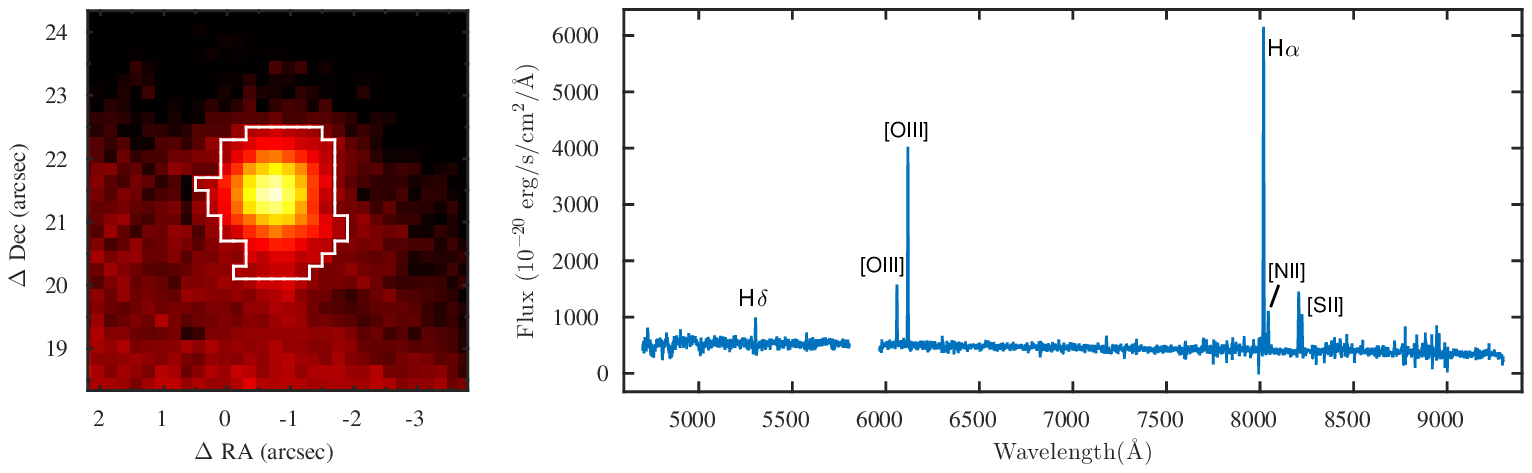}
\caption{Background galaxy (BgGx1) detected to the north and its flux-calibrated spectrum. The most prominent features are marked.}
\label{Figure:RegionR19_spec} 
\end{figure*}

\begin{figure*}
\centering
\includegraphics[angle=0, width=\linewidth]{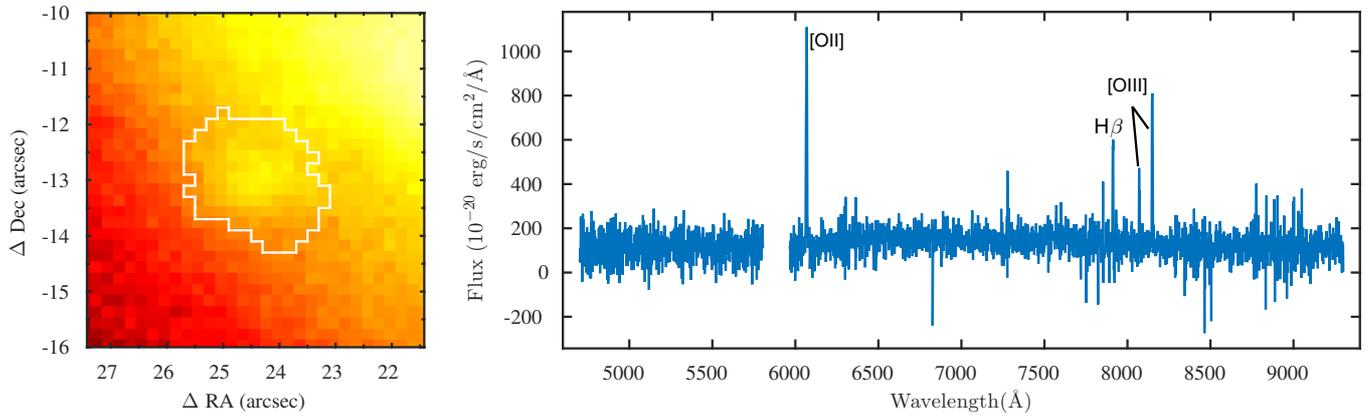}
\caption{Background galaxy (BgGx2) detected to the southwest and its flux-calibrated spectrum. The most prominent features are marked.}
\label{Figure:RegionB41_spec} 
\end{figure*}

Two of the clumps in the continuum were identified as background galaxies.
The first one (BgGx1) is placed to the north  and the second one (BgGx2) to the southeast of the main galaxy body (see Figure~\ref{Figure:Haro14_broadband}).  
We derived their physical parameters and they are shown in Table~\ref{tab:bkgalaxy}. 
Figures~\ref{Figure:RegionR19_spec} and \ref{Figure:RegionB41_spec} show
their continuum maps and spectra.

\begin{table}
\begin{center}
\caption{Basic parameters for the two background galaxies in the observed field. 
\label{tab:bkgalaxy}}
\begin{tabular}{lcc}
\hline
Parameter   & BgGx1 & BgGx2 \\ 
\hline
\\
$\Delta$RA,$\Delta$Dec                              &  (-0.6$\arcsec$, 21.2$\arcsec$) &    (24.4$\arcsec$, -13.2$\arcsec$) \\
m$_{V}$                                                     &  22.09 & 23.71 \\
M$_{V}$                                                      &   -18.20 & -19.24 \\
(V-I)                                                              &  1.06 &  1.39 \\
Redshift                                                        &  0.222    & 0.628   \\
 \hline
\end{tabular}
\end{center}
\small Notes: (1) $\Delta$RA,$\Delta$Dec are offsets from the position of the continuum peak. 
\end{table}

\end{appendix}

\end{document}